%% file: falcon_paper.tex
\documentclass[twocolumn]{svjour3}          

\smartqed  

\usepackage{pgfplots}
\usetikzlibrary{arrows}
\usepackage{verbatim}
\usetikzlibrary{arrows, decorations.markings}
\usetikzlibrary{positioning}
\usetikzlibrary{patterns}

\usepackage{xcolor}
\definecolor{chartblue}{RGB}{0,163,255} 
\definecolor{chartred}{RGB}{255,0,0} 
\definecolor{chartgreen}{RGB}{47, 227, 129}

\pgfplotscreateplotcyclelist{barplot cycle list}{%
	{black,fill=chartblue},
    {black,fill=chartred}, 
    {black,fill=chartgreen},
	{black,fill=white,postaction={pattern=north east lines,pattern color=chartblue}},
    {black,fill=white,postaction={pattern=north west lines,pattern color=chartred}},
    {black,fill=white,postaction={pattern=crosshatch,pattern color=chartgreen}},
    {black,fill=white,postaction={pattern=crosshatch dots,pattern color=black}},
}
\usepackage{tcolorbox}
\usepackage{algorithm}
\usepackage{algorithmic}
\usepackage{subfigure}
\usepackage{graphicx}
\usepackage{multirow}
\usepackage{xspace}
\usepackage{import}
\usepackage{transparent}
\usepackage{array}
\usepackage{listings}
\usepackage{float}
\usepackage{subfloat}
\usepackage{todonotes}
\usepackage{enumitem}    
\usepackage{enumerate}
\usepackage{paralist}
\usepackage{url}

\usepackage{colortbl}

\usepackage{tikz}
\usepackage{pgfplots}
\usetikzlibrary{pgfplots.groupplots}
\usetikzlibrary{patterns}
\usetikzlibrary{matrix,positioning}
\usepgfplotslibrary{groupplots}
\usepgfplotslibrary{statistics}

\usepackage{wrapfig} 
\pgfdeclarelayer{background}
\pgfdeclarelayer{foreground}
\pgfsetlayers{background,main,foreground}

\usepackage{flushend}

\input{listing_color_line_background.tex}

\newcommand{\barchartheight}[0]{0.15\textwidth}
\newcommand{\barchartfontsize}[0]{\scriptsize}
\definecolor{applegreen}{rgb}{0.0, 0.5, 0.0}
\definecolor{myblue}{RGB}{0,163,255}
\definecolor{myblue2}{RGB}{91,196,255}
\definecolor{mygreen}{RGB}{167,243,188}

\newcommand{\setdslcolor}[0]{\color{blue}}

\input{grouped_aggregation_query_1.csv.tex}
\input{grouped_aggregation_query_2_many_groups.csv.tex}
\input{projection-query-1-medium-selectivity.csv.tex}
\input{projection-query-2-high-selectivity.csv.tex}
\input{query_ssb11.csv.tex}
\input{query_ssb21.csv.tex}
\input{query_ssb31.csv.tex}
\input{ssb41.csv.tex}
\input{tpch_query1.csv.tex}
\definecolor{darkgreen}{RGB}{0,128,0}
\newcommand{\markchange}[0]{\color{black}}
\newcommand{\reviewer}[1]{}
\newcommand{\resolved}[1]{}

\newcommand{\ch}[0]{\color{black}}

\newcommand{\cogadb}[0]{CoGaDB\xspace}

\begin{document}
%
\title{Generating Custom Code for Efficient Query Execution on Heterogeneous Processors}


\author{\mbox{Sebastian Bre\ss\ \boldmath$\cdot$ Bastian K{\"o}cher \boldmath$\cdot$
        Henning Funke      \boldmath$\cdot$
        Tilmann Rabl       \boldmath$\cdot$
        Volker Markl}
}


\institute{Sebastian Bre\ss \at
              DFKI GmbH and TU Berlin
              \email{sebastian.bress@dfki.de}
           \and
           Bastian K{\"o}cher \at
           TU Berlin
           \email{bastian.koecher@tu-berlin.de}               
           \and
           Henning Funke\at
           TU Dortmund
           \email{henning.funke@tu-dortmund.de}     
           \and
           Tilmann Rabl \at
           TU Berlin and DFKI GmbH
           \email{rabl@tu-berlin.de}                    
           \and
           Volker Markl \at
           TU Berlin and DFKI GmbH
           \email{volker.markl@tu-berlin.de}  
}

%

\maketitle




\begin{abstract}
{
Processor manufacturers build increasingly specialized processors to mitigate the effects of the power wall to deliver improved performance. 
Currently, database engines are manually optimized for each processor: A costly and error prone process.

In this paper, we propose concepts to enable the database engine to perform per-processor optimization automatically.
Our core idea is to create variants of generated code and to learn a fast variant for each processor.  
We create variants by modifying parallelization strategies, specializing data structures, and applying different code transformations.

Our experimental results show that the performance of variants may diverge up to two orders of magnitude. 
Therefore, we need to generate custom code for each processor to achieve peak performance.
We show that our approach finds a fast custom variant for multi-core CPUs, GPUs, and MICs.

}
\end{abstract}


\newpage
\section{Introduction}



\begin{figure}[t!]
\begin{center}
 \includegraphics[width=\linewidth]{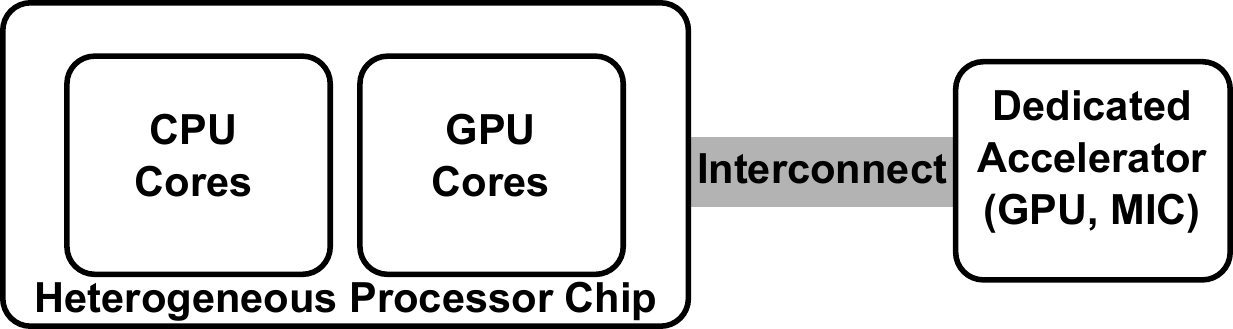}
\end{center}
\caption{Modern processors expose heterogeneity in the form of heterogeneous cores located on the same processor chip or specialized accelerator cards.}
\label{pic:future_heterogeneous_processor}
\end{figure}

{ 
The design of modern processors is primarily limited by a fixed energy budget per chip. This \emph{power wall} forces vendors to explore new processor designs to stay in the energy budget~\cite{borkar2011future,Esmaeilzadeh:2011:DSE:2000064.2000108}. 
%
%
One trend integrates heterogeneous processor cores on the same chip, e.g., combining CPU and GPU cores on the same chip as in Intel's processors with HD Graphics and AMD's Accelerated Processing Units (APUs).
Another trend is \emph{specialization}: processors are optimized for certain tasks, which already became commodity in the form of \emph{Graphics Processing Units} (GPUs), \emph{Multiple Integrated Cores} (MICs), or \emph{Field-Program\-mable Gate Arrays} (FPGAs). 
These accelerators promise large performance gains because of their additional computational power and memory bandwidth. 
As a direct consequence of the power wall, current machines are built with a set of heterogeneous processors. 
Thus, from a processor design perspective, the \emph{homogeneous many core age} ends~\cite{borkar2011future,Zahran:2017:HCH:3055102.3024918}. 
The upcoming \emph{heterogeneous many core age} forces data\-base systems to embrace processor heterogeneity to ach\-ieve peak performance. 
We show such a heterogeneous processor system in Figure~\ref{pic:future_heterogeneous_processor}.
}


{
Parallel programming APIs such as OpenCL allow us to run single operators on a wide range of processors. However, we still need to customize the operator implementation for peak performance~\cite{rosenfeld:2015:variant_selection}. 
In general, we need to select the most suitable parallelization strategies, data structures and code transformations for best performance. 


Previous solutions either focused on generating highly efficient code for a single processor~\cite{Neumann:2011:ECE:2002938.2002940,Wu:2012:KWA:2457472.2457490} or allowed database operators to run on multiple processors using the same operator code~\cite{ocelot2013,Zhang:2013:OTP:2536274.2536319}. 
Code generation approaches suffered from a high compilation time, or {\ch{}were confined} to a single processor by generating low-level machine code (e.g., LLVM~\cite{Neumann:2011:ECE:2002938.2002940}). 
Hardware-oblivious approaches suffer from {\ch{}limited} performance portability~\cite{rosenfeld:2015:variant_selection}. As of now, we need to manually adapt the database system to every new processor (e.g., for Intels MIC architecture) to achieve peak performance. 


{
Our long term goal is to enable database systems to automatically generate efficient code for any processor without \emph{any a priori} hardware knowledge. To achieve this goal, we propose Hawk, a hardware-adaptive query compiler, which can generate variants of generated code. 
By executing different variants of a compiled query, Hawk can adapt to a wide range of different processors without any manual tuning.
By compiling queries to OpenCL kernels, Hawk achieves low compilation times and can run {\color{black}queries} on any OpenCL-capable processor.

}
In this paper, we make the following contributions: 
{
\begin{compactenum}
\item We introduce \emph{pipeline programs}, a new form of physical query plan. {Pipeline programs store operations and implementation properties of a pipeline and are the basis for code generation.}
 Furthermore, we demonstrate how we can systematically generate variants of pipeline programs (cf. Section 3).
\item We discuss the dimensions in which we can vary the generated code of \emph{pipeline programs} (cf. Section 4, 5, 6) and show the impact of these variations on common processors and co-processors.
\item We compile all operators of a pipeline to kernel programs. These kernels can be compiled to a broad set of heterogeneous processors using parallel programming libraries such as OpenCL (cf. Section 7).
\item We present a learning strategy to automatically derive efficient variant configurations and incorporate them into a query optimizer (cf. Section 8).
\item We show the potential of a database system that rewrites its code until it runs efficiently on the underlying heterogeneous processor hardware (cf. Section 9).
\end{compactenum}
}

\section{Background}

{ 
In this section, we discuss the background required for the remainder of the paper. {Since we directly build on the produce/consume model of Neumann~\cite{Neumann:2011:ECE:2002938.2002940}, we first provide an overview of query compilation using the produce/consume model}. Second, we briefly discuss OpenCL, as it is the target language for our code generation.
}


\begin{figure}
 \begin{center}
    \includegraphics[width=0.8\linewidth]{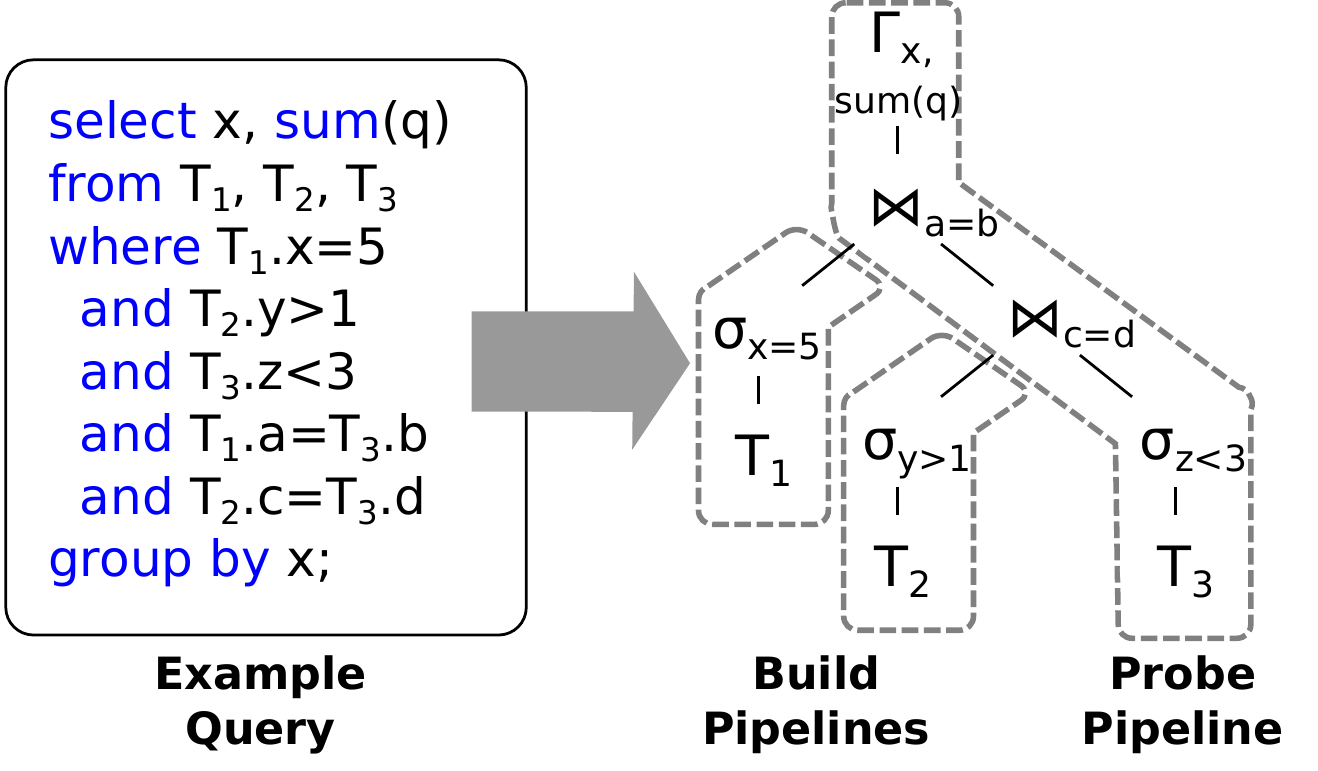}
 \end{center}
 \caption{Example for produce/consume style query compilation. The query plan is partitioned into three operator pipelines, two for building join hash tables, and one for probing both join hash tables.} 
 \label{fig:produce_consume_example}
\end{figure}

\subsection{Query Compilation}

{
For our kernel compilation approach, we build on the produce/consume model for code generation by Neumann~\cite{Neumann:2011:ECE:2002938.2002940}. 
{\color{black}
The goal of the produce/consume model is to partition a query into pipelines and to merge the operators belonging to the same pipe\-line into a single code fragment that implements the pipeline. 

This code fragment iterates in a tight for-loop over all tuples of the input relation. Each tuple is pushed through all operators of the pipeline, before the next tuple is processed. 
This code generation achieves excellent data locality by keeping the tuple in the processor registers.
}

In the produce/consume model, each operator needs to provide a produce and a consume function. The \emph{produce} function traverses the query plan top down from the root operator and creates a new pipeline for every pipeline breaking operator. When produce reaches a scan, the consume function of succeeding operators is called bottom up and generates the code for each operator in the current pipeline until a pipeline boundary is reached. Then, the code for the next pipeline is generated. Thus, the produce functions essentially partition the query plan into pipelines, whereas the consume functions fill the pipelines with operators and generate the code. 
In this paper, we refer to pipelines that were filled with operators by their consume functions as \emph{operator pipelines}. 



We illustrate the { produce/consume style query compilation} in Figure~\ref{fig:produce_consume_example}. We evaluate a query with two joins using hash joins. The hash tables {\color{black}are built on the left sub plans. As} building a hash table is a pipeline breaker, the produce/consume model creates two build pipelines on table T$_1$ and T$_2$ and one probe pipeline on table T$_3$. 
}

{ 
\subsection{Overview of OpenCL}
\label{sec:opencl}

The \emph{Open Compute Language} (OpenCL) is a framework for massively parallel computing, which supports processors with different architectures to achieve functional portability~\cite{gaster2012heterogeneous}. Functional portability means that an OpenCL program developed for any OpenCL-capable processor will run on any other OpenCL-capable processor (e.g., a OpenCL program written for a GPU can also run on a CPU). 
OpenCL abstracts all processors as so-called devices. The CPU that executes the OpenCL API functions is called the host.
All computations are expressed in special functions called \emph{kernels}, which are then compiled just-in-time for all devices. 
The just-in-time compilation abilities make OpenCL especially interesting for our work, as it provides a native mechanism to compile generated code to a device. Furthermore, the functional portability allows us to run any variant of generated code on any device.

}

\begin{figure}
 \begin{center}
    \includegraphics[width=0.8\linewidth]{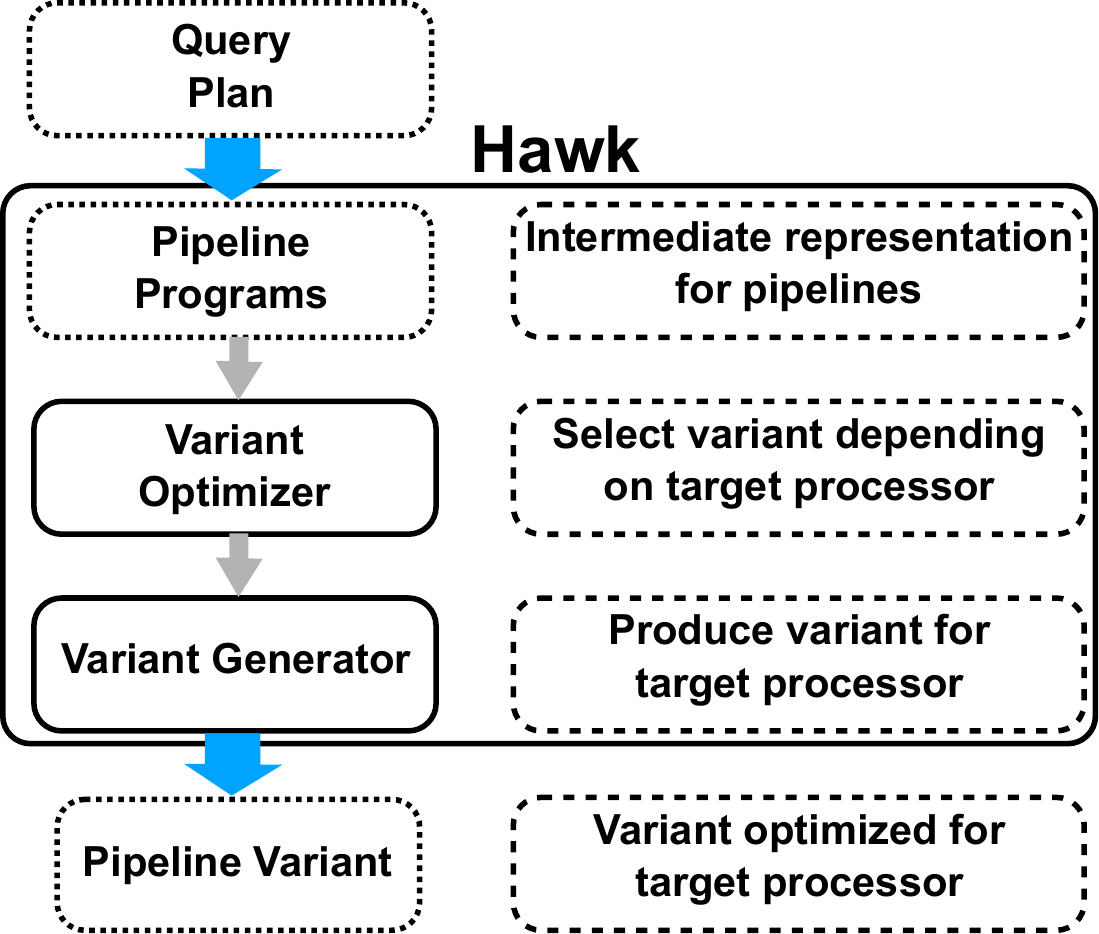}
 \end{center}
 \caption{Core concepts of  Hawk and their role in the system.} 
 \label{fig:falcon_architecture}
\end{figure}

\section{Hawk Compiler Architecture}

\resolved{ Improve the presentation to emphasize key novel contributions, and implementation lessons separately.}

{

In this section, we discuss how Hawk generates custom code for each processor by producing \emph{variants} of operator pipelines (cf. Figure~\ref{fig:falcon_architecture}). 
As a core concept, Hawk represents operator pipelines of queries as \emph{pipeline programs}. We use these programs as an intermediate representation that defines the semantics of a pipeline and serves as basis for target code generation.

Hawk produces variants along three major dimensions: execution strategy, data structure optimization, and code transformations. Variations in these dimensions are implemented either by transforming or by re-parame\-terizing pipeline programs. We can freely combine individual variations, this way spanning a large space of potential variants. 
The \emph{variant optimizer} selects a variant configuration from this space. 
Then, the \emph{variant generator} produces the code variant selected by the optimizer for the specified processor.

This architecture allows Hawk to create custom per-processor code for each pipeline in a query plan.

}

\begin{table*}
\scriptsize
\begin{tabular}{|m{5cm}|m{11.5cm}|} \hline
Pipeline Operation & Description \\ \hline\hline
{\setdslcolor LOOP}($T$; \textbf{step}, \textbf{s}, \textbf{e})  & {\raggedright Iterate over input tuples of table $T$, using a loop increment of \textbf{step}, and a loop start index \textbf{s} and end index \textbf{e}} \\
\hline
{\setdslcolor FILTER}($F_{\sigma}$; \textbf{m}, \textbf{o}) & Apply a filter condition $F_{\sigma}$ using a branching mode \textbf{m} and an element access offset \textbf{o}\\
\hline
{\setdslcolor HASH\_PUT}($A$; \textbf{h}, \textbf{p}) &  Insert tuple in a hash table for attribute set $A$ using hash table \textbf{h} with parameters \textbf{p}\\
\hline
{\setdslcolor HASH\_PROBE}($A$; \textbf{h}, \textbf{p}, \textbf{m}, \textbf{o})  & Probe hash table for attribute set $A$ using hash table \textbf{h} with parameters \textbf{p}, branching mode \textbf{m}, and element access offset \textbf{o}\\                      
\hline
{\setdslcolor CROSS\_JOIN}($T$) & Loop over additional input table $T$ and compute cross product of current tuple and $T$\\

\hline
{\setdslcolor ARITHMETIC}($f$; \textbf{o})   & Apply a computation $f: A \times B \rightarrow C$ of attributes $A$, $B$, $C$ using element access offset \textbf{o})  \\ 
\hline
{\setdslcolor AGGREGATE}($F$; \textbf{m}, \textbf{o})  & Apply a non-grouping aggregation with aggregation expression $F=(f_1, f_2, \cdots, f_n$), branching mode \textbf{m}, and element access offset \textbf{o} \\ 
\hline
{\setdslcolor HASH\_AGGREGATE}($G$,$F$; \textbf{h}, \textbf{p}, \textbf{m}, \textbf{o}) &  Apply an aggregation with grouping attributes $G$, aggregation expression $F=(f_1, f_2, \cdots, f_n$), using hash table \textbf{h} with parameters \textbf{p}, branching mode \textbf{m}, and element access offset \textbf{o} \\ 
\hline
{\setdslcolor PROJECT}($A$; \textbf{m}, \textbf{o}) & Materialize tuples to output relation projecting attributes for attribute set $A$, branching mode \textbf{m}, and element access offset \textbf{o}\\
\hline
\end{tabular}
\caption{Overview of pipeline operations in a pipeline program.}
\label{tab:dsl_pipeline_operations}
\end{table*}

\subsection{Pipeline Programs}

Our goal is to create variants of a pipeline that can take into account various aspects of the underlying hardware.
For this, we provide a precise definition of the semantics of an operator pipeline.

The produce/consume model fills a pipeline with operations by calling their consume functions during code generation. 
As such, the ordered sequence of operations added by the consume function defines the semantics of the pipeline. The calls to consume consist of \emph{pipeline operations}, such as filtering, inserting a tuple into a hash table, probing a hash table, or aggregating tuples. Pipeline operations often have no direct equivalent in relational algebra, but are used to implement them. 
{
We briefly introduce the {\color{black}pipeline operations used} in this work in Table~\ref{tab:dsl_pipeline_operations}. 

{\color{black}
Operations accept two categories of parameters:
\begin{compactenum}
 \item \textbf{Regular parameters:} These parameters encode the semantics of the operation, such as the table scanned in LOOP or the filter predicate in FILTER. We format these parameters \emph{italic}.
 \item \textbf{Code generation modes:} These parameters define which code variant is generated by the operation, such as the hash table implementation used in HASH\_PUT. We format these parameters \textbf{bold}.
\end{compactenum}
}

We store the sequence of pipeline operations and their parameters as a \emph{pipeline program}. We illustrate the process for the example query in Figure~\ref{fig:produce_consume_example} and show the pipeline programs produced in Table~\ref{tab:pipeline_program_example}. The query contains two joins, which forces the produce/consume model to create a new pipeline for each hash table build. The build pipelines iterate over their input tables (T$_1$ and T$_2$), apply their filters, insert the matching key into a hash table and materialize the result on the required attributes.
The probe pipeline iterates over table T$_3$, applies its filter, probes the previously {\color{black}built} hash tables, and performs the aggregation. 
}

\begin{table}
 \begin{center}
\setlength{\tabcolsep}{2pt}
{\scriptsize
\begin{tabular}{ |m{2.4cm}|m{2.3cm}|m{3cm}| }
  \hline
  Build Pipeline 1  &   Build Pipeline 2 & Probe Pipeline\\
  \hline			
  {\setdslcolor LOOP}(T$_1$, ..) &   {\setdslcolor LOOP}(T$_2$, ..) & {\setdslcolor LOOP}(T$_3$, ..) \\
  {\setdslcolor FILTER}(x=5, ..) & {\setdslcolor FILTER}(y$>$1, ..) & {\setdslcolor FILTER}(z$<$3, ..) \\  
  {\setdslcolor HASH\_PUT}(a, ..) & {\setdslcolor HASH\_PUT}(b, ..) & {\setdslcolor HASH\_PROBE}(a=c, ..) \\ 
  {\setdslcolor PROJECT}(a, x, ..) & {\setdslcolor PROJECT}(b, ..) & {\setdslcolor HASH\_PROBE}(b=d, ..) \\  
   & & {\setdslcolor HASH\_AGGREGATE}(x, sum(q), ..)\\
  \hline    
\end{tabular}
}
 \end{center}
\caption{Pipeline programs created for example query plan from Figure~\ref{fig:produce_consume_example}. Each pipeline program belongs to one pipeline. }
\label{tab:pipeline_program_example}
\end{table}

\newpage
\subsection{Pipeline Variants}

{\ch{}We define a \emph{variation} as a systematic modification of a pipeline program that changes the generated target code or run-time parameters. 
One example is the predication mode, where we can either use conditional expressions or software predication to evaluate a selection.
We define a pipeline variant (in short \emph{variant}) by the set of variations applied to a pipeline program. 
From an implementation perspective, the set of program parameters and the sequence of transformation steps determine the variant.} 
Program parameters are global properties of the pipeline program, such as the number of threads used or the execution strategy. 
Many variations require a fine-grained adaption of pipeline operations, such as software predication, loop unrolling, and vectorization. 
These variants are created by one transformation pass per variation. 
{\ch{}Each pass modifies} the pipeline program and, thus, changes the produced result code. 
{\ch{}Therefore}, a sequence of transformation steps defines {\ch{the code that Hawk generates}}.

\subsection{Dimensions of Variant Generation}

We now discuss how we capture hardware properties in a generic way. 
We differentiate between execution strategies, data structure optimization, and code transformations.

\textbf{Execution Strategy.} The execution strategy defines how a pipeline is executed. As we will discuss in Section~\ref{sec:kernel_exec_strat}, different strategies are optimal for various processors and have a strong impact on performance. Thus, a hardware-adaptive query compiler needs to cope with different execution strategies. For example, on CPUs, synchronization overhead is still cheap {\color{black}compared to co-processors}: we have usually only tens of threads, so a single pass over the data is most efficient. However, on processors similar to GPUs or MICs (co-processors), it is much faster to make multiple passes over the data to avoid synchronization cost.


{\color{black}
\textbf{Data Structure Optimization.} Depending on data and query characteristics, data structures with a certain parametrization are optimal. 
One critical case are hash tables, where we can choose between different hashing techniques~\cite{Richter:2015:SAH:2850583.2850585} and specialized hash tables for a certain query~\cite{Shaikhha:2016:AQC:2882903.2915244}.
Thus, an efficient query compiler needs to be able to exchange the data structures used in a query to generate efficient code.
} 


\textbf{Code Transformations.} 
Optimizing code for certain processors usually involves many low-level code transformations. For example, we need to decide on the optimal memory access pattern or the predication mode. In general, we do not achieve the best performance by just applying all available optimizations. Therefore, a hardware-adaptive query compiler {\ch{}must} be flexible enough to apply a certain subset of code transformations to the generated code.

In the following sections, we discuss how we can apply these variations to pipeline programs. Furthermore, describe how we keep each variation orthogonal to other variations. For example, the execution strategy should {\ch{}not depend on} the hash tables or memory access pattern used.

{



\subsection{Pipeline Variant Generation}

\resolved{Tone down claims in terms of encapsulation or explain in more detail how their approach achieves encapsulation and why encapsulation would be harder to achieve in a system like, say, Hyper.}   

We generate variants of a pipeline program in two steps: transformation and code generation. 
At first, we define which variant of the pipeline is compiled, because the variant determines which transformation passes need to be executed.

In the \emph{transformation step}, we execute a sequence of transformation passes. These passes can modify the pipeline program in two ways. First, they can set the global properties of the pipeline program (e.g., the memory access pattern). Second, they can (re-)configure a single pipeline operation (e.g., set the hash table in HASH\_PUT).

In the \emph{code generation step}, we create an interpreter for the selected execution-strategy. The interpreter traverses the pipeline program and calls the code generator for each pipeline operation. Depending on the number of kernels used by an execution strategy, the generated code is injected in one or more kernels (cf. Figure~\ref{fig:multiple_execution_strategies}). We illustrate the process in Figure~\ref{variant_generation_process}. We discuss target code generation in detail in Section~\ref{sec:code_generation}.

Note that this code generation algorithm allows us to freely combine variations in the same pipeline program. For example, it is possible to generate a variant that uses coalesced memory access, software predication, with a fine-grained multi-pass execution strategy. Thus, it is a very flexible and powerful mechanism, which keeps the possible variations {orthogonal to} each other.

\begin{figure}
    \begin{center}
    \includegraphics[width=\linewidth]{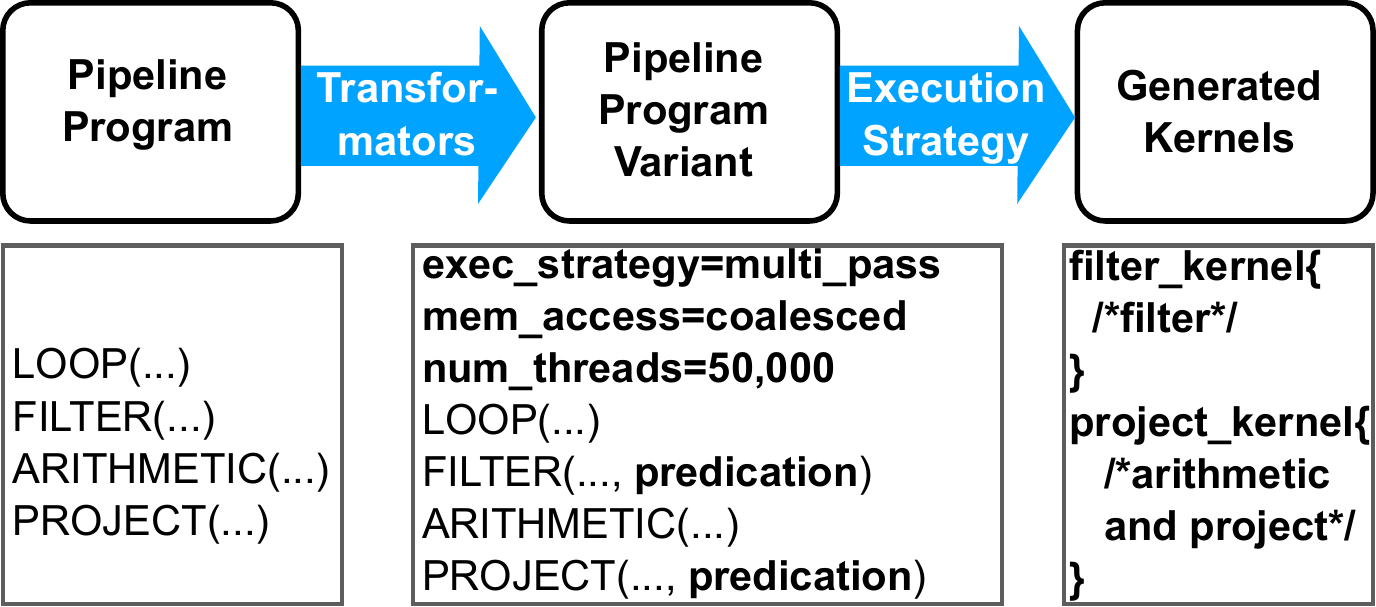}
    \end{center}
    \caption{Pipeline Variant Generation.}
    \label{variant_generation_process}
\end{figure}

}

\section{Execution Strategies}
\label{sec:kernel_exec_strat}

{
Efficient code generation for heterogeneous processors needs to trade-off two basic design dimensions: the degree of parallelism and synchronization overhead. 
Coarse-grained  parallelism is usually sufficient on CPUs (e.g., spawning one thread per processor core). 
In contrast, co-processors require fine-grained parallelism (e.g., using all available SIMD lanes in a GPUs streaming multi processor and having enough thread blocks to hide memory latencies). 
{\ch{}Fined-grained parallelism requires ten thousand and more threads. Thus,} synchronization among threads is very expensive compared to thread synchronization {\ch{}among} tens of threads for coarse-grained parallelism. 

{

Algorithms optimized for co-processors avoid synchronization by not writing the result directly. 
Instead, the algorithms first compute unique write positions for each result tuple. 
Then, they repeat the computation and write the result tuples in parallel without any synchronization. 
Thus, these algorithms need to perform multiple passes over the data~\cite{He:2009:RQC:1620585.1620588}. 
}
Due to the high memory bandwidth of co-processors, these multi-pass algorithms still achieve excellent performance. However, on CPUs, a multi-pass strategy introduces overhead. {\ch{}This is because} a single pass over the data is typically more efficient{\ch{}, as} the synchronization costs are moderate. 

Depending on the type of operator pipeline, we generate different kernels. We differentiate between projection pipelines and aggregation pipelines.
}

\begin{figure}
 \begin{center}
    \includegraphics[width=\linewidth]{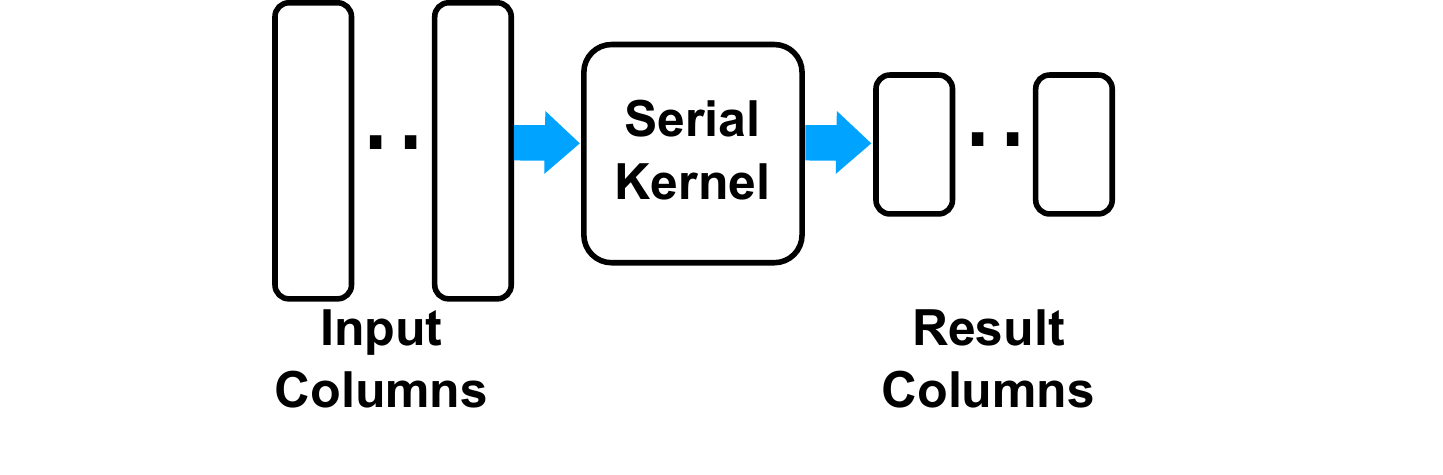}
 \end{center}
 \caption{ Coarse-grained parallelism. One serial kernel is generated per pipeline. We parallelize by executing one kernel per physical core on different blocks of the input.} 
 \label{fig:coarse_grained_parallelism}
 \begin{lstlisting}[
    language=SQL,
    showspaces=false,
    basicstyle=\ttfamily,
    keywordstyle=\color{blue}\bf,
    breaklines=true,
    frame=single,
    label=lab:projection_query_1,  
    captionpos=t,  
    abovecaptionskip=-0.4cm,
%    belowcaptionskip=-0.4cm,
%    belowskip=-0.4cm,
    caption={Projection Query 1}.
        ]
select lo_linenumber, lo_quantity, lo_revenue
from lineorder where lo_quantity<25;
\end{lstlisting}
\end{figure}





\subsection{Projection Pipelines}

A projection pipeline is an operator pipeline in a query plan that does not perform aggregations. {\color{black}Thus, it} projects matching tuples (filters and hash probes) in an output buffer.  
We show a simple query that creates a single projection pipeline in Listing \ref{lab:projection_query_1}. It consists of one filter predicate and projects three attributes. 

\textbf{Coarse-Grained Parallelism.}
On CPUs, it is common to generate a single for-loop per pipeline. {\ch{}This loop} processes all input tuples and writes result tuples to the output buffers (cf. Figure~\ref{fig:coarse_grained_parallelism}). 
{\ch{}The coarse-grained strategy parallelizes query processing by concurrently executing the same pipeline on different chunks of the input relation~\cite{Leis:2014:MPN:2588555.2610507}. 

{
\textbf{Fine-Grained Parallelism.}
On processors with many light-weight cores (e.g., GPUs or MICs), the coarse-grained parallelization cannot utilize all cores. {\ch{}In this case, we need fine-grained parallelism.} 
%
We illustrate this trade-off in Figure~\ref{pic:coarse_grained_vs_fine_grained_parallelism}, where we execute Projection Query 2 (cf. Listing~\ref{lab:projection_query_2}) with the coarse-grained and the fine-grained strategy on different processors. 
{We describe our detailed experimental setup in Section~\ref{sec:experimental_setup}.}
The coarse-grained strategy outperforms the fine-grained strategy on CPUs by a factor of 2.8. The fine-grained strategy outperforms the coarse-grained strategy on a GPU by a factor of 148 and a MIC by a factor of 3.19.
}

Algorithms that use fine-grained parallelism avoid latching at all cost and are typically multi-pass strategies, consisting of three phases~\cite{He:2009:RQC:1620585.1620588}. In the first phase, the operator is {\color{black}executed} and all matching tuples are marked in a flag array. In the second phase, per-thread write positions are computed using an exclusive prefix sum. Finally, the operator is {\color{black}executed} again, but this time, the threads can lookup globally unique write positions and can write their result. 

We generalize this three-step processing technique to operator pipelines as follows. We generate two kernels, a filter and a projection kernel. In the first step, the \emph{filter kernel} performs all operations that reduce the number of result tuples. These are essentially filter and hash probes (e.g., to conduct joins). All matching tuples are marked in a flag array. The second step computes the write positions for each thread by performing a prefix sum on the flag array. In the third step, the \emph{projection kernel} repeats the hash probes to obtain the payload of matching join tuples. The projection kernel also performs arithmetic instructions and writes the result to the computed write positions. We illustrate the algorithm in Figure~\ref{fig:fine_grained_parallelism}.

\input{coarse_vs_fined_grained_parallelism.tex}

\begin{figure}[t]
    \begin{center}
    \includegraphics[width=\linewidth]{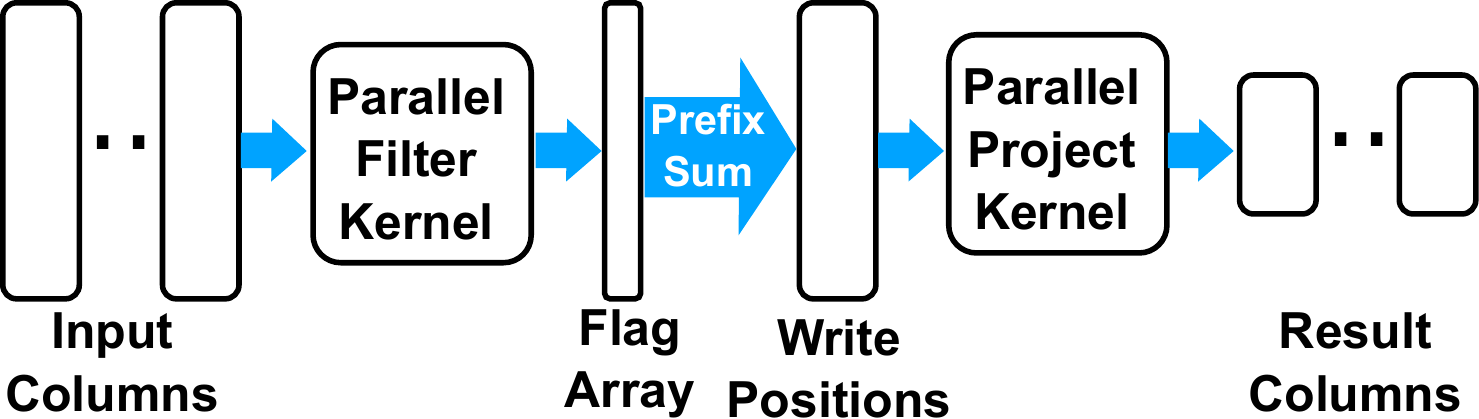}
    \end{center}
    \caption{Fine-grained parallelism. We generate two kernels that are executed massively parallel.}
    \label{fig:fine_grained_parallelism}
\end{figure}

\begin{figure}[t]
\begin{lstlisting}[
    language=SQL,
    showspaces=false,
    basicstyle=\ttfamily,
    keywordstyle=\color{blue}\bf,
    breaklines=true,
    frame=single,
    label=lab:projection_query_2,
    captionpos=t,          
    caption={Projection Query 2}.
        ]
select lo_linenumber, lo_quantity, lo_revenue
from lineorder where lo_quantity<25 and lo_discount<=3 and lo_discount>=1 and lo_revenue>4900000;
\end{lstlisting}
\end{figure}

\begin{figure}[t]
    \begin{center}
    \includegraphics[width=\linewidth]{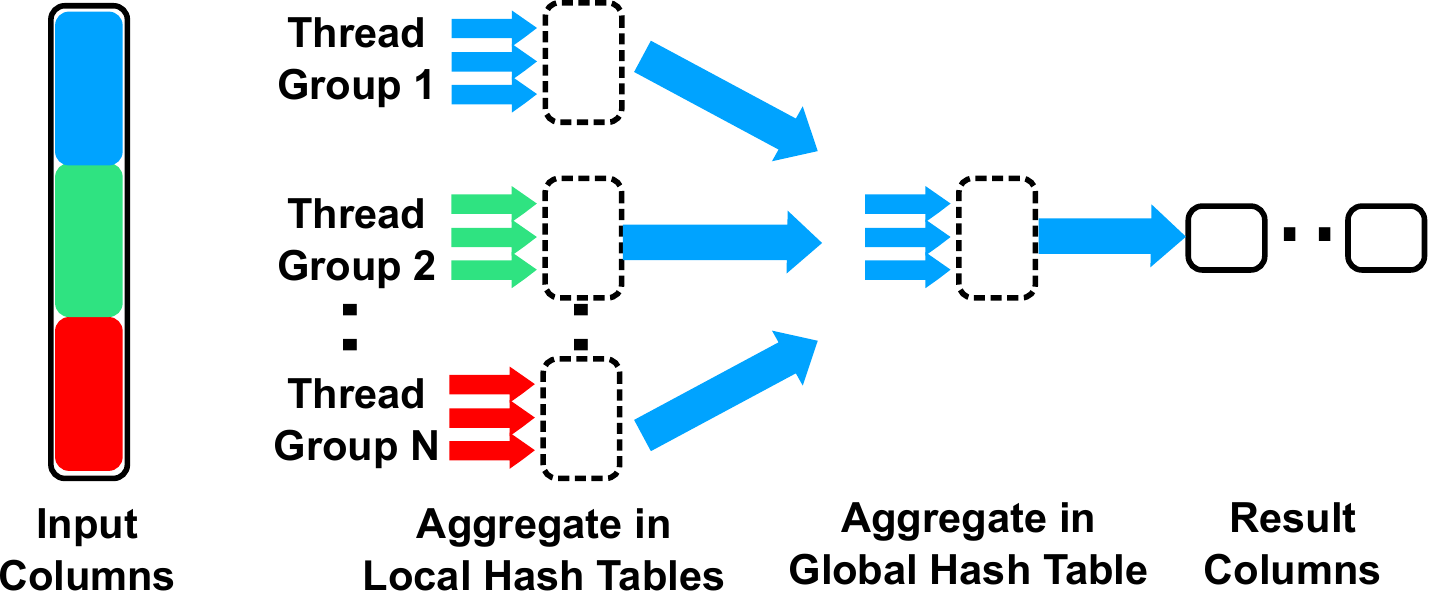}
    \end{center}
    \caption{Local hash table execution strategy for aggregation pipelines. For each of the $N$ hash tables, $M$ threads perform the aggregation.}
    \label{pic:aggregation_pipeline_fine_grained_parallelism}
\end{figure}

\subsection{Aggregation Pipelines}

An aggregation pipeline is a pipeline where the last operator is an aggregation operator. Here, we materialize the result in a hash table and, therefore, we do not need to compute write positions in an output buffer. 

\subsubsection{Execution Strategies}

Depending on the number of result groups, we use different aggregation strategies. 

\textbf{Local Hash Table Aggregation.}
{\ch{}If we expect few result groups, we perform the aggregation in two steps. First, we pre-aggregate the result in parallel in multiple local hash tables. Second, we merge the local hash tables into a global result hash table.} 
We call this \emph{local aggregation} and illustrate the principle in Figure~\ref{pic:aggregation_pipeline_fine_grained_parallelism}. 
For each of the $N$ hash tables, $M$ threads perform the aggregation. 
The number of hash tables and threads per hash table are thus important tuning parameters (cf. Section 9). 
{\ch{}We synchronize concurrent operations on the aggregates using OpenCL's atomics.}

\textbf{Global Hash Table Aggregation.}
{\ch{}If we expect many result groups, we aggregate into a single global hash table. 
In this case, synchronization overhead is small and cost for merging large partial results high.}
We refer to this as \emph{global aggregation}, which is a special case of local aggregation with a single local hash table. 
Thus, we only need to tune the number of threads per hash table.

\subsubsection{Supporting different degrees of Parallelism}

We implement \emph{coarse-grained parallelism} by using the local hash table aggregation with one thread per hash table. Furthermore, we set the number of hash tables to the number of OpenCL compute units (e.g., the number of CPU cores). 

Each thread group is responsible for one hash table. 
Thus, if we increase the number of threads per thread group, we achieve \emph{fine-grained parallelism}. 



\section{Data Structure Optimization}
\label{sec:data_structure_specializaton}

{\color{black}
Besides a well-selected execution strategy, high performance implementations require optimized data structures. 
A prominent example in a database context are hash tables, where different implementations are optimal depending on data and query characteristics~\cite{Richter:2015:SAH:2850583.2850585}. 
We can also improve cache efficiency by removing unnecessary payloads, when we specialize hash tables for a certain query~\cite{Shaikhha:2016:AQC:2882903.2915244} (e.g., hash tables used for aggregations). A query compiler allows us to directly include these optimizations on a per-query basis.
}

\begin{figure}
    \begin{center}
    \includegraphics[width=0.8\linewidth]{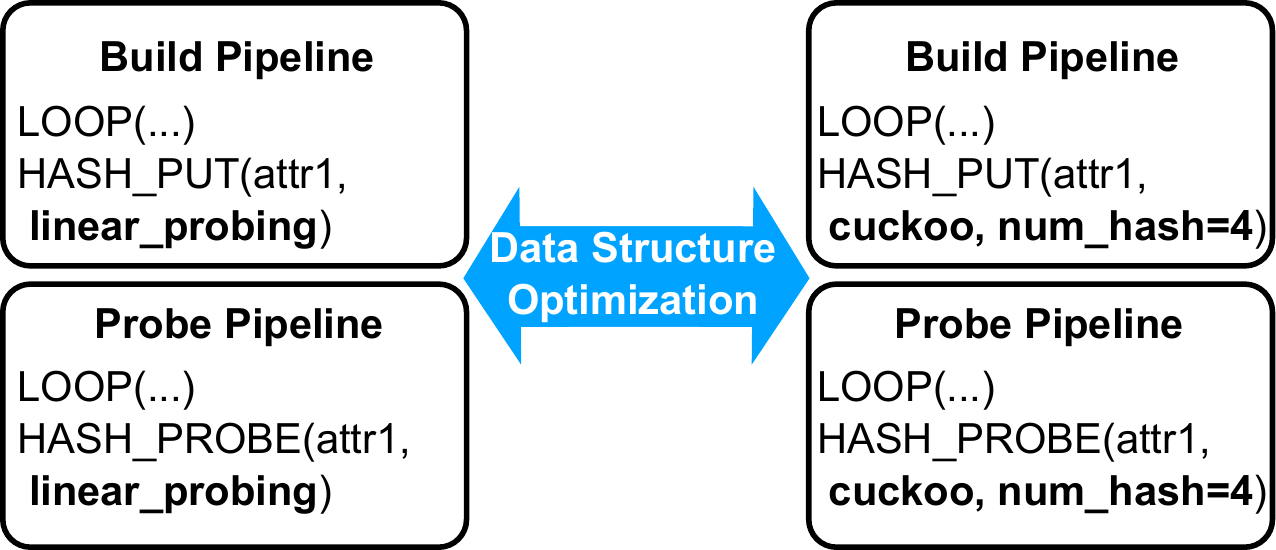}
    \end{center}
    \caption{Data Structure Optimization on the example of exchanging hash table implementations.}
    \label{fig:data_structure_specialization}
\end{figure}

\subsection{Support of Different Hash Tables}

We configure each HASH\_PUT and HASH\_PROBE operation with a hash table and its parameters, as we illustrate in Figure~\ref{fig:data_structure_specialization}. Here, we exchange the linear probing hash table with a Cuckoo hash table. Note that we can also change the parametrization of a hash table (e.g., we can set the number of hash functions of Cuckoo hashing). 

Build and probe pipeline operations need to work with the same hash table (and same parametrization). This introduces a dependency between pipeline programs. Thus, the query processor needs to ensure that corresponding HASH\_PUT and HASH\_PROBE operations use the same data structure.





\section{Code Transformations}

We now discuss how a query compiler can capture different ways to exploit the hardware at the level of traditional code transformations (e.g., the memory access pattern).

\subsection{Transformations}

\begin{figure}[t]
    \begin{center}
    \includegraphics[width=\linewidth]{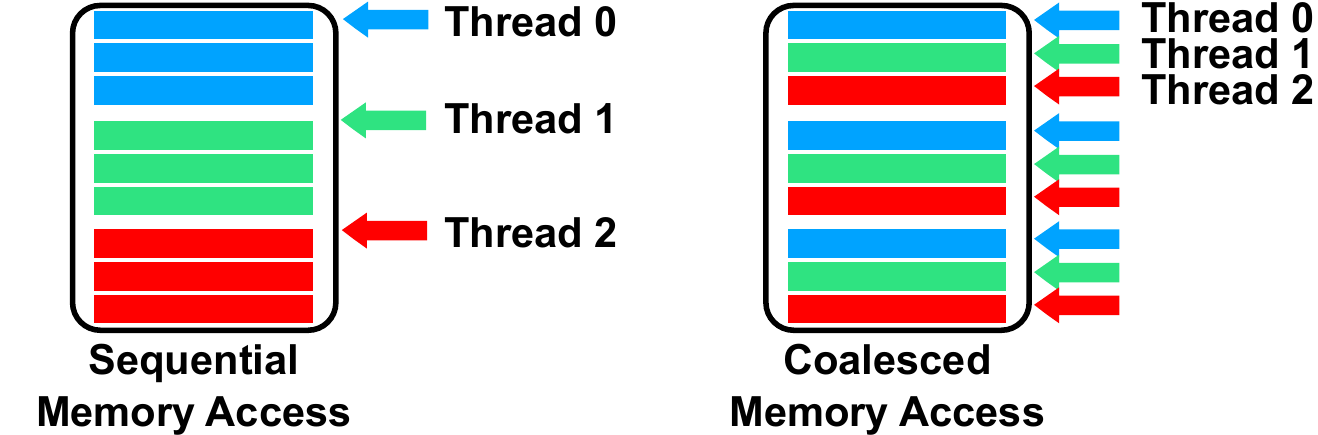}
    \end{center}
    \caption{Visualizing different memory access strategies.}
    \label{fig:code_opt_mem_access}
\end{figure}

\input{sequential_vs_coalesced_mem_access.tex}

\begin{figure}
    \begin{center}
    \includegraphics[width=\linewidth]{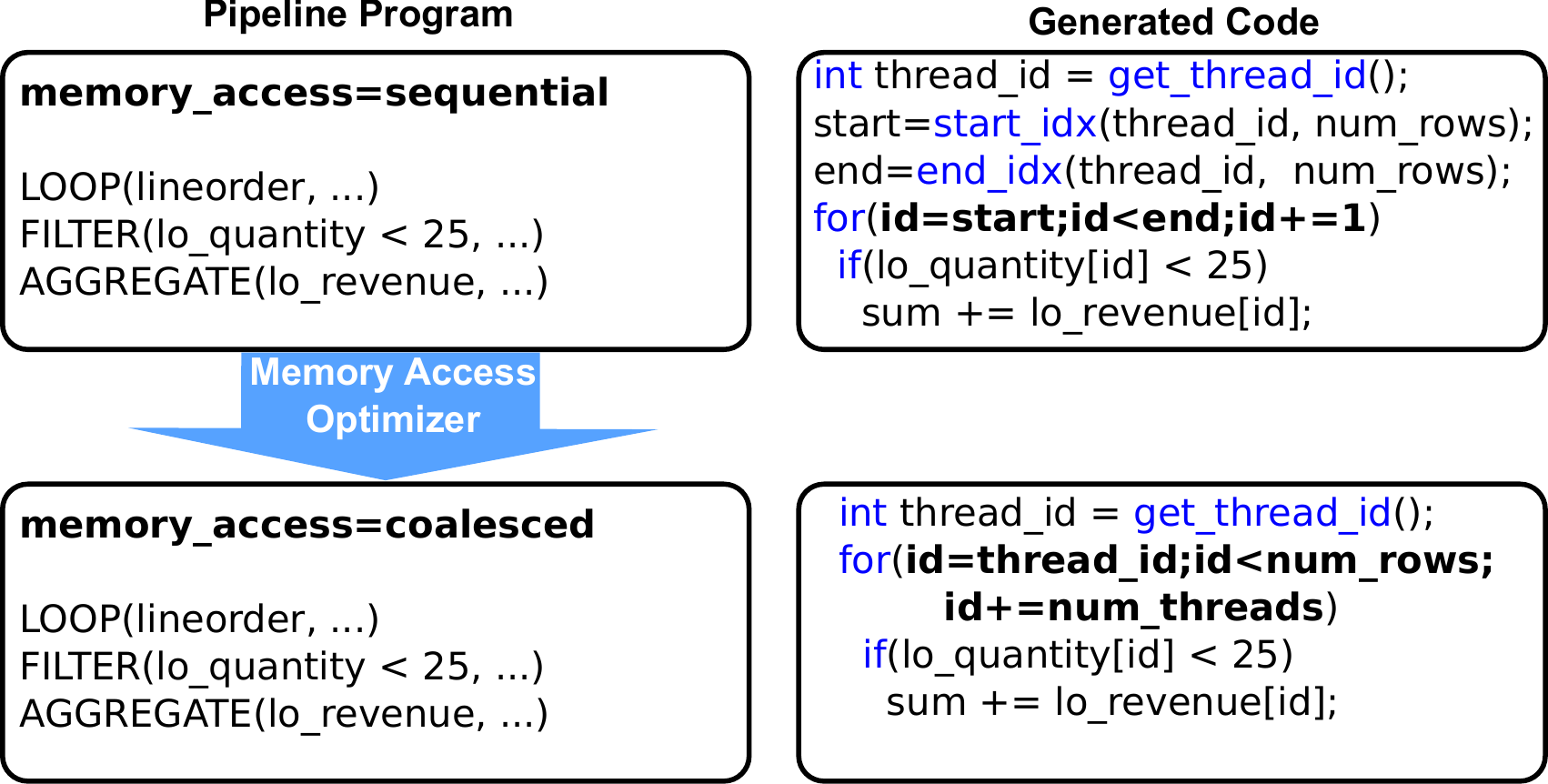}
    \end{center}
    \caption{Effect of memory access pattern on generated code.}
    \label{fig:code_opt_mem_access_code_rewrite}
\end{figure}



{

\textbf{Adjusting the memory access pattern.}
Different processors prefer different ways of {\color{black}accessing} the memory. In sequential access, each thread processes a continuous chunk of tuples, very similar to horizontal range partitioning.
In coalesced memory access, every thread reads a neighbored location relative to other threads. We illustrate the principle of sequential and coalesced memory access in Figure~\ref{fig:code_opt_mem_access}.

We show the performance impact of the memory access pattern in Figure~\ref{pic:sequential_vs_coalesced_mem_access}. On a CPU, sequential access outperforms coalesced memory access by a factor of 1.6. On a GPU, coalesced memory access outperforms sequential memory access by a factor of 1.8. In this measurement, we see no significant difference for the MIC processor. 

We rewrite the memory access pattern in a pipeline program by setting the memory access property. 
We show the impact on the generated code in Figure~\ref{fig:code_opt_mem_access_code_rewrite}.

\begin{figure}
    \begin{center}
    \includegraphics[width=\linewidth]{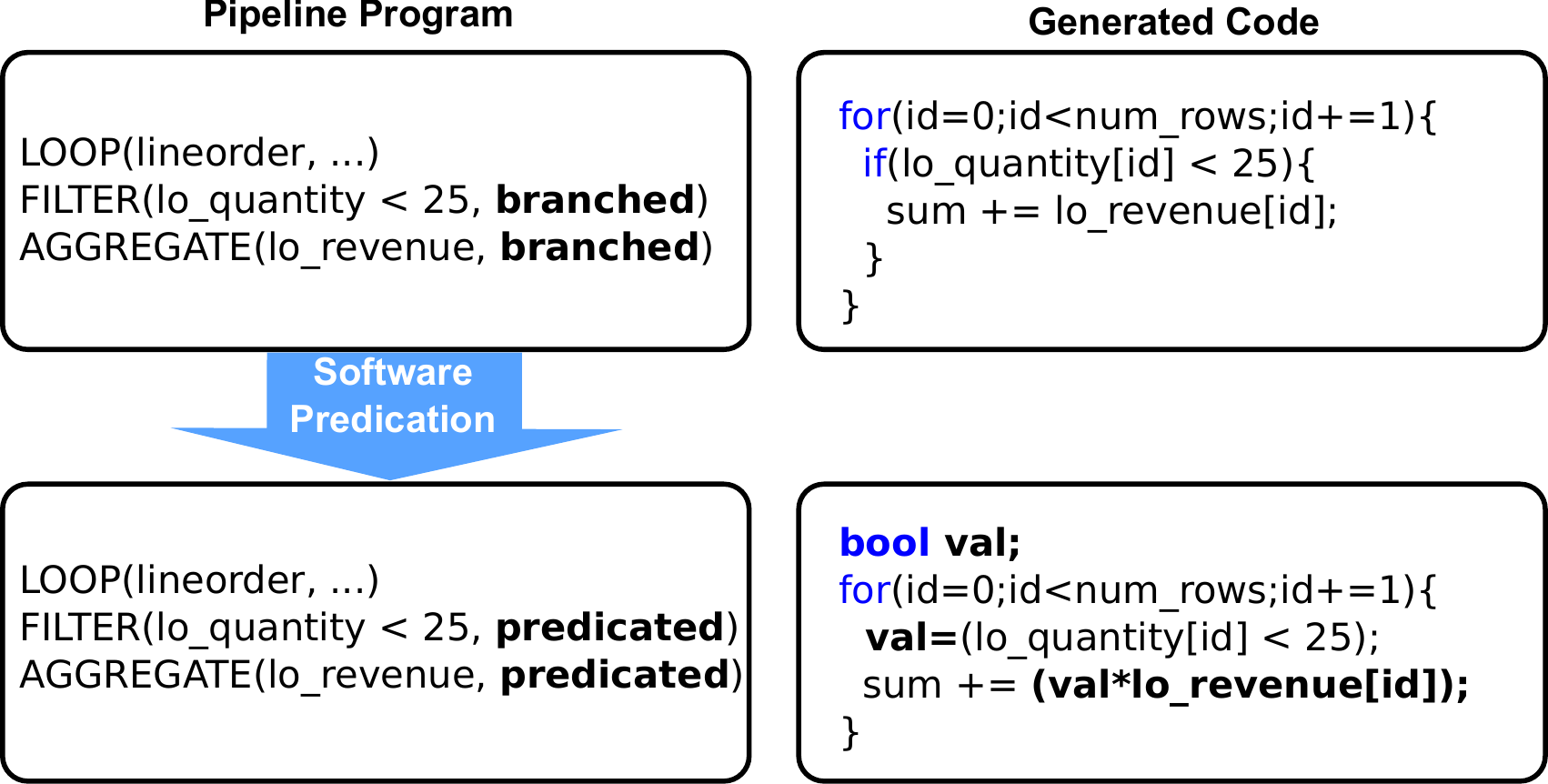}
    \end{center}      
    \caption{Applying software predication transformation to a pipeline program.}
    \label{fig:code_opt_software_predication}  
\end{figure}

\begin{figure}
    \begin{center}
    \includegraphics[width=\linewidth]{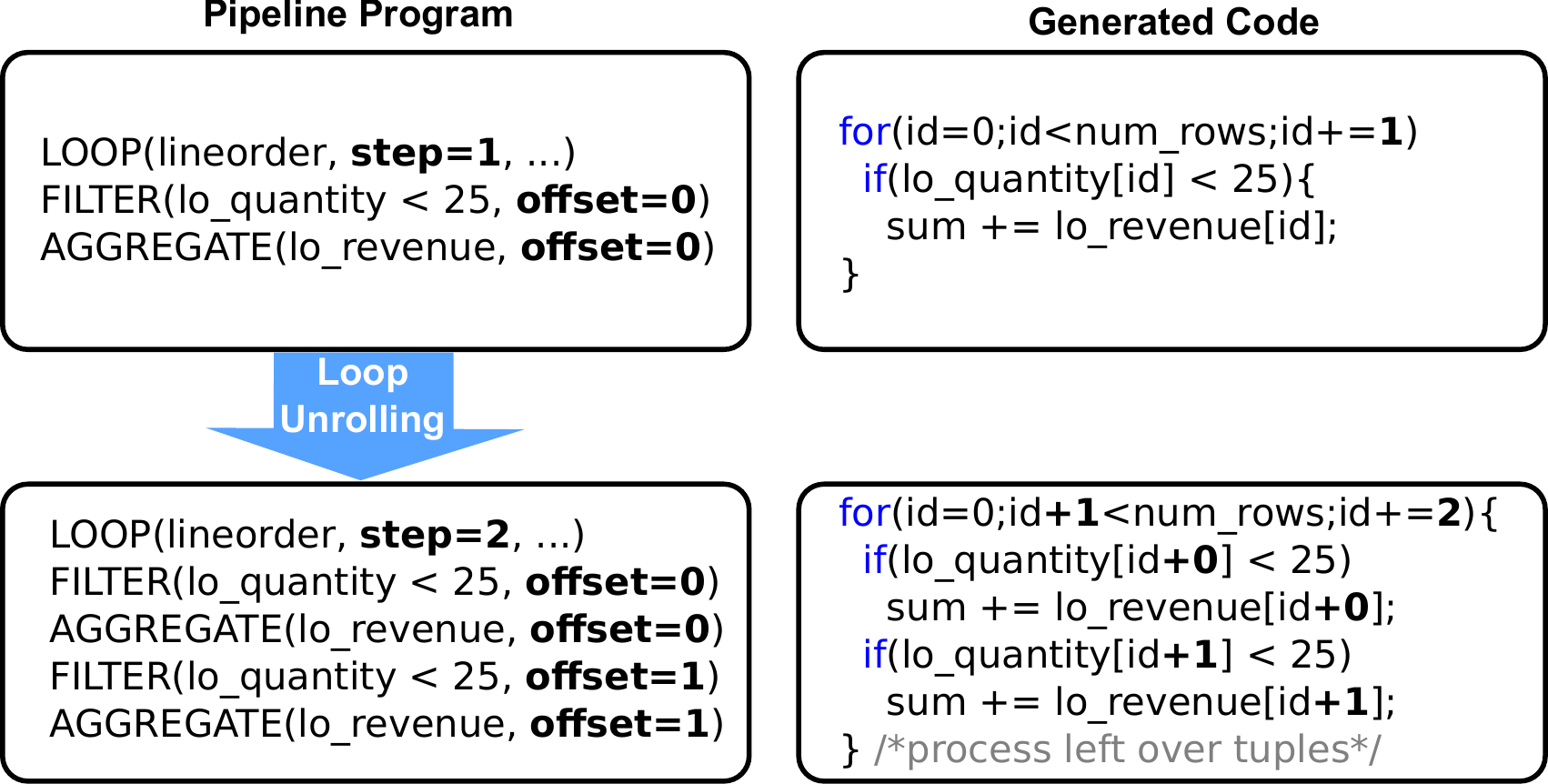}
    \end{center}
    \caption{Applying loop unrolling transformation to a pipeline program.}
    \label{fig:code_opt_loop_unrolling}
\end{figure}

\textbf{Applying software predication.}
Software predication is a common technique to avoid branch misprediction penalties. 
To support predication, each pipeline operation has a flag that determines whether code with branching (if statements) or with predication should be generated. 
In the predicated mode, the result of predicate evaluations is stored in a result value. This value is either added to the variable storing the result size (projection pipeline) or multiplied to the input values before an aggregation (aggregation pipeline). We illustrate the principle in Figure~\ref{fig:code_opt_software_predication}, where we switch a simple aggregation pipeline from branched to predication mode. 
{
Note that all pipeline operations have to be in the same mode. Otherwise, the result becomes incorrect. 
Thus, either the complete pipeline program uses predicated execution or not. 
}

\textbf{Other optimizations.}
We can also apply more complex code transformations, such as loop unrolling or vectorization. 
We exemplary show how loop unrolling can be supported by pipeline programs in Figure~\ref{fig:code_opt_loop_unrolling}.
{\color{black} Loop unrolling affects the original pipeline program beyond the choice of code generation parameters. However, loop unrolling does not limit the combinations with other variations.}

{\ch{}The key point is that a pipeline program, which physically represents a pipeline, is a highly flexible representation. 
It stores low-level code transformations that are hard to represent in a physical query plan.




}



\begin{figure}
\begin{center}
\begin{tikzpicture}
  \centering
  \begin{axis}[
    ybar=0pt,
    ymin=0,
    ymax=120,
    xmin=-1,
    xmax=3,
    enlarge y limits=false,
    width=6cm, 
    height=4cm,
    legend style={at={(0.38,-0.7)},
    anchor=south,legend columns=2},
    ylabel style={align=center},
    ylabel = {Compile Time\\ (in ms)},
    xtick=\empty,
    xticklabels={},
    bar width=20pt,
    nodes near coords={\pgfmathprintnumber[precision=1,zerofill,fixed]{\pgfplotspointmeta}},
    cycle list name=barplot cycle list,
    area legend
    ]
    \addplot coordinates {(1,13)};    
    \addplot coordinates {(1,25.20862353)}; 
    \addplot coordinates {(1,39.25192619)}; 
    \addplot coordinates {(1,96.66036471)}; 
    \addplot coordinates {(1,81.45652059)}; 
    \addplot coordinates {(1,55.22430952)}; 
    
    \legend{HyPer v0.5-222-g04766a1, OpenCL CPU (Intel), OpenCL CPU (AMD), OpenCL MIC (Intel), OpenCL GPU (NVIDIA),  OpenCL GPU (AMD)}
  \end{axis}
\end{tikzpicture}
\end{center}
\caption{Query compilation times of a simple Projection Query (cf. Listing~\ref{lab:projection_query_2}). OpenCL kernel compilation times for CPUs are in the same order of magnitude as HyPer's LLVM IR code generation.}
\label{pic:query_compilation_times_proj}
\end{figure}

\section{Target Code Generation}
\label{sec:code_generation}

{
In this section, we discuss the target code generation of Hawk. 
We reason why we use OpenCL as target language and discuss how we generate kernels by fragment generation and assembly. 
Furthermore, we discuss how one can extend the code generator by new data structures and algorithms. Finally, we discuss implementation details of Hawk.
}

\resolved{Provide a more detailed discussion on merits of OpenCL. For instance, discuss manipulating code chunks as kernels and what kernel manipulation (e.g. fusion?) are essential to the framework. Discuss if such mechanisms would be available should SPIR be used instead (or complementary).}

\resolved{Add (perhaps in an appendix) more complete algorithms used for code fragment generations}

\subsection{Target Code: OpenCL Kernel}

The drawback of generating high-level code is usually high compilation time~\cite{DBLP:conf/icde/KrikellasVC10,Neumann:2011:ECE:2002938.2002940}. 
{
By compiling pipeline programs to OpenCL kernels, Hawk benefits from the JIT compilation capabilities and the performance portability of OpenCL. The latter allows Hawk to run any variant on any OpenCL-capable processor.
}  

{\color{black}In Figure \ref{pic:query_compilation_times_proj}, we show query compilation times for a simple query (cf. Listing~\ref{lab:projection_query_2}) for compiling OpenCL kernels for an Intel CPU, an AMD CPU, an Intel MIC, a NVIDIA GPU, and an AMD GPU. As reference, we also show the compilation time of HyPer~\cite{hyper:icde:2011} (v0.5-222-g04766a1), a state-of-the-art system for query compilation.} Compiling OpenCL kernels for CPUs is in the same order of magnitude (slower by a factor of 2 to 3 for Intel and AMD OpenCL SDKs) as the LLVM IR query compilation used by HyPer~\cite{Neumann:2011:ECE:2002938.2002940}. Furthermore, we observe that compilation for GPUs and MICs is up to a factor of 4.3 to 7.4 slower compared to LLVM IR query compilation. 
The compilation times are consistently below 100ms. Thus, we conclude that query compilation using OpenCL is sufficiently efficient for compiling database queries to support interactive querying on GPUs and MICs. Note that the OpenCL compilation times can be further reduced. 
We can disable certain optimization passes, trading off runtime and optimization time, similar to optimization levels in some commercial database engines. 

\begin{figure}[t]
    \begin{center}
    \includegraphics[width=\linewidth]{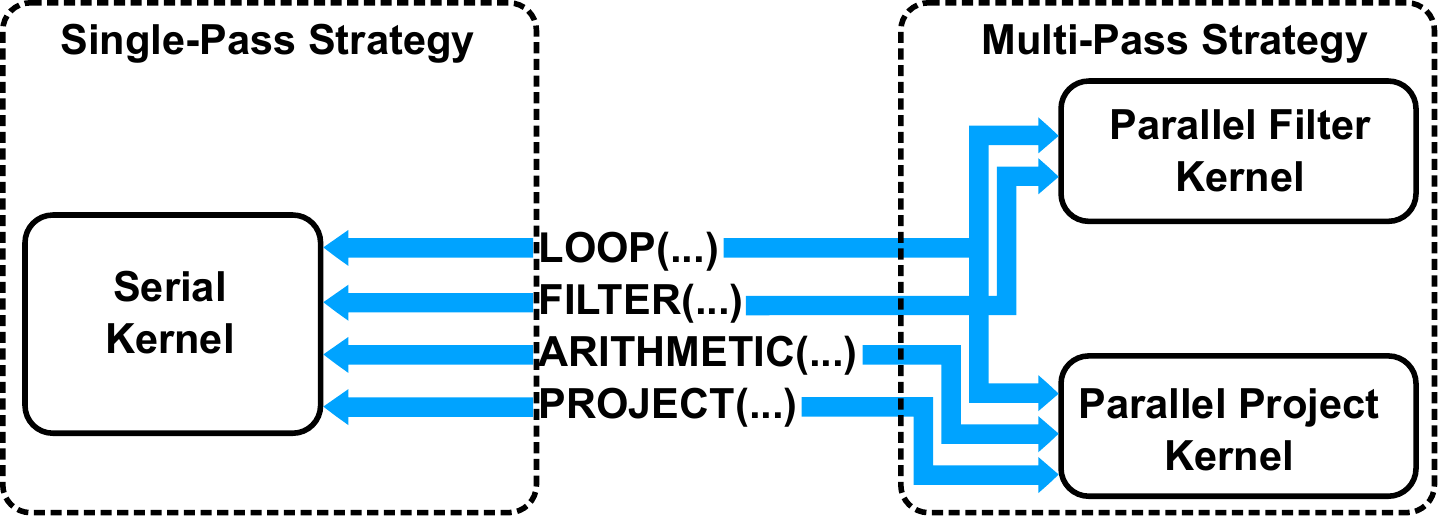}
    \end{center}
    \caption{Supporting multiple execution strategies. Each strategy is designed to act as an interpreter for a pipeline program, which routes generated code to the appropriate kernels.}
    \label{fig:multiple_execution_strategies}
\end{figure}

\subsection{Fragment Generation and Assembly}

{\color{black}
We now discuss how we can generate code for projection and aggregation pipelines from pipeline programs.
The code generation follows a two step scheme: \emph{fragment generation}, followed by \emph{fragment assembly}.

\subsubsection{Fragment Generation}
{
A code fragment (in short \emph{fragment}) consists of six segments: host variable declarations, host initialization code, host cleanup code, kernel variable declarations, kernel code top, and kernel code bottom.
These fine-grained separations allow us to route fragments into different kernels.

Each pipeline operation produces a fragment that implements its semantic. 
We retrieve the fragment for each pipeline operation to create all fragments. 

Each operation can generate code for any part in the target source code (e.g., body of the for-loop, declarations, or cleanup operations). 
Furthermore, the fragment produced by a pipeline operation depends on the code generation modes: these modes allow us to adapt the fragment by re-parameterizing the pipeline operations or global properties of the pipeline program.
Using this code generation scheme, it is straightforward to create systematic variants of a pipeline program to adapt to the underlying hardware.

}


\begin{figure*}

\begin{center}

\includegraphics[width=\linewidth]{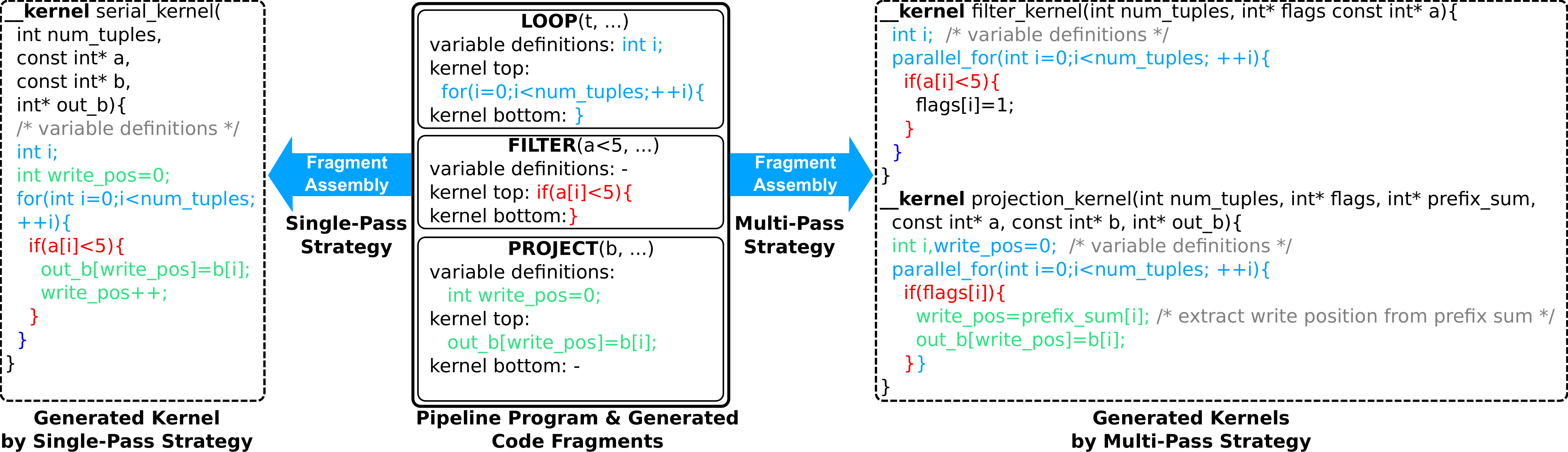}

\end{center}


\caption{Example for fragment generation and fragment assembly: Each pipeline operation generates fragments, which are then assembled to kernels. {The single-pass strategy generates one kernel that includes all operations from the fragments. The multi-pass strategy generates a filter and a projection kernel which include different fragments.}}

\label{fig:fragment_assembly}

\end{figure*}

\subsubsection{Fragment Assembly}

{
We combine fragments by assembling them into a single fragment. 
Note that this \emph{fragment assembly} is essentially a string concatenation of code segments.

Our guiding idea is as follows. 
We provide an \emph{interpreter} for pipeline-programs for each execution strategy. 
Each interpreter knows how many kernels are required for the strategy. 
The interpreter assigns the fragments, depending on the pipeline operation, to one or more kernels. 

We illustrate this process in Figure~\ref{fig:multiple_execution_strategies}. 
For the single-pass strategy, all fragments belong to the same kernel.
The multi-pass strategy routes fragments from different pipeline operations to different kernels.
Thus, a fragment can be part of multiple kernels, such as LOOP or HASH\_PROBE.

For each kernel used by the execution strategy, the interpreter assembles all fragments assigned to the kernel to a result fragment. 
We create the final kernel from this result fragment. 
Note that Hawk's code generator is conceptionally not limited to OpenCL kernels, but could also produce code for frameworks such as CUDA.



Since we implement execution strategies as interpreters, we can apply different strategies to pipeline programs.
Our design keeps the execution strategies orthogonal to any other variation on the pipeline program.
%
}
}

{%

\subsection{Example: Fragment Generation and Assembly}
We now present an example that illustrates the code generation process.
Consider the query \emph{select b from t where a$<$5}, which will result in a pipeline program with three operations: LOOP, FILTER, and PROJECT. We show the generated fragments of the pipeline operations in Figure~\ref{fig:fragment_assembly}.
The generated fragments can add code to two parts of the kernel: the variable declaration and initialization code block, and the for-loop.
Code can be inserted into a for-loop at two positions: at the top position we insert the actual code; at the bottom position we insert closing brackets and perform operations (e.g., increasing counters) after an iteration. 
Generated code of succeeding operations is nested inside the brackets of previous operations. For example, the final projection is nested in the generated code of the filter operation.




\begin{figure}
\begin{lstlisting}[
    language=C++,
    showspaces=false,
    basicstyle=\small\ttfamily,
%    numbers=left,    
    keywordstyle=\color{blue}\bf,
    stringstyle=\color{myblue},     
    breaklines=true,
    frame=single,
    showstringspaces=false,
    escapeinside={(*}{*)},
    label=lab:code_gen_loop,  
    linebackgroundcolor = {\btLstHL{1,3-5,7-8}},    
    captionpos=t,  
%    caption={Loop Fragment Generation: LOOP(table, memory\_access\_pattern).}
    caption={[Fragment Generation]%
    \begin{tabular}[t]{@{}l@{}}
                 Loop Fragment Generation: \\
                 LOOP(table, memory\_access\_pattern). \\[.5\normalbaselineskip]
            \end{tabular}}
        ]        
<thr_id = get_thread_id()>        
if(memory_access_pattern==SEQUENTIAL){
  <start=start_idx(thr_id,num_rows)>
  <end=end_idx(thr_id,num_rows)>
  <for(id=start;id<end;id+=1)>
}else if(memory_access_pattern==COALESCED){
  <for(id=thr_id;id<num_rows;id
  +=num_threads)>
}
\end{lstlisting}
\vspace{-0.5cm}
\begin{lstlisting}[
    language=C++,
    showspaces=false,
    basicstyle=\small\ttfamily,
%    numbers=left,    
    keywordstyle=\color{blue}\bf,
    stringstyle=\color{myblue},     
    breaklines=true,
    frame=single,
    showstringspaces=false,
    escapeinside={(*}{*)},
    label=lab:code_gen_filter,  
    linebackgroundcolor = {\btLstHL{2,4}}, 
    captionpos=t,  
%    caption={Filter Fragment Generation: FILTER(condition, predication\_mode).}
    caption={[Fragment Generation]%
    \begin{tabular}[t]{@{}l@{}}
      Filter Fragment Generation: \\
      FILTER(condition, predication\_mode). \\[.5\normalbaselineskip]
    \end{tabular}}
        ]        
if(predication_mode==BRANCHED){
  <if(condition)>
}else if(predication_mode==PREDICATED){
  <result_increment=(condition)>
}
\end{lstlisting}
\vspace{-0.8cm}
\begin{center}
{\color{black}Generated Code: \colorbox{codeblue}{$<$code$>$}}
\end{center}
\end{figure}

\newcommand{\createcolorbox}[0]{\pgfmathsetmacro{\myinnersep}{1}
\begin{tikzpicture}\node[rectangle,fill=lightgray,minimum width=10pt, outer sep=0,minimum height=\heightof{Cap}] {generated code};\end{tikzpicture}\xspace}

{
\subsection{Fragment Generation and Assembly Algorithms}
\label{sec:fragment_generation_and_assembly}

We now discuss algorithms for fragment generation. 
We show pseudo code for each algorithm and highlight generated code by surrounding it with angle brackets and by coloring the background 
({\color{black}\colorbox{codeblue}{$<$generated code$>$}}).
 
The LOOP operation generates code that iterates over every input tuple of a table in parallel. 
We can either iterate sequentially or interleaved over the tuples, which leads to sequential or coalesced memory access (cf. Listing~\ref{lab:code_gen_loop}).
In case of sequential access, we compute the start and end offset of the partition that each thread processes. 
In case of coalesced access, each thread starts the iteration on its unique thread identifier and advances by adding the number of threads to the loop variable \emph{id}. 

The FILTER operation generates code that evaluates a selection predicate. It either generates an if-statement (no predication) or stores the result of the predicate evaluation in the variable \emph{result\_increment} (predication), as we illustrate in Listing~\ref{lab:code_gen_filter}. 

The PROJECT operation generates code that copies the values of each projected attribute and writes them to the write position \emph{write\_pos} in the projection attribute's output array (cf. Listing~\ref{lab:code_gen_project}). 
The generated code depends on the predication mode. 
If predication is disabled, we know the tuple passed all previous filters. 
Thus, we increment the write position after writing the tuple into the output buffer. 
If predication is enabled, we always write the result tuple but add the variable \emph{result\_increment} to \emph{write\_pos}. 
If the tuple passed all previous filters, \emph{result\_increment} is one and the write position is advanced by one row. 
In case the tuple did not match all filters, \emph{result\_increment} is zero and the write position is not changed, which discards the current tuple.  



\begin{figure}
\begin{lstlisting}[
    language=C++,
    showspaces=false,
    basicstyle=\small\ttfamily,
%    numbers=left,    
    keywordstyle=\color{blue}\bf,
    stringstyle=\color{myblue},     
    breaklines=true,
    frame=single,
    showstringspaces=false,
    escapeinside={(*}{*)},
    label=lab:code_gen_project,  
    linebackgroundcolor = {\btLstHL{1,3-4,7, 9}}, 
    captionpos=t,  
%    caption={Fragment Generation: PROJECT(proj\_attributes, predication\_mode).}
    caption={[Fragment Generation]%
    \begin{tabular}[t]{@{}l@{}}
      Project Fragment Generation: \\
      PROJECT(proj\_attributes, predication\_mode). \\[.5\normalbaselineskip]
    \end{tabular}}
        ]        
<declare variable write_pos=0>
for(attribute in proj_attributes){
  <copy value of attribute to result
  column at position write_pos>
}
if(predication_mode==BRANCHED){
  <write_pos++>
}else if(predication_mode==PREDICATED){
  <write_pos+=result_increment>
}
\end{lstlisting}
\vspace{-0.5cm}
\begin{lstlisting}[
    language=C++,
    showspaces=false,
    basicstyle=\small\ttfamily,
%    numbers=left,    
    keywordstyle=\color{blue}\bf,
    stringstyle=\color{myblue},     
    breaklines=true,
    frame=single,
    showstringspaces=false,
    escapeinside={(*}{*)},
    label=lab:code_gen_hash_probe,  
    linebackgroundcolor = {\btLstHL{2,4}},     
    captionpos=t,  
%    caption={Fragment Generation: HASH\_BUILD(attr, hash\_table) and HASH\_PROBE(attr, hash\_table).}
    caption={[Fragment Generation]%
    \begin{tabular}[t]{@{}l@{}}
      Join Fragment Generation: \\
      \hspace{-1.5cm}HASH\_BUILD(attr, hash\_table) and \\
      \hspace{-1.5cm}HASH\_PROBE(attr, hash\_table). \\[.5\normalbaselineskip]
    \end{tabular}}
        ]        
if(hash_table==CUCKOO){
  <insert/lookup attr in cuckoo HT>
}else if(hash_table==LINEAR_PROBING){
  <insert/lookup attr in lin. probing HT>
}
\end{lstlisting}
\vspace{-0.5cm}
\begin{lstlisting}[
    language=C++,
    showspaces=false,
    basicstyle=\small\ttfamily,
%    numbers=left,    
    keywordstyle=\color{blue}\bf,
    stringstyle=\color{myblue},     
    breaklines=true,
    frame=single,
    showstringspaces=false,
    escapeinside={(*}{*)},
    label=lab:code_gen_aggregate,  
    linebackgroundcolor = {\btLstHL{1,3,5}},     
    captionpos=t,  
%    caption={Fragment Generation: AGGREGATE(attr, SUM, predication\_mode).}
    caption={[Fragment Generation]%
    \begin{tabular}[t]{@{}l@{}}
      Aggregate Fragment Generation: \\
      AGGREGATE(attr, SUM, predication\_mode). \\[.5\normalbaselineskip]
    \end{tabular}}
        ]  
<declare variable aggregate=0>           
if(predication_mode==BRANCHED){
  <aggregate+=attr[id]>
}else if(predication_mode==PREDICATED){
  <aggregate+=(attr[id]*result_increment)>
}
\end{lstlisting}
\vspace{-0.8cm}
\begin{center}
{\color{black}Generated Code: \colorbox{codeblue}{$<$code$>$}}
\end{center}
\end{figure}

The HASH\_PUT and HASH\_PROBE operations generate code that insert/lookup tuples into/from a certain hash table (cf. Listing~\ref{lab:code_gen_hash_probe}). 
{
For a linear probing hash table, we generate code that uses a single hash function. A Cuckoo hash table has a variable number of hash functions, which directly affects the generated code. 
We omit the detailed code for the sake of brevity.
}


The AGGREGATE operation generates code that computes the aggregates. 
The generated code depends on the predication mode. 
If predication is disabled, we evaluate the aggregate expression without any further modifications.
In case of enabled predication, we need to take special care to not include a filtered out tuple in the aggregation. 
Therefore, we need to ensure that the aggregate is not changed in case the variable \emph{result\_increment} is zero. 
For example, for the count or sum aggregation functions, we multiply the tuple value with the \emph{result\_increment} before applying the aggregation function (cf. Listing~\ref{lab:code_gen_aggregate}). 
This way, the aggregate stays unchanged if and only if \emph{result\_increment} is zero.

}

}

{
\subsection{Extending the Code Generator}

\resolved{While the whole framework is very powerful, it is not immediately clear how to introduce specialized optimizations such as scan strategies (e.g. in [36]) or the join strategies for a given architecture.}

We now discuss how we can extend the code generator to support new data structures and algorithms.

\subsubsection{Support for Different Data Structures}

For each hash table supported, the query compiler requires the code templates for initializing, accessing, and modifying a data structure. 
These code templates may also be generated at run-time (e.g., to adapt the hash function depending on the data properties).
The pipeline operations encapsulate the code generation for different data structures. For example, the HASH\_PUT operation implicitly generates code for the hash table specified in its parameter. This variability allows us to select different hash table implementations depending on certain data characteristics~\cite{Richter:2015:SAH:2850583.2850585}. 
For example, to add a robin hood hash table, we need to extend the existing HASH\_PUT and HASH\_PROBE operations by the respective code templates. 
Finally, we need to introduce a pipeline operation (e.g., an index scan). 

\subsubsection{Support for Different Algorithms}

For each new algorithm, we extend the code generator by the data structures the algorithm uses. 
Then, the algorithm needs to be registered to the variant generator. We either add a new pipeline operation or include the algorithm in an existing pipeline operation. 
Finally, we provide the respective code templates. 

{
For example, we can extend Hawk with the scan of Zhou and Ross~\cite{Zhou:2002:IDO:564691.564709}, which uses SIMD instructions to check the predicate of multiple tuples at once.
To support this SIMD scan in Hawk, we need to add a new code generation mode to the FILTER operation. 
}
The mode parameter allows us to select either the scalar or the SIMD code template. Furthermore, the code template for the SIMD scan has to be added to the FILTER operation. The same procedure applies for SIMD support for other pipeline operations supported by Hawk. 


}

\subsection{Hawk Implementation Details}

\resolved{Mention what open-source DBMS was used (subject to double blind restrictions), and explain how deep is this integration and what operators are currently supported. Be more upfront about what is implemented already and what is left for future work.}

{
We implemented Hawk as a prototype that targets main-memory database engines that store data in a column-major format.
We show the viability of our approaches on the example of \cogadb~\cite{cogadb14,BressT15}, as it fulfills our requirements and resulted in the smallest integration effort for us. Note that we can apply our concepts to any other system having an in-memory column store, including commercial systems such as SAP HANA~\cite{DBLP:journals/debu/FarberMLGMRD12}, DB2 BLU~\cite{Raman:2013:DBA:2536222.2536233}, or the Apollo engine of SQL Server~\cite{Larson:2015:RAP:2824032.2824071}. The main changes to the database engine consists of the extension of the query plan interface by the produce/consume code generation along with our proposed approach for variant generation. Furthermore, the execution engine has to be replaced by a run-time for the compiled queries. 


{

Hawk supports all pipeline operations discussed in Table~\ref{tab:dsl_pipeline_operations}, which allows for producing code for selections, projections, joins, and aggregations. Aggregations are currently limited to distributive and algebraic aggregation functions (e.g., holistic aggregation functions such as the median are currently not supported). 
}
}



{\color{black}

\section{Optimizing Pipeline Variants}
\label{sec:variant_selection}

\resolved{The variant search and heuristics in Section 8 seem naive, should be improved.  In particular, they seem to do a bad job on MIC, and you should explore what is causing the
problems.  Further, you should discuss the execution time of this phase and discuss the possibility of parameter space pruning.}

\resolved{R3/D5: The optimization algorithm varies on one parameter at a time and does multiple sweeps over the dimensions of variations where each iteration starts with the best configuration found in the previous iteration. In this algorithm, the order of dimensions does effect the result and it seems that the choices for the earlier dimensions can limit the choices for further dimensions. It is clear that the full search space is to large, but why not use meta-heuristics here (e.g., simulated annealing or genetic algorithms).}

Hawk can generate a large number of variants to adapt {\ch{}code} to various processors. Consequently, we face a large optimization space that is the cross product of all values of all variation dimensions. 
Exploring the search space is very expensive for two reasons. 
First, we pay query compilation cost for each generated variant.  
Second, the execution time of variants may be significantly slower than the optimal variant.
It is especially problematic when we explore variants that are very slow on a certain processor (e.g., a serial implementation on a GPU).

In this section, we discuss how we can automatically find a fast-performing variant configuration for each processor for a given query workload. 

{
\renewcommand{\algorithmicrequire}{\textbf{Input:}}
\renewcommand{\algorithmicensure}{\textbf{Output:}}
\renewcommand{\algorithmicdo}{\textbf{do}}

\begin{algorithm}
\color{black}
 \begin{algorithmic}[1]
	\REQUIRE dimensions of variations: $D=\{D_1, \cdots, D_n\}$
	\REQUIRE workload of $k$ queries: $W=\{Q_1, \cdots, Q_k\}$
	\ENSURE variant configuration $v$
	\STATE $v=(v_1, \cdots, v_n) \in D_1 \times \cdots \times D_n$
	\FOR{$(iter=0;iter<q;iter++)$}
	\STATE last\_variant=$v$	
	\FOR{$D_i \in D$}
	\STATE execution\_time=$\infty$
	\STATE best\_dimension\_value=$\emptyset$
	\FOR{$d \in D_i$}
	\STATE $v'=v;$ 		
	\STATE $v_i'=d;$
	\STATE execution\_time$'$ = executeQueries($W,v'$);
        \IF{execution\_time$'$<execution\_time} 
           \STATE {execution\_time=execution\_time$'$;} 
           \STATE {best\_dimension\_value=$d$;}            
        \ENDIF	
	\ENDFOR
	\STATE /* Update configuration */
	\STATE {$v_i=$best\_dimension\_value;}
	\ENDFOR
        \IF{$v==$last\_variant} 
	\RETURN $v$;        
        \ENDIF
	\ENDFOR	
	\RETURN $v$
 \end{algorithmic}
 \caption{Learning an efficient variant configuration for a processor. 
}
 \label{alg:learn_variant_configuration}

\end{algorithm}
}

\subsection{Navigating the Optimization Space}


The key idea is that we explore the search space for a processor offline by executing a workload of representative test queries. We compile different variants of each query in the workload, and systematically explore which variations work best on a particular processor.  
We use the same strategy as we would in a structured experiment. {We present the strategy in Algorithm~\ref{alg:learn_variant_configuration}.} 
Initially, we have no knowledge about the performance behavior of the processor. 
We start from a base configuration (Line~1), which we initialize with the first parameter value in each variant dimension. 
We change one parameter at a time (Line~4--10), and select the parameter value with the best performance (Line~11--14). 
We perform this step for every variant dimension (e.g., execution strategy or memory access pattern). 
The best parameter values are stored in the variant configuration (Line~16--17). 
{A variant configuration is the abstract representation of a variant in the search space.}

Note that different variations may influence each other. 
This means that a previously optimal parameter value of a variation may be sub-optimal in the new configuration. 
To make sure that our algorithm finds a fast performing variant, we repeat the core of the algorithm (Line~4--18) iteratively. 
The algorithm terminates in case we have not found any faster variant configuration (Line~3, 19--21) or reach a maximum number of iterations $q$ (Line~2).

{
\subsection{Reducing Variant Optimization Time}
As the variant exploration requires us to execute the variants, very slow variants increase exploration time significantly. 
We can reduce the exploration time by applying \emph{early termination} and \emph{feature ordering}.

\textbf{Early Termination:}
Our learning strategy allows us to systematically gain knowledge over the complete variant space. 
We can terminate the search early, when we reach a local optimum during an iteration.
This \emph{early termination} saves additional exploration time, but we may not reach the global optimum. 

\textbf{Feature Ordering:}
We can further optimize the search of the variant space when we take into account which variations typically have the most impact on performance. 
In this case, we explore the parameter values of the most critical variations first to find an efficiently performing variant faster. These variations are: the execution strategy, the number of threads, and the memory access pattern. 


}







\subsection{Building a Heuristic Query Optimizer}

We now discuss how we can build a heuristic optimizer using the variant configurations learned. We learn variant configurations for a representative query workload. Thus, the resulting variant configuration is a \emph{heuristic} that performs well for a workload. 
While the heuristic delivers good performance for the given queries, it may not be optimal, as query-dependent parameters can influence the optimal variant. 

To avoid high overhead during query processing, we execute the learning algorithm before query processing.
We use the best found variant (heuristic) of a processor to produce a custom variant of the generated code as discussed in Section~\ref{sec:code_generation}.
%
%
%





}

\section{Evaluation}

{
In this section, we discuss our experimental setup and design, present our results on hardware adaption, and discuss the implications of these results.
}

\subsection{Experimental Setup}
\label{sec:experimental_setup}


\begin{table}
\begin{center}
\begin{tabular}{ c | c | c | c }
Processor & Short & Architecture & Vendor\\
\hline
A10-7850K (CPU) & CPU  & Kaveri& AMD\\
A10-7850K (GPU) & iGPU  & CGN 1.1 & AMD\\
Tesla K40M & dGPU & Kepler & Nvidia\\
Xeon Phi 7120 & MIC & Knights Corner & Intel\\
\end{tabular}
\caption{Processors used in evaluation.}
\label{tab:hardware}
\end{center}
\end{table}

{
As evaluation platform, we use two machines that have several heterogeneous
processors installed. In total, we consider four different processor types with
varying architectures: a CPU, an integrated GPU (iGPU), a dedicated GPU (dGPU),
and a MIC, as shown in Table~\ref{tab:hardware}. All machines run Ubuntu 16.04
LTS (64bit). Depending on the processor's vendor, we have to use a certain
OpenCL SDK and driver to compile and run our kernels. {\color{black} For CPU and iGPU from
AMD, we use the AMD APP SDK version 3.0. For the MIC,
  we use the Intel OpenCL SDK Version 4.5.0.8. For the dGPU from Nvidia, we use
  the CUDA SDK version 8.}

For processors with dedicated main memory, we cache all input data before running the variants to avoid biased observations because of PCIe transfers. 
{ Our goal is to evaluate the performance of queries on heterogeneous processors, rather than bottlenecks in current interconnects.}
We run each variant of a pipeline program 5 times and report the mean and the standard deviation. 
{ We prune the variant space if we detect a very slow variant (execution time greater than one second) to keep the run-time of the benchmark in a reasonable time frame. }

{\markchange{}
 {\color{black} 
As evaluation datasets, we use the Star Schema Benchmark~\cite{ssb:2009} and the
TPC-H Benchmark~\cite{tpc-h}. 
We use Scale Factor 1 for the experiments including a full exploration of all variants. With larger scale factors, poor performing variants would not finish in reasonable time.} As OpenCL does not provide any mechanism to abort a kernel, we have to wait until the kernel finishes.
{\color{black}For all other experiments, we use a scale factor of 10.
  The main memory of the iGPU usable by OpenCL is limited to 2.2GB and thus, we can not use a
  larger scale factor.
}
}


\subsection{Experimental Design}

{
We now discuss our {\color{black}evaluation workload} and the variants we generate for our test queries. 
}

\begin{figure}
\begin{lstlisting}[
    language=SQL,
    showspaces=false,
    basicstyle=\ttfamily,
    keywordstyle=\color{blue}\bf,
    breaklines=true,
    frame=single,
    label=lab:aggregation_query_1,
    captionpos=t,      
    caption={Grouped Aggregation Query 1}
        ]
select lo_shipmode, sum(lo_quantity) from lineorder group by lo_shipmode;
\end{lstlisting}

\begin{lstlisting}[
    language=SQL,
    showspaces=false,
    basicstyle=\ttfamily,
    keywordstyle=\color{blue}\bf,
    breaklines=true,
    frame=single,
    label=lab:aggregation_query_2,
    captionpos=t,      
    caption={Grouped Aggregation Query 2}
        ]
select lo_partkey, sum(lo_quantity) from lineorder group by lo_partkey;
\end{lstlisting}
\end{figure}

{
\subsubsection{Queries}
\label{sec:micro_benchmark_queries}


{%
All SQL queries can be split in a series of projection and aggregation pipelines. Thus, we evaluate our approaches for pipeline variant generation and optimization on simple queries representing a single pipeline. These single-pipeline queries allow for unbiased observation of hardware adaption using pipeline variations. 
Additionally, we validate the results on complex benchmark queries.
}



{
\textbf{Projection Pipelines.}
We take as representatives for projection pipelines one query with 50\ \% selectivity (Projection Query 1, cf. Listing~\ref{lab:projection_query_1}) and one filter query with very high selectivity ($<$0.01\ \%, Projection Query 2, cf. Listing~\ref{lab:projection_query_2}). While Projection Query 1 is read and write intensive, Projection Query 2 is read intensive.

\textbf{Aggregation Pipelines.}
As representatives for aggregation pipelines, we use one query with few result groups (Aggregation Query 1, cf. Listing~\ref{lab:aggregation_query_1}) and one query with many result groups, e.g., several hundred thousand (Aggregation Query 2, cf. Listing~\ref{lab:aggregation_query_2}). 
The first query is common for the final group by in an OLAP query. The second query is common in sub-queries using a group by (e.g., TPC-H Query 15). 
Having so many groups, Aggregation Query 2 is latency bound. 
}

{
\textbf{TPC-H Q1 and SSB Q4.1.}
We perform a full variant exploration for TPC-H Q1 and SSB Q4.1. The TPC-H query is a compute intensive aggregation query. 
It consists of a single pipeline with a FILTER, several ALGEBRA operations and a grouped aggregation with multiple aggregation functions. 
The SSB query is a join dominated query (four joins), consisting of four projection pipelines and one aggregation pipeline. 
The projection pipelines build the hash tables, whereas the aggregation pipeline probes each hash table.
}

{\color{black}
\textbf{Other Queries.}
We also evaluate the performance of other Star Schema Benchmark and TPC-H queries. Due
to current implementation restrictions of our prototype system (e.g., a missing
LIKE operator), we limit the evaluation queries to a representative subset. For the Star Schema
Benchmark, we use the queries Q1.1-Q1.3, Q3.2-Q3.4 and Q4.1-Q4.3. For the TPC-H benchmark, we use the queries Q5, Q6 and Q7.

\begin{figure*}[t]
\begin{center}
\input{kernel_compilation_times2.tex}
\end{center}
\end{figure*}


\subsubsection{Variant Space of Generated Variants}
\label{sec:variant_space}

For all pipeline types, we vary the memory access pattern (sequential and
coalesced) and the branch evaluation mode (branched predicate evaluation and
software predication). 
{The total number of variants multiply with each new variant dimension. 
We encode the number of variants in brackets [x variants].}

\textbf{Projection Pipelines.}
{\color{black}
For projection pipelines, we additionally vary the execution strategy (single
pass for coarse-grained parallelism and multi pass for
fine-grained parallelism) [2 variants]. For the single-pass strategy, we set the
number of parallel running pipelines to the number of \emph{maximal compute
  units} of the OpenCL device [1 variant]. Thus, we generate 4 variants that use the
single-pass strategy. 
The multi-pass strategy uses a multiplier (1, 8, 64, 256, 1024, 16384, 65536)
that is multiplied with the number of \emph{maximal compute units} of the OpenCL
device to calculate the number of threads [7 variants]. We generate 28 variants that use
the multi-pass strategy. In total, we generate 32 variants for a projection pipeline.
}

\textbf{Aggregation Pipelines.}
{\color{black}
For aggregation pipelines, we additionally vary the aggregation execution strategy, the hash table
implementation, and the hash function. For the hash table implementation, we vary
between linear probing and Cuckoo hashing [2 variants]. The hash function is either Murmur hashing
or Multiply-Shift hashing [2 variants]. For the execution strategy we vary between local and global
aggregation.}
{\markchange{}\color{black} In case of a local aggregation, we optimize the number of hash
  tables (1, 8, 64, 256, 1024, 16384, 65536) as a multiplier of the number of
  \emph{maximal compute units} of the OpenCL device
  to test different levels of thread over-subscription. We also
  optimize the number of threads per hash table (16, 32, 64, 128, 256, 512,
  1024) to find the best configuration between high parallelism and
  synchronization overhead [7x7 variants].} 
In case of global aggregation, we optimize the number of threads per hash table,
which configurations are identical to local aggregation [7 variants].
{\color{black} We generate [2x2x2x2x7x7 variants] for the local aggregation and
  [2x2x2x2x7 variants] variants for the global aggregation.}
In total, we generate {\color{black} 896 variants} for an aggregation pipeline.

}

\begin{figure*}[t]
\subfigure[Execution times of variants optimized for CPU, iGPU, dGPU, and MIC for Projection Query 1.]{\label{lab:projection_query_1_full_measurement}\input{projection_query_1_medium_selectivity.tex}}
\subfigure[Execution times of variants optimized for CPU, iGPU, dGPU, and MIC for Projection Query 2.]{\label{lab:projection_query_2_full_measurement}\input{projection_query_2_high_selectivity.tex}}
\subfigure[Execution times of variants optimized for CPU, iGPU, dGPU, and MIC for Aggregation Query 1.]{\label{lab:grouped_aggregation_query_1_full_measurement}\input{grouped_aggregation_query_1.tex}}
\caption{Execution times  of variants optimized for CPU, iGPU, dGPU, and MIC for different queries, executed on all processors. 
}
\end{figure*}

\subsection{Results}

{
We validate our concepts as follows. First, we evaluate kernel compilation times for all generated kernels. 
Second, we evaluate all variants on representative queries: two projection queries, two aggregation queries, TPC-H Query 1, and SSB Query 4.1. 
We determine the optimal variant of a pipeline program by performing a full search. {\color{black}This means} that we generate all possible variants for a pipeline and execute them multiple times. The in average fastest variant is reported in the plots as CPU, iGPU, dGPU, and MIC optimized.
Furthermore, we evaluate our learning strategy for automatic hardware adaption. We report the learned variant configurations and the query execution times on different processors.
}

\reviewer{Run a few experiments with more complex queries to demonstrate that compilation times, optimization times, and efficiency translate from the microbenchmark to more realistic queries.}

\reviewer{(If time permits) Explore how sensitive the choice of the best variant for an architecture is to the query workload. This would motivate the need (or lack thereof) for optimizing the compilation for a specific workload instead of only for a given architecture.}

\subsubsection{Compilation Times}
\label{sec:eval_compilation_times}






In Figure~\ref{pics:query_compilation_times_boxplot}, we show for each of our evaluation queries and processors the compilation time of all variants in a box plot. The boxes include 50\ \% of the observations, whereas the upper and lower whiskers mark a 99\ \% confidence interval. It becomes clearly visible that all compilation times for kernels for the projection query are below 70ms for CPU, iGPU, and dGPU, and below 100\ ms for the MIC processor. For the aggregation query, we observe that except for the MIC processor, kernel compilation times are either 100ms, or below. 
{ As we need to generate more code for aggregation pipelines compared to projection pipelines, the compilation time increases.}

{
Compiling TPC-H Query 1 takes longer compared to the aggregation queries. This is because the TPC-H query results in a larger kernel due to many additional computations. We observe 66ms on the CPU, 113ms on the iGPU, 216ms on the dGPU and 4.9s on the MIC. 
Compiling SSB Query 4.3 is even more time intensive, as we have to compile four projection and one aggregation pipelines.
We observe 245ms on the CPU, 380ms on the iGPU, 818ms on the dGPU and 1.8s on the MIC. Note that we can compile multiple pipelines in parallel to reduce the compilation time.

}

Compiling for the MIC is very expensive, and may take longer than a second, even for a single pipeline. However, this is the only processor {\color{black}where} we observed this behavior. We repeated our experiments on other machines using NVIDIA GPUs and Intel CPUs, and measured similar kernel compilation times reported here for CPU, iGPU, and dGPU. Thus, we assume that the high compilation time for the MIC is an implementation artifact, which we expect will be resolved in future versions of the Intel OpenCL SDK.

\begin{figure*}[t]
\begin{center}
\subfigure[Execution times of variants optimized for CPU, iGPU, dGPU, and MIC for Aggregation Query 2.]{\label{lab:grouped_aggregation_query_2_full_measurement}\input{grouped_aggregation_query_2.tex}}
\subfigure[Execution times of variants optimized for CPU, iGPU, dGPU, and MIC for TPC-H Query 1.]{\label{lab:tpch_query1_full_measurement}\input{tpch_query1.tex}}   
\subfigure[Execution times of variants optimized for CPU, iGPU, dGPU, and MIC for SSB Query 4.1.]{\label{lab:ssb41_full_measurement}\input{query_ssb41.tex}} 
\caption{Execution times  of variants optimized for CPU, iGPU, dGPU, and MIC for different queries, executed on all processors. 
}
\end{center}
\end{figure*}

\subsubsection{Full Variant Exploration}
\label{subsec:eval_full_exploration}


{
We show the run-time of all variants optimized for a particular processor and query. 
We show these variants for the following queries: the projection queries (Listing~\ref{lab:projection_query_1} and~\ref{lab:projection_query_2}), the aggregation queries (Listing~\ref{lab:aggregation_query_1} and~\ref{lab:aggregation_query_2}), TPC-H Query 1, and SSB Query 4.3. 
}

{
\textbf{Observations Projection Query 1}.
We show in Figure~\ref{lab:projection_query_1_full_measurement} that the
CPU-optimized variant outperforms the variants optimized for the iGPU, dGPU, and
MIC by a factor of {\color{black}25, 30, and 3.5}, respectively. However, we see
that the same implementation performs more slowly compared to optimized
variants on the iGPU, dGPU, and MIC by a factor of up to {\color{black}81, 237,
  and 3.1}, respectively. The large performance difference between CPU and the
other processors is mainly due to the execution strategy: CPUs prefer the
single-pass strategy using coarse-grained parallelism, whereas GPUs, and MICs
prefer the multi-pass strategy using fined-grained parallelism. On the iGPU, we
observe that the variant optimized for iGPU {\color{black}outperforms the variant optimized
for MIC by a factor of {\color{black}4.3}. For the dGPU optimized variant on the
iGPU, we observe that the performance is equal to the iGPU optimized variant.}
{\markchange{}The main difference among the variants optimized for iGPU, dGPU, and MIC is in the optimal number of threads. Furthermore, the MIC prefers sequential memory access, similar to CPUs, whereas the GPU variants prefer coalesced memory access.}
We do not observe further performance gaps on the dGPU and MIC processor. 
{\color{black}Our learning strategy found a configuration that performs closely to the optimal variant on CPU, iGPU and dGPU. On the MIC, the found variant is by a factor of 1.8 slower than the optimum.}

{ 
\textbf{Observations Projection Query 2}.
In Figure~\ref{lab:projection_query_2_full_measurement}, we make the same basic
observation for Projection Query 2 (very high selectivity, $<$0.001\ \%) as
for Projection Query 1 (50\ \% selectivity). The CPU-optimized variant
outperforms variants optimized for iGPU, dGPU, and MIC by a factor of
12, 12, and 14, respectively. On the other processors, the
CPU-optimized variant is slower by a factor of 118, 345, 2.4 on
the iGPU, dGPU, and MIC, respectively.
Our learning strategy found a configuration that performs closely to the optimal variant on CPU, iGPU, dGPU, and MIC.

}

\begin{table}[t]
\color{black}
\begin{center}
{
\begin{tabular}{ | c | c | c | c | }
\hline
Processor &  Backtracking & Feature-Wise & Factor\\
 &    (in seconds) & (in seconds) & Improved\\
\hline
CPU &  197,139 & 479 & 411\\
iGPU & 78,388 & 1,219 & 64.3\\
dGPU & 52,897 & 1,036 & 51\\
MIC & 177,914 & 3,390 & 52.5\\
\hline
\end{tabular}
}
\end{center}       
	\caption{Variant exploration times for SSB Q4.1 on SF1. Our learning strategy outperforms the backtracking search by up to two orders of magnitude.}   
        \label{table:exploration_time}
\end{table}

{\color{black}
\setlength{\tabcolsep}{1pt}
\begin{table*}
\color{black}
\begin{center}
{
\begin{tabular}{ | c | c | c | c | c | }
\hline
Variation Dimension &  \multicolumn{3}{c|}{Learned Optimizers} \\
                                   & cpu-o            & dgpu-o         & mic-o\\
\hline
Execution Strategy (Projection)    & single-pass    & multi-pass     & multi-pass\\
Execution Strategy (Aggregation)   & local hash table   & local hash table & local hash table \\
Memory Access Pattern              & sequential     & coalesced      & coalesced\\
Hash Table Implementation          & linear probing & Cuckoo hashing & Cuckoo hashing\\
Predication Mode (Query Dependent) & branched       & branched     & branched\\
Thread Multiplier (Projection)     & 1 & 16384 & 65536\\
Thread Multiplier (Aggregation)    & 1 & 1 & 1\\
Work Group Size (Aggregation)      & 256 & 1024 & 64\\ 
\hline
\end{tabular}
}
\end{center}       
	\caption{Per processor variant configurations identified by the variant learning strategy for the SSB and TPC-H workload.} 
        \label{table:heuristics}
\end{table*}

}

{
\textbf{Observations Aggregation Query 1}.
We show in Figure~\ref{lab:grouped_aggregation_query_1_full_measurement} that
{\color{black}on the CPU} the CPU-optimized variant outperforms variants optimized for iGPU, dGPU, and MIC
by a factor of {\color{black}16, 24.4, and 7.6, respectively}. 
However, we see that the same implementation performs significantly slower compared to optimized variants on the iGPU, dGPU, and MIC by a factor of up to {\color{black}606, 861, and 24}, respectively.
We also see significant differences between the variants optimized for iGPU, dGPU, and MIC: On the iGPU, the iGPU-optimized variant outperforms the variants optimized for dGPU and MIC by a factor of {\color{black}1.2, and 6}, respectively. On the dGPU, the dGPU-optimized variant outperforms variants optimized for iGPU and MIC by a factor of {\color{black}1.2, and 6}. On the MIC, the MIC-optimized variant outperforms variants optimized for iGPU and dGPU by a factor of {\color{black}7.6 and 7.6}, respectively.
{\color{black}Our learning strategy found a configuration that performs closely to the optimal variant on CPU, iGPU and dGPU. On the MIC, the found variant is by a factor of 3.4 slower than the optimal variant.}
  
{  
\textbf{Observations Aggregation Query 2}.
We make the same basic observation as for Aggregation Query~1 (cf. Figure~\ref{lab:grouped_aggregation_query_2_full_measurement}): On the CPU, the CPU-optimized variant outperforms variants optimized for iGPU, dGPU, and MIC by a factor of {\color{black}1.7, 1.7, and 1.6}, respectively. The same CPU-optimized variant is significantly slower compared to variants optimized for iGPU, dGPU, and MIC by a factor of up to {\color{black}94, 94, and 23}. Note that for this query, the optimal variant of iGPU and dGPU is the same, thus we will report numbers only once (GPU). On the GPUs, the GPU-optimized variant outperforms the MIC by a factor of {\color{black}1.1} (iGPU) and {\color{black}1.2} (dGPU). 
On the MIC, the MIC-optimized variant achieves the same performance as the GPU variant. Our learning strategy found a configuration that performs closely to the optimal variant on CPU, iGPU, dGPU and MIC.
}

{

\textbf{Observations on Complex Queries.} 
We show that the variant exploration has the same impact on more complex queries. 
Thus, we present the result of the variant exploration for two OLAP queries: TPC-H Q1 (cf. Figure~\ref{lab:tpch_query1_full_measurement}) and SSB Q4.1 (cf. Figure~\ref{lab:ssb41_full_measurement}). 
For every processor, we observe similar factors between the optimal variant and the other variants. 
We also see that on the MIC, the variance of execution time for some variants is very high. We do not observe this issue on any other processor.
}


{
\subsubsection{Optimization Time}
\label{sec:eval_optimization_time}

\reviewer{There's no comment about how long it takes in the "installation" process -- uncovering what parametrization works best for a give scenario -- that sectoin 8.1 suggests.}

We now investigate how long the variant exploration itself takes. We show in Table~\ref{table:exploration_time} the time to explore the best variant for SSB Query 4.1. 
We compare backtracking (executing every possible variant and choosing the fastest) with our learning strategy. 
We observe that our strategy improves the search time by up to a factor of 411. 
While the longest exploration took more than two days, our strategy finished within an hour. 
Thus, we can run our calibration benchmarks offline (e.g., as part of the database installation process). 
}

{
\subsubsection{Hardware Adaption on Full Queries}
\label{sec:eval_hardware_adaption}


\textbf{Learned Variant Configurations.}
We derive pro\-cessor-specific optimizers using our learning strategy from Section~\ref{sec:variant_selection}. We explore the same variant space as the full exploration, which we discuss in Section~\ref{sec:variant_space}.
As training workload, we use the Query Groups 1, 3, and 4 of the Star Schema Benchmark and Queries 5, 6, 7 from the TPC-H benchmark. In this experiment, we use a scale factor of 10 for both benchmarks.
We show the learned variant configurations optimized for the CPU (cpu-o), for the dGPU (dgpu-o) and for the MIC (mic-o) in Table~\ref{table:heuristics}. 

The learning strategy correctly identifies that for projection pipelines, CPUs prefer single-pass strategies with coarse-grained parallelism, whereas the GPUs and the MIC prefer multi-pass strategies with fine-grained parallelism. For aggregation pipelines, the CPU, GPU, and MIC prefer local hash table aggregation. The learning strategy also found that the CPU prefers sequential memory access, whereas the GPUs and MIC prefer coalesced memory access. The preferred aggregation hash table for CPUs is linear probing, whereas the GPUs and the MIC are more efficient when using Cuckoo hashing. For the query workload, the learning strategy found that the branched predication mode (evaluation using if-statements) outperforms variants that use software predication. 

We implement the number of threads as a multiplier of the number of OpenCL compute units (``cores''), as the multiplier quantifies the degree of over-subscription required for a processor. 
CPUs prefer no over-subscription (one thread per core), whereas the GPUs and the MIC need a large multiplier (over-subscription) to have enough thread blocks ready to hide memory access latencies.
Additionally, we need to specify the work group size for aggregation pipelines, which also strongly differs between the different processors. 
}


{\markchange{}

\input{table_ssbm_sf10_measurements_generic.tex}

{
\textbf{Performance.}
We execute for each query a variant optimized for CPU, dGPU, and MIC and measure the execution times on CPU, dGPU, and MIC without compilation times. 
{ Note that each variant is optimized for a complete workload (cpu-o, dgpu-o, and mic-o). We call these variants per-workload variants. We include measurements of a per-query optimized variant for each query (q-o) to show additional optimization potential compared to the per-workload variants.}

We illustrate the results in Table~\ref{table:selected_ssb_queries} and include measurements of HyPer (v0.5-222-g04766a1) with the same queries on the same dataset on the CPU.\footnote{Note that this comparison is not intended to be an end-to-end measurement of system performance.} We observe that the code generated by Hawk on a CPU is in the same order of magnitude as an optimized state-of-the-art query compiler. 
%
%
%


Most queries are executed faster when we use the per-workload variant of the target processor. 
On the CPU, the performance of a CPU-optimized variants outperforms GPU and MIC-optimized variants by up to a factor of 5.5 (SSB Query 3.4). On the GPU, the GPU-optimized variant outperforms the other per-workload variants by up to a factor of 9 (SSB Query 3.2). On the MIC, the MIC-optimized variant outperforms the other per-workload variants by up to a factor of 1.12 (SSB Query 3.2). 
The reason for this low factor is that GPU variants are typically also fast on a MIC (but not the other way around). However, we can still improve the performance with a custom variant for the MIC.
We occasionally observe a better performance of another variant for some queries, such as TPC-H Q5 and Q6 for the CPU. The reason for this is that a variant optimized for several queries may be sub-optimal for a particular query.
We conclude that we achieve the best performance when we use a variant optimized for a target processor.

If we additionally tune the variant to a particular query, we observe for our workload speedups on the CPU by up to a factor of 1.02, on the GPU by up to a factor of 27 (TPC-H Query 5), and on the MIC by up to a factor of 1.43 (SSB Query 3.2).
The per-query variants differ mainly in the thread multipliers, as different degrees of parallelism are optimal for different queries on GPU and MIC. Additionally, some queries are faster with enabled predication or prefer global instead of local hash table aggregation specific queries, such as TPC-H Q5 on GPUs and MICs. We conclude that the optimal variant is query-dependent on GPU and MIC. On the CPU, a per-query variant provides small benefit over a generic per-workload variant.
}

\textbf{Summary.}
In general, we conclude that we need to create custom variants per processor in order to reach peak performance. 
Furthermore, we observe that the algorithm derives an efficient configuration, but the optimal variant is to some extend query dependent, e.g., number of threads, branching mode~\cite{Raducanu:2013:MAV:2463676.2465292} and hash table implementation~\cite{Richter:2015:SAH:2850583.2850585}.
This limitation can be lifted by adding a run-time optimizer that performs per-query variant optimization, similar to the work of Raducanu~\cite{Raducanu:2013:MAV:2463676.2465292} and Zeuch~\cite{DBLP:journals/pvldb/ZeuchPF16}.
}

\subsection{Discussion}

{\color{black}
In our experiments, we observed that most compilation times for single pipelines are very fast ($<100$\ ms). OpenCL could compile even complex queries in several hundred milliseconds, if we disregard vendor-specific artifacts.  
We conclude that efficient query compilation is possible using OpenCL. This ensures that the database engine still allows for interactive querying despite using query compilation. 
}

Furthermore, we observe large performance differences among variants optimized for a CPU, a GPU, and a MIC by up to two orders of magnitude. 
Thus, we conclude that a hardware-adaptive query compiler can achieve high performance gains. 
This is because it can optimize for various processors of different architectures with previously unknown performance behavior \emph{without any manual tuning}.

The diversity of the optimized variants shows that we need to support the discussed dimensions of variant generation and their individual variations.  
%

Finally, we find that our learning strategy detected all major preferences of all processors. 
The strategy derived efficient per-processor variants without having to explore all variants. 
In future work, we will research a run-time optimizer that also captures query-dependent parameters. 

We observed on the MIC that the execution time of some variants have a high variance (up to a factor of four). 
{We could not pinpoint the exact cause. 
We run the same OpenCL code on different CPUs and GPUs with different OpenCL vendors and observe this behavior with the MIC only. Thus, the cause is likely to be an implementation artifact of the Intel OpenCL SDK. 
The unreliable performance behavior of the MIC processor makes it difficult to find a fast variant on the MIC for any search strategy. 
We mitigate the problem using outlier detection on the execution times. 
However, with growing randomness of the processors performance, optimization gets increasingly difficult.
}


\section{Related Work}



In this section, we discuss related work on query compilation, compiling programs to heterogeneous processors, data processing on heterogeneous processors, and automatic optimization of variants.

\subsection{Query Compilation}

Query compilation goes back to System R~\cite{Chamberlin:1981:HES:358769.358784}, and was re-investigated in the 80s~\cite{Freytag:1989:TRQ:62032.62033}. 
With the upcoming of main-memory databases, query compilation received new attention as reducing main-memory traffic and executed CPU instructions became increasingly important.
Rao and others generated query-specific code using the just-in-time compilation capabilities of Java~\cite{Rao:2006:CQE:1129754.1129884}.
Krikellas and others used a template-based code generation approach to compile queries to C code, which was then compiled by a C compiler to machine code~\cite{DBLP:conf/icde/KrikellasVC10}. 
Neumann introduced the produce/consume model, which provides a systematic way to generate code that allows for data-centric query processing by fusing all operators in an operator pipeline. Additionally, Neumann proposed to generate LLVM IR code instead of C code to achieve low compilation times~\cite{Neumann:2011:ECE:2002938.2002940}.
Leis and others proposed the morsel framework, which introduces NUMA-aware parallelization of compiled operator pipelines~\cite{Leis:2014:MPN:2588555.2610507}.

Sompolski and others carefully studied vectorized and compiled execution~\cite{Sompolski:2011:VVC:1995441.1995446}. They observe that compilation is not always superior to vectorization and conclude that compilation should always be combined with block-wise query processing.
Dees and Sanders compiled the 22 TPC-H queries by hand to C code and showed large performance potentials for query compilation~\cite{conf_icde_DeesS13}.
{ Nagel and others investigated query compilation in the context of language-integrated que\-ries in managed run-times~\cite{DBLP:journals/pvldb/NagelBV14}.
Amad and others developed DBToaster, which uses code generation to compile view maintenance queries to efficient machine code~\cite{Ahmad:2009:DSC:1687553.1687592}.}
Query compilation also found its way into commercial products such as Hekaton~\cite{DBLP:journals/debu/FreedmanIL14} and Impala~\cite{wanderman2014runtime}.

{
Weld is a run-time that efficiently executes data-intensive applications~\cite{weld_2017}. The key idea is to compile code to a common intermediate representation. 
Weld removes data movement between functions in a workflow and generates efficient parallel code for CPUs. 
In contrast to Hawk, Weld cannot generate custom code for different heterogeneous processors. 
However, Welds code-generation backend can be enriched by the variant generation concepts introduced in this paper to efficiently support GPUs and MICs.
}

\subsection{Query Compilation for CPUs and GPUs}

Wu and others proposed Kernel Weaver, a compiler framework that can automatically fuse the kernels of relational operators and kernels of other domains~\cite{Wu:2012:KWA:2457472.2457490}. In contrast to kernel weaver, Hawk uses our concept of execution strategies to generate a minimal number of kernels. We see Kernel Fusion as a complementary building block. 
{ Another key difference is that Kernel Weaver targets GPUs only, whereas Hawk executes efficiently on CPUs, GPUs, and MICs.}

Rauhe and others proposed the compute/accumulate model to compile queries to GPU code \cite{DBIS:Rauhe2013adbis}. Here, query operations are put in three kernels. The compute phase generates a local compute and a local accumulate kernel. The accumulate phase generates a global accumulate kernel. 
{In contrast to Hawk, they cannot generate highly tailored code for CPUs and GPUs.}

A new line of research focuses on writing database systems in a high-level language~\cite{DBLP:journals/debu/Koch14}. The LegoBase system uses generative programming to generate efficient low-level C code for a database implementation in a high-level language~\cite{Klonatos:2014:BEQ:2732951.2732959}. Shaikhha and others further refine this principle in DBLAB~\cite{Shaikhha:2016:AQC:2882903.2915244} by introducing a stack of multiple Domain Specific Languages (DSLs) that differ in the levels of abstraction. Here, high-level code is progressively lowered to low-level code, by compiling code in multiple stages, where each stage compiles to a DSL of lower abstraction level, until the final code is generated. 

{

Pirk and others propose the Voodoo framework, which consists of an intermediate algebra representation based on vectors and a code generator for OpenCL~\cite{Pirk:2016:VVA:3007328.3007336}. Based on the algebra, Voodoo is capable of generating code for different processors, including CPUs and GPUs. The voodoo algebra and our pipeline programs are conceptually similar, albeit on different abstraction levels. Note that pipeline programs and voodoo algebra are complementary, as we could generate voodoo algebra from pipeline programs. The key difference between Hawk and Voodoo is that Hawk creates a large space of potential variants and can systematically fine-tune the generated code to the underlying processors. 
}

In summary, existing query compilation approaches generate efficient code for a single processor. Hawk is the first hardware-adaptive query compiler that can produce variants of code to run efficiently on different processors.

\subsection{Compilers}


Brown and others developed Delight, a framework that allows to build, compile and execute DSLs which enable users to program at a high-abstraction level~\cite{Brown:2011:HPF:2120965.2121410}. 
{
The key idea is to compile domain-specific languages to a common intermediate representation. From the intermediate representation, Delight generates code for CPUs and GPUs. 
However, Delight does not optimize for heterogeneous processors to the degree Hawk does, such as changing execution strategies. The concepts of Delight and Hawk complement each other.
}

Dandelion is a general purpose compiler based on .NET LINQ that compiles data-parallel programs to multiple heterogeneous processors, such as CPUs, GPUs, and FPGAs and automatically distributes data processing on different processors, be it in a single machine or a cluster \cite{Rossbach:2013:DCR:2517349.2522715}. 
{ While Dandelion uses cross compilation to support GPUs, Hawk profits from the functional portability of OpenCL, which allows Hawk to run code on any OpenCL-capable processor. At the same time, Hawk can generate custom code for different processors using the same code generator.}

{ Jacc is a compiler framework that can compile and run Java code on GPUs. Jacc compiles Java byte code directly to NVIDIA's PTX code and manages computation and data transfers transparantly to the user~\cite{Clarkson2017}. In contrast to Jacc, Hawk generates variants of code to find a high-performance implementation.}

\subsection{Databases on Heterogeneous Hardware}


Balkesen and others studied efficient hash joins~\cite{DBLP:conf/icde/BalkesenTAO13} and sort-merge joins on multi-core CPUs~\cite{DBLP:journals/pvldb/BalkesenATO13}. 
He and others developed efficient algorithms for joins~\cite{He:2008:RJG:1376616.1376670,He:2013:RCH:2536206.2536216} and other relational operators~\cite{He:2009:RQC:1620585.1620588} on GPUs. 
{ He and others also studied efficient co-processing on APUs~\cite{DBLP:journals/pvldb/HeZH14}}.
{\color{black}Pirk and others studied common database operations on the Intel Xeon Phi (MIC) and compared them to GPUs~\cite{Pirk:2015:FSY:2771937.2771944}.}
Jha and others investigated hash joins on the Intel Xeon Phi~\cite{Jha:2015:IMM:2735703.2735704}.

{Paul and others investigated the effect of pipelining between multiple GPU kernels using the channel mechanism provided by OpenCL 2.0 pipes~\cite{Paul:2016:GGP:2882903.2915224}.

Meraji and others implemented support for GPU acceleration into DB2 with BLU acceleration and observed significant performance gains using GPUs for query processing~\cite{Meraji:2016:THD:2882903.2903735}.} 

{
Karnagel and others analyzed hash-based grouping and aggregation on GPUs~\cite{DBLP:conf/vldb/KarnagelML15}. This work was the basis for Hawk's execution strategies for grouped aggregation.
}
{M{\"{u}}ller and others studied database query processing on FPGAs~\cite{DBLP:journals/pvldb/MuellerTA09} and developed Glacier, a query compiler that generates logic circuits for queries to accelerate stream processing~\cite{DBLP:journals/pvldb/MullerTA09}.}

Many database prototypes were developed to study different aspects of query processing on CPUs and GPUs, such as GDB~\cite{He:2009:RQC:1620585.1620588}, GPUDB~\cite{Yuan:2013:YYP:2536206.2536210}, OmniDB~\cite{Zhang:2013:OTP:2536274.2536319}, Ocelot~\cite{ocelot2013}, CoGaDB~\cite{BressT15}, and HeteroDB~\cite{DBLP:journals/jcst/ZhangCDHLLWYZ15}.


{
Pirk and others introduced the approximate and refine technique, which only keeps the higher bits of a value in GPU memory, which allows to keep more data on GPU. The lossily compressed data is stored on the GPU~\cite{Pirk:2014:ICDE}. Queries on the GPU return an approximate result, which has to be refined on the CPU using the uncompressed data. 
}

Heimel and others showed the feasibility of building a database engine in OpenCL, which allows to run a database engine with the same operator code base on any OpenCL-capable processor~\cite{ocelot2013}. To achieve high efficiency on CPUs and GPUs, they left the memory access pattern configurable so they can adapt it to the processor type. 
{\color{black}The core difference between Ocelot and Hawk is that Ocelot provides the same operator implementations for each processor, while Hawk can generate custom per-processor variants for each query.}

{ All the aforementioned techniques improve the absolute performance of a database system running on heterogeneous processors. These optimizations are orthogonal to the concepts presented in this paper.
}

\subsection{Variant Optimization}

Raducanu and others propose Micro Adaptivity, a framework that provides alternative function implementations called \emph{flavors} (equivalent to our term variant)~\cite{Raducanu:2013:MAV:2463676.2465292}. Micro Adaptivity exploits the vector-at-a-time processing model and can potentially exchange a flavor at each function call, which allows for finding the best implementation for a certain query and data distribution.

Rosenfeld and others showed for selection and aggregation operations that many operator variants can be generated and that different code transformations are optimal for a particular processor \cite{rosenfeld:2015:variant_selection}. 

Zeuch and others propose to use performance counters of modern CPUs for progressive optimization. They introduce cost models for cache accesses and branch mispredictions and derive selectivities of predicates at query run-time to re-optimize predicate evaluation orders~\cite{DBLP:journals/pvldb/ZeuchPF16}. 

{The techniques for variant optimization from Raducanu~\cite{Raducanu:2013:MAV:2463676.2465292}, Rosenfeld~\cite{rosenfeld:2015:variant_selection}, and Zeuch~\cite{DBLP:journals/pvldb/ZeuchPF16} are orthogonal to the variant generation of this paper.}


\section{Summary}


{\color{black}
In this paper, we describe a hardware-adaptive query compiler that can generate code for a wide range of heterogeneous processors. 
Through hardware-tailored implementations, our query compiler produces fast code \emph{without manual tuning} for a specific processor.


Our key findings are as follows. 
Our abstraction of pipeline programs allows us to flexibly produce variants of pipelines while keeping a clean interface and a maintainable code base.
Pipeline variants optimized for a particular processor can result in performance differences of up to two orders of magnitude on the same processor. 
Therefore, it is crucial to optimize the query compiler to each processor.
Consequently, we proposed a learning strategy that automatically derives an efficient variant configuration for a processor.
Based on this algorithm, we derived efficient variant configurations for three common processors. 
Finally, we incorporated the variant configurations into a heuristic query optimizer. 
}

\section*{Acknowledgment}
{\footnotesize
We thank Manuel Renz, Tobias Fuchs, Martin Kiefer, and Viktor Rosenfeld from Technische Universit{\"a}t Berlin for helpful feedback and discussions.
The work has received funding from the European Union's Horizon2020 Research \& Innovation Program under grant agreement 671500 (project ``SAGE''), from the German Ministry for Education and Research as Berlin Big Data Center BBDC (funding mark\\ 01IS14013A), and from the German Research Foundation (DFG), Collaborative Research Center SFB 876, project C5 and the DFG Priority Program ``Scalable Data Management for Future Hardware''.
}



%

\normalsize

\end{document}

%% file: listing_color_line_background.tex
   \usepackage{ucs}
    \usepackage{pgf}
    \usepackage{pgffor}

    \usepackage{mdframed}
    \definecolor{bggray}{rgb}{0.85, 0.85, 0.85}
    \BeforeBeginEnvironment{lstlisting}{\begin{mdframed}[
  topline=false,
  bottomline=false,  
  rightline=false,
  leftline=false]\vskip-.5\baselineskip}
    \AfterEndEnvironment{lstlisting}{\end{mdframed}}

    \makeatletter
        \newcount\bt@rangea
        \newcount\bt@rangeb

        \newcommand\btIfInRange[2]{%
            \global\let\bt@inrange\@secondoftwo%
            \edef\bt@rangelist{#2}%
            \foreach \range in \bt@rangelist {%
                \afterassignment\bt@getrangeb%
                \bt@rangea=0\range\relax%
                \pgfmathtruncatemacro\result{ ( #1 >= \bt@rangea) && (#1 <= \bt@rangeb) }%
                \ifnum\result=1\relax%
                    \breakforeach%
                    \global\let\bt@inrange\@firstoftwo%
                \fi%
            }%
            \bt@inrange%
        }

        \newcommand\bt@getrangeb{%
            \@ifnextchar\relax%
            {\bt@rangeb=\bt@rangea}%
            {\@getrangeb}%
        }

        \def\@getrangeb-#1\relax{%
            \ifx\relax#1\relax%
                \bt@rangeb=100000
            \else%
                \bt@rangeb=#1\relax%
            \fi%
        }

\definecolor{codeblue}{RGB}{100, 234, 161}

        \newcommand{\btLstHL}[1]{%
            \btIfInRange{\value{lstnumber}}{#1}%
            {\color{codeblue}}%
            {\def\lst@linebgrd}%
        }%

        \lst@Key{numbers}{none}{%
            \def\lst@PlaceNumber{\lst@linebgrd}%
            \lstKV@SwitchCases{#1}{%
                none&\\%
                left&\def\lst@PlaceNumber{\llap{\normalfont
                \lst@numberstyle{\thelstnumber}\kern\lst@numbersep}\lst@linebgrd}\\%
                right&\def\lst@PlaceNumber{\rlap{\normalfont
                \kern\linewidth \kern\lst@numbersep
                \lst@numberstyle{\thelstnumber}}\lst@linebgrd}%
            }{%
                \PackageError{Listings}{Numbers #1 unknown}\@ehc%
            }%
        }

        \lst@Key{linebackgroundcolor}{}{%
            \def\lst@linebgrdcolor{#1}%
        }
        \lst@Key{linebackgroundsep}{0pt}{%
            \def\lst@linebgrdsep{#1}%
        }
        \lst@Key{linebackgroundwidth}{\linewidth}{%
            \def\lst@linebgrdwidth{#1}%
        }
        \lst@Key{linebackgroundheight}{\ht\strutbox}{%
            \def\lst@linebgrdheight{#1}%
        }
        \lst@Key{linebackgrounddepth}{\dp\strutbox}{%
            \def\lst@linebgrddepth{#1}%
        }
        \lst@Key{linebackgroundcmd}{\color@block}{%
            \def\lst@linebgrdcmd{#1}%
        }

        \newcommand{\lst@linebgrd}{%
            \ifx\lst@linebgrdcolor\empty\else
                \rlap{%
                    \lst@basicstyle
                    \color{-.}
                    \lst@linebgrdcolor{%
                        \kern-\dimexpr\lst@linebgrdsep\relax%
                        \lst@linebgrdcmd{\lst@linebgrdwidth}{\lst@linebgrdheight}{\lst@linebgrddepth}%
                    }%
                }%
            \fi
        }

%% file: coarse_vs_fined_grained_parallelism.tex
\begin{figure}[t]
        \begin{tikzpicture}[every text node part/.style={align=center}, font=\barchartfontsize,        
       		bar width=10pt,
]
                \begin{groupplot}[group style={group name=my plots,group size= 3 by 1,vertical sep=1.5cm },height=\barchartheight,width=3cm,cycle list name=barplot cycle list,
		area legend, 
		ybar=0pt]
                        \nextgroupplot[legend style={at={(0.38,-1)}, anchor=south}, ylabel={Execution\\ Time in s}, xtick=\empty, xticklabels=\empty,xmin=-3, xmax=3, ymin=0, bar width=10pt, cycle list name=barplot cycle list,   area legend, 
        ymin=0, 
        ymax=0.5,
                         nodes near coords={\pgfmathprintnumber[precision=2,zerofill,fixed]{\pgfplotspointmeta}},  
                         every node near coord/.append style={rotate=90, anchor=west, /pgf/number format/precision=2, xshift=0.00cm},  
]
    \addplot+[
        error bars/.cd,
        y dir=both,
        y explicit,
    ]
    table[
            x=id,
            y=Mean,
            y error=Stdev,
            col sep=semicolon
    ]
    {fine-vs-coarse_parallelism_cpu_coarse_grained.csv};	 \label{plots:coarse_grained_parallelism}
    \addplot+[
        error bars/.cd,
        y dir=both,
        y explicit
    ]
    table[
            x=id,
            y=Mean,
            y error=Stdev,
            col sep=semicolon
    ]
    {fine-vs-coarse_parallelism_cpu_fine_grained.csv}; \label{plots:fine_grained_parallelism}	
                                \coordinate (top) at (rel axis cs:0,1);
                        \nextgroupplot[xtick=\empty, xticklabels=\empty,xmin=-3, xmax=3, ymin=0, bar width=10pt, cycle list name=barplot cycle list,   area legend,
        ymin=0, 
        ymax=9,
                         nodes near coords={\pgfmathprintnumber[precision=2,zerofill,fixed]{\pgfplotspointmeta}},  
                         every node near coord/.append style={rotate=90, anchor=west, /pgf/number format/precision=2, xshift=0.00cm},  
    ]
    \addplot+[
        error bars/.cd,
        y dir=both,
        y explicit
    ]
    table[
            x=id,
            y=Mean,
            y error=Stdev,
            col sep=semicolon
    ]
    {fine-vs-coarse_parallelism_dgpu_coarse_grained.csv};	 
    \addplot+[
        error bars/.cd,
        y dir=both,
        y explicit
    ]
    table[
            x=id,
            y=Mean,
            y error=Stdev,
            col sep=semicolon
    ]
    {fine-vs-coarse_parallelism_dgpu_fine_grained.csv}; 
                        \nextgroupplot[xtick=\empty, xticklabels=\empty,xmin=-3, xmax=3, ymin=0, bar width=10pt, cycle list name=barplot cycle list,   area legend,
                         ymax=1.9,
                         nodes near coords={\pgfmathprintnumber[precision=2,zerofill,fixed]{\pgfplotspointmeta}},  
                         every node near coord/.append style={rotate=90, anchor=west, /pgf/number format/precision=2, xshift=0.00cm},  
    ]
    \addplot+[
        error bars/.cd,
        y dir=both,
        y explicit
    ]
    table[
            x=id,
            y=Mean,
            y error=Stdev,
            col sep=semicolon
    ]
    {fine-vs-coarse_parallelism_phi_coarse_grained.csv};	 
    \addplot+[
        error bars/.cd,
        y dir=both,
        y explicit
    ]
    table[
            x=id,
            y=Mean,
            y error=Stdev,
            col sep=semicolon
    ]
    {fine-vs-coarse_parallelism_phi_fine_grained.csv}; 
\coordinate (bot) at (rel axis cs:1,0);
                \end{groupplot}
                \node[below = 0.1cm of my plots c1r1.south] {CPU};
                \node[below = 0.1cm of my plots c2r1.south] {dGPU};
                \node[below = 0.1cm of my plots c3r1.south] {MIC};               
        \path (top|-current bounding box.south)--
                    coordinate(legendpos)
                    (bot|-current bounding box.south);
        \matrix[
                matrix of nodes,
                anchor=south,
                draw,
                inner sep=0.2em,
                draw
            ]at([xshift=-2ex, yshift=-4ex]legendpos)
            {
                \ref{plots:coarse_grained_parallelism}  & Coarse-Grained Parallelism  &[5pt]
                \ref{plots:fine_grained_parallelism} & Fine-Grained Parallelism\\};
        \end{tikzpicture}
        \caption{Impact of different execution strategies on Projection Query 2 (cf. Listing~\ref{lab:projection_query_2}).}
\label{pic:coarse_grained_vs_fine_grained_parallelism}
\end{figure}

%% file: sequential_vs_coalesced_mem_access.tex
\begin{figure}[t]
        \begin{tikzpicture}[every text node part/.style={align=center}, font=\barchartfontsize,        
       		bar width=10pt,
]
                \begin{groupplot}[group style={group name=my plots,group size= 3 by 1,vertical sep=1.5cm },height=\barchartheight,width=3cm,cycle list name=barplot cycle list,
		area legend, 
		ybar=0pt]
                        \nextgroupplot[legend style={at={(0.38,-1)}, anchor=south}, ylabel={Execution\\ Time in s}, xtick=\empty, xticklabels=\empty,xmin=-3, xmax=3, ymin=0, bar width=10pt, cycle list name=barplot cycle list,   area legend, 
        ymin=0, 
        ymax=0.55,
                         nodes near coords={\pgfmathprintnumber[precision=2,zerofill,fixed]{\pgfplotspointmeta}},  
                         every node near coord/.append style={rotate=90, anchor=west, /pgf/number format/precision=2, xshift=0.00cm},  
]
    \addplot+[
        error bars/.cd,
        y dir=both,
        y explicit,
    ]
    table[
            x=id,
            y=Mean,
            y error=Stdev,
            col sep=semicolon
    ]
    {memory_access_pattern_cpu_sequential.csv};	 \label{plots:sequential_mem_access}
    \addplot+[
        error bars/.cd,
        y dir=both,
        y explicit
    ]
    table[
            x=id,
            y=Mean,
            y error=Stdev,
            col sep=semicolon
    ]
    {memory_access_pattern_cpu_coalesced.csv}; \label{plots:coaledced_mem_access}	
                                \coordinate (top) at (rel axis cs:0,1);
                        \nextgroupplot[xtick=\empty, xticklabels=\empty,xmin=-3, xmax=3, ymin=0, bar width=10pt, cycle list name=barplot cycle list,   area legend,
        ymin=0, 
        ymax=0.1,
        yticklabels={0, 0.05, 0.1},
	ytick={0, 0.05, 0.1},   
                         nodes near coords={\pgfmathprintnumber[precision=2,zerofill,fixed]{\pgfplotspointmeta}},  
                         every node near coord/.append style={rotate=90, anchor=west, /pgf/number format/precision=2, xshift=0.00cm},  
    ]
    \addplot+[
        error bars/.cd,
        y dir=both,
        y explicit
    ]
    table[
            x=id,
            y=Mean,
            y error=Stdev,
            col sep=semicolon
    ]
    {memory_access_pattern_dgpu_sequential.csv};	 
    \addplot+[
        error bars/.cd,
        y dir=both,
        y explicit
    ]
    table[
            x=id,
            y=Mean,
            y error=Stdev,
            col sep=semicolon
    ]
    {memory_access_pattern_dgpu_coalesced.csv}; 
                        \nextgroupplot[xtick=\empty, xticklabels=\empty,xmin=-3, xmax=3, ymin=0, bar width=10pt, cycle list name=barplot cycle list,   area legend,
                         ymax=0.5,
                         nodes near coords={\pgfmathprintnumber[precision=2,zerofill,fixed]{\pgfplotspointmeta}},  
                         every node near coord/.append style={rotate=90, anchor=west, /pgf/number format/precision=2, xshift=0.00cm},  
    ]
    \addplot+[
        error bars/.cd,
        y dir=both,
        y explicit
    ]
    table[
            x=id,
            y=Mean,
            y error=Stdev,
            col sep=semicolon
    ]
    {memory_access_pattern_phi_sequential.csv};	 
    \addplot+[
        error bars/.cd,
        y dir=both,
        y explicit
    ]
    table[
            x=id,
            y=Mean,
            y error=Stdev,
            col sep=semicolon
    ]
    {memory_access_pattern_phi_coalesced.csv}; 
\coordinate (bot) at (rel axis cs:1,0);
                \end{groupplot}
                \node[below = 0.1cm of my plots c1r1.south] {CPU};
                \node[below = 0.1cm of my plots c2r1.south] {dGPU};
                \node[below = 0.1cm of my plots c3r1.south] {MIC};               
        \path (top|-current bounding box.south)--
                    coordinate(legendpos)
                    (bot|-current bounding box.south);
        \matrix[
                matrix of nodes,
                anchor=south,
                draw,
                inner sep=0.2em,
                draw
            ]at([xshift=-2ex, yshift=-4ex]legendpos)
            {
                \ref{plots:sequential_mem_access}  & Sequential  &[5pt]
                \ref{plots:coaledced_mem_access} & Coalesced\\};
        \end{tikzpicture}
        \caption{Impact of memory access pattern on Projection Query 2 (cf. Listing~\ref{lab:projection_query_2}).}
\label{pic:sequential_vs_coalesced_mem_access}
\end{figure}

%% file: kernel_compilation_times2.tex
        \begin{tikzpicture}[every text node part/.style={align=center}, font=\barchartfontsize,        
       		bar width=10pt,
]
                \begin{groupplot}[group style={group name=my plots,group size= 4 by 2,vertical sep=1.5cm },height=\barchartheight,width=4.5cm,cycle list name=barplot cycle list,
		area legend, 
		ybar=0pt]
                        \nextgroupplot[xticklabels={CPU, iGPU, dGPU, MIC},
xtick={1, 2, 3, 4},
ymin=0,
ymax=0.12,
ytick={0.03, 0.1},
yticklabels={0.03, 0.1},
xtick pos=left,
boxplot/draw direction=y,
xticklabel style={rotate=45},
ylabel={Compilation\\ Time in s},
scaled y ticks=false,
every boxplot/.append style={draw=black, fill=gray},
        log ticks with fixed point,
        ymode=log
]
\addplot+[ 
    boxplot prepared={
       median=0.038561,
upper quartile=0.0391952,
lower quartile=0.0380005,
upper whisker=0.04626606,
lower whisker=0.03322905599999999
    },
    ] coordinates {};
\addplot+[ 
    boxplot prepared={
       median=0.0579665,
upper quartile=0.0585229,
lower quartile=0.0548057,
upper whisker=0.064995408,
lower whisker=0.047644292
    },
    ] coordinates {};
\addplot+[ 
    boxplot prepared={
       median=0.058842599999999995,
upper quartile=0.0607799,
lower quartile=0.055569799999999996,
upper whisker=0.066272684,
lower whisker=0.044648631999999994
    },
    ] coordinates {};
\addplot+[ 
    boxplot prepared={
       median=0.091602,
upper quartile=0.0950053,
lower quartile=0.0748187,
upper whisker=0.1032378,
lower whisker=0.050497892
    },
    ] coordinates {};
    \draw[densely dotted] (axis cs:0,0.1) |- (axis cs:20,0.1) node [right] {};
                                \coordinate (top) at (rel axis cs:0,1);
                        \nextgroupplot[xticklabels={CPU, iGPU, dGPU, MIC},
xtick={1, 2, 3, 4},
ymin=0,
ymax=2,
boxplot/draw direction=y,
xtick pos=left,
xticklabel style={rotate=45},
every boxplot/.append style={draw=black, fill=gray},
   log ticks with fixed point,
        ymode=log
]
\addplot+[ 
    boxplot prepared={
       median=0.0479158,
upper quartile=0.049089,
lower quartile=0.0468597,
upper whisker=0.055735627999999995,
lower whisker=0.039088364
    },
    ] coordinates {};
\addplot+[ 
    boxplot prepared={
       median=0.07330249999999999,
upper quartile=0.0957295,
lower quartile=0.0671716,
upper whisker=0.11195124000000001,
lower whisker=0.057495528
    },
    ] coordinates {};
\addplot+[ 
    boxplot prepared={
       median=0.07096439999999998,
upper quartile=0.098731,
lower quartile=0.0675763,
upper whisker=0.1095602,
lower whisker=0.055301052
    },
    ] coordinates {};
\addplot+[ 
    boxplot prepared={
       median=0.587757,
upper quartile=0.970535,
lower quartile=0.520644,
upper whisker=1.1161316,
lower whisker=0.27379696
    },
    ] coordinates {};
    \draw[densely dotted] (axis cs:0,0.1) |- (axis cs:20,0.1) node [right] {};
    \nextgroupplot[xticklabels={CPU, iGPU, dGPU, MIC},
xtick={1, 2, 3, 4},
ymin=0.05,
ymax=6,
boxplot/draw direction=y,
xtick pos=left,
xticklabel style={rotate=45},
every boxplot/.append style={draw=black, fill=gray},
        log ticks with fixed point,
        ymode=log
]
\addplot+[ 
    boxplot prepared={
       median=0.066682,
upper quartile=0.0695702,
lower quartile=0.064473875,
upper whisker=0.07491107,
lower whisker=0.048611645
    },
    ] coordinates {};
\addplot+[ 
    boxplot prepared={
       median=0.113023,
upper quartile=0.15349925,
lower quartile=0.1068335,
upper whisker=0.16339483,
lower whisker=0.074596655
    },
    ] coordinates {};
\addplot+[ 
    boxplot prepared={
       median=0.21622799999999998,
upper quartile=0.23543100000000003,
lower quartile=0.20356725,
upper whisker=0.25804069999999996,
lower whisker=0.13308891
    },
    ] coordinates {};
\addplot+[ 
    boxplot prepared={
       median=4.862865,
upper quartile=5.104355,
lower quartile=4.643095,
upper whisker=5.3234743,
lower whisker=1.7800147
    },
    ] coordinates {};
   \nextgroupplot[xticklabels={CPU, iGPU, dGPU, MIC},
xtick={1, 2, 3, 4},
ymin=0.1,
ymax=3,
boxplot/draw direction=y,
xtick pos=left,
xticklabel style={rotate=45},
every boxplot/.append style={draw=black, fill=gray},
        log ticks with fixed point,
        ymode=log
]
\addplot+[ 
    boxplot prepared={
       median=0.2449365,
upper quartile=0.2543485,
lower quartile=0.23672375,
upper whisker=0.27495462,
lower whisker=0.21334211
    },
    ] coordinates {};
\addplot+[ 
    boxplot prepared={
       median=0.388459,
upper quartile=0.4062935,
lower quartile=0.355781,
upper whisker=0.45483552,
lower whisker=0.30693132
    },
    ] coordinates {};    
\addplot+[ 
    boxplot prepared={
       median=0.818864,
upper quartile=0.83858225,
lower quartile=0.79305475,
upper whisker=0.88027207,
lower whisker=0.6838772399999999
    },
    ] coordinates {};
\addplot+[ 
    boxplot prepared={
       median=1.81658,
upper quartile=2.258165,
lower quartile=1.6043575,
upper whisker=2.5110194,
lower whisker=1.1585401
    },
    ] coordinates {};
\coordinate (bot) at (rel axis cs:1,0);
                \end{groupplot}
                \node[below = 0.9cm of my plots c1r1.south] {Projection Query 1};
                \node[below = 0.9cm of my plots c2r1.south] {Aggregation Query 1};
                \node[below = 0.9cm of my plots c3r1.south] {TPC-H Q1};
                \node[below = 0.9cm of my plots c4r1.south] {SSB Q4.1};               
        \end{tikzpicture}
        \caption{Compilation times for all generated kernel variants for each processor and query pipeline. Most kernels can be compiled in less than 100ms, which allows for fast query compilation.\label{pics:query_compilation_times_boxplot}}  

%% file: projection_query_1_medium_selectivity.tex
    \begin{tikzpicture}[every text node part/.style={align=center}, font=\barchartfontsize,        
      bar width=10pt,
      ]
      \begin{groupplot}[group style={group name=my plots,group size= 4 by 2,vertical sep=1.5cm },height=\barchartheight,width=4.5cm,cycle list name=barplot cycle list,
        area legend, 
        ybar=0pt]
        \nextgroupplot[legend style={at={(0.38,-1)}, anchor=south}, ylabel={Execution\\ Time in s}, xtick=\empty, xticklabels=\empty,xmin=-3, xmax=3,
        ymin=0, ymax=100, bar width=10pt, cycle list name=barplot cycle list, area legend,
        nodes near coords={\pgfkeys{/pgf/fpu}\pgfmathparse{pow(e,\pgfplotspointmeta)}\pgfmathprintnumber[precision=2,zerofill,fixed]{\pgfmathresult}}, 
        log ticks with fixed point,
        ymode=log,
        log origin=infty,                           
        every node near coord/.append style={rotate=90, anchor=west, /pgf/number format/precision=2, xshift=0.00cm},  
        ]
        \addplot+[
        error bars/.cd,
        y dir=both,
        y explicit
        ]
        table[
        x=id,
        y=Mean,
        y error=Stdev,
        col sep=semicolon
        ]
        {cpu_processor_cpu_optimized.projection-query-1-medium-selectivity.report.csv};	 \label{plots:cpu_optimized}
        \addplot+[
        nodes near coords={{\pgfkeys{/pgf/fpu}\pgfmathparse{pow(e,\pgfplotspointmeta)}\pgfmathprintnumber[precision=1,zerofill,fixed]{\pgfmathresult}}},
        error bars/.cd,
        y dir=both,
        y explicit
        ]
        table[
        x=id,
        y=Mean,
        y error=Stdev,
        col sep=semicolon
        ]
        {cpu_processor_igpu_optimized.projection-query-1-medium-selectivity.report.csv}; \label{plots:igpu_optimized}	
        \addplot+[
        nodes near coords={{\pgfkeys{/pgf/fpu}\pgfmathparse{pow(e,\pgfplotspointmeta)}\pgfmathprintnumber[precision=1,zerofill,fixed]{\pgfmathresult}}},
        error bars/.cd,
        y dir=both,
        y explicit
        ]
        table[
        x=id,
        y=Mean,
        y error=Stdev,
        col sep=semicolon
        ]
        {cpu_processor_dgpu_optimized.projection-query-1-medium-selectivity.report.csv}; \label{plots:dgpu_optimized}
        \addplot+[
        error bars/.cd,
        y dir=both,
        y explicit
        ]
        table[
        x=id,
        y=Mean,
        y error=Stdev,
        col sep=semicolon
        ]
        {cpu_processor_phi_optimized.projection-query-1-medium-selectivity.report.csv}; \label{plots:phi_optimized}
        \addplot+[
        error bars/.cd,
        y dir=both,
        y explicit
        ]
        coordinates
        {
          (0,0)
        };
        \addplot+[
        error bars/.cd,
        y dir=both,
        y explicit
        ]
        table[
        x=id,
        y=Mean,
        y error=Stdev,
        col sep=semicolon
        ]
        {cpu_processor_feature_wise.projection-query-1-medium-selectivity.report.csv}; \label{plots:feature_wise_optimized}
        \coordinate (top) at (rel axis cs:0,1);
        \nextgroupplot[xtick=\empty, xticklabels=\empty,xmin=-3, xmax=3, ymin=0, ymax=100, bar width=10pt, cycle list name=barplot cycle list,   area legend,
        nodes near coords={\pgfkeys{/pgf/fpu}\pgfmathparse{pow(e,\pgfplotspointmeta)}\pgfmathprintnumber[precision=2,zerofill,fixed]{\pgfmathresult}}, 
        log ticks with fixed point,
        ymode=log,
        log origin=infty,  
        every node near coord/.append style={rotate=90, anchor=west, /pgf/number format/precision=2, xshift=0.00cm},  
        ]
        \addplot+[
        nodes near coords={\rotatebox{-90}{\pgfkeys{/pgf/fpu}\pgfmathparse{pow(e,\pgfplotspointmeta)}\pgfmathprintnumber[precision=1,zerofill,fixed]{\pgfmathresult}}},
        error bars/.cd,
        y dir=both,
        y explicit
        ]
        table[
        x=id,
        y=Mean,
        y error=Stdev,
        col sep=semicolon
        ]
        {igpu_processor_cpu_optimized.projection-query-1-medium-selectivity.report.csv};
        \addplot+[
        error bars/.cd,
        y dir=both,
        y explicit
        ]
        table[
        x=id,
        y=Mean,
        y error=Stdev,
        col sep=semicolon
        ]
        {igpu_processor_igpu_optimized.projection-query-1-medium-selectivity.report.csv};
        \addplot+[
        error bars/.cd,
        y dir=both,
        y explicit
        ]
        table[
        x=id,
        y=Mean,
        y error=Stdev,
        col sep=semicolon
        ]
        {igpu_processor_dgpu_optimized.projection-query-1-medium-selectivity.report.csv};  
        \addplot+[
        error bars/.cd,
        y dir=both,
        y explicit
        ]
        table[
        x=id,
        y=Mean,
        y error=Stdev,
        col sep=semicolon
        ]
        {igpu_processor_phi_optimized.projection-query-1-medium-selectivity.report.csv};
        \addplot+[
        error bars/.cd,
        y dir=both,
        y explicit
        ]
        coordinates
        {
          (0,0)
        };
        \addplot+[
        error bars/.cd,
        y dir=both,
        y explicit
        ]
        table[
        x=id,
        y=Mean,
        y error=Stdev,
        col sep=semicolon
        ]
        {igpu_processor_feature_wise.projection-query-1-medium-selectivity.report.csv};
        \nextgroupplot[xtick=\empty, xticklabels=\empty,xmin=-3, xmax=3, ymin=0, ymax=100, bar width=10pt, cycle list name=barplot cycle list,   area legend,
        nodes near coords={\pgfkeys{/pgf/fpu}\pgfmathparse{pow(e,\pgfplotspointmeta)}\pgfmathprintnumber[precision=2,zerofill,fixed]{\pgfmathresult}}, 
        log ticks with fixed point,
        ymode=log,
        log origin=infty,                           
        every node near coord/.append style={rotate=90, anchor=west, /pgf/number format/precision=2, xshift=0.00cm},  
        ]
        \addplot+[
        nodes near coords={\rotatebox{-90}{\pgfkeys{/pgf/fpu}\pgfmathparse{pow(e,\pgfplotspointmeta)}\pgfmathprintnumber[precision=1,zerofill,fixed]{\pgfmathresult}}},
        error bars/.cd,
        y dir=both,
        y explicit
        ]
        table[
        x=id,
        y=Mean,
        y error=Stdev,
        col sep=semicolon
        ]
        {dgpu_processor_cpu_optimized.projection-query-1-medium-selectivity.report.csv};
        \addplot+[
        error bars/.cd,
        y dir=both,
        y explicit
        ]
        table[
        x=id,
        y=Mean,
        y error=Stdev,
        col sep=semicolon
        ]
        {dgpu_processor_igpu_optimized.projection-query-1-medium-selectivity.report.csv};
        \addplot+[
        error bars/.cd,
        y dir=both,
        y explicit
        ]
        table[
        x=id,
        y=Mean,
        y error=Stdev,
        col sep=semicolon
        ]
        {dgpu_processor_dgpu_optimized.projection-query-1-medium-selectivity.report.csv};  
        \addplot+[
        error bars/.cd,
        y dir=both,
        y explicit
        ]
        table[
        x=id,
        y=Mean,
        y error=Stdev,
        col sep=semicolon
        ]
        {dgpu_processor_phi_optimized.projection-query-1-medium-selectivity.report.csv};
        \addplot+[
        error bars/.cd,
        y dir=both,
        y explicit
        ]
        coordinates
        {
          (0,0)
        };
        \addplot+[
        error bars/.cd,
        y dir=both,
        y explicit
        ]
        table[
        x=id,
        y=Mean,
        y error=Stdev,
        col sep=semicolon
        ]
        {dgpu_processor_feature_wise.projection-query-1-medium-selectivity.report.csv};
        \nextgroupplot[xtick=\empty, xticklabels=\empty, xmin=-3, xmax=3, ymin=0, ymax=10, bar width=10pt, cycle list name=barplot cycle list,   area legend,
        nodes near coords={\pgfkeys{/pgf/fpu}\pgfmathparse{pow(e,\pgfplotspointmeta)}\pgfmathprintnumber[precision=2,zerofill,fixed]{\pgfmathresult}}, 
        log ticks with fixed point,
        ymode=log,
        log origin=infty,  
        every node near coord/.append style={rotate=90, anchor=west, /pgf/number format/precision=2, xshift=0.00cm},  
        ]
        \addplot+[
        nodes near coords={{\pgfkeys{/pgf/fpu}\pgfmathparse{pow(e,\pgfplotspointmeta)}\pgfmathprintnumber[precision=2,zerofill,fixed]{\pgfmathresult}}},
        error bars/.cd,
        y dir=both,
        y explicit
        ]
        table[
        x=id,
        y=Mean,
        y error=Stdev,
        col sep=semicolon
        ]
        {phi_processor_cpu_optimized.projection-query-1-medium-selectivity.report.csv};
        \addplot+[
        error bars/.cd,
        y dir=both,
        y explicit
        ]
        table[
        x=id,
        y=Mean,
        y error=Stdev,
        col sep=semicolon
        ]
        {phi_processor_igpu_optimized.projection-query-1-medium-selectivity.report.csv};
        \addplot+[
        error bars/.cd,
        y dir=both,
        y explicit
        ]
        table[
        x=id,
        y=Mean,
        y error=Stdev,
        col sep=semicolon
        ]
        {phi_processor_dgpu_optimized.projection-query-1-medium-selectivity.report.csv};  
        \addplot+[
        error bars/.cd,
        y dir=both,
        y explicit
        ]
        table[
        x=id,
        y=Mean,
        y error=Stdev,
        col sep=semicolon
        ]
        {phi_processor_phi_optimized.projection-query-1-medium-selectivity.report.csv}; 
        \addplot+[
        error bars/.cd,
        y dir=both,
        y explicit
        ]
        coordinates
        {
          (0,0)
        };
        \addplot+[
        error bars/.cd,
        y dir=both,
        y explicit
        ]
        table[
        x=id,
        y=Mean,
        y error=Stdev,
        col sep=semicolon
        ]
        {phi_processor_feature_wise.projection-query-1-medium-selectivity.report.csv};
        \coordinate (bot) at (rel axis cs:1,0);
      \end{groupplot}
      \node[below = 0.1cm of my plots c1r1.south] {Executed on CPU};
      \node[below = 0.1cm of my plots c2r1.south] {Executed on iGPU};
      \node[below = 0.1cm of my plots c3r1.south] {Executed on dGPU};
      \node[below = 0.1cm of my plots c4r1.south] {Executed on MIC};               
      \path (top|-current bounding box.south)--
      coordinate(legendpos)
      (bot|-current bounding box.south);
      \matrix[
      matrix of nodes,
      anchor=south,
      draw,
      inner sep=0.2em,
      draw
      ]at([yshift=-5ex]legendpos)
      {
        \ref{plots:cpu_optimized}  & CPU Optimized  &[5pt]
        \ref{plots:igpu_optimized} & iGPU Optimized &[5pt]
        \ref{plots:dgpu_optimized} & dGPU Optimized &[5pt]
        \ref{plots:phi_optimized}  & MIC Optimized  &[5pt]
        \ref{plots:feature_wise_optimized} & Learned\\};
    \end{tikzpicture}

%% file: projection_query_2_high_selectivity.tex
    \begin{tikzpicture}[every text node part/.style={align=center}, font=\barchartfontsize,        
      bar width=10pt,
      ]
      \begin{groupplot}[group style={group name=my plots,group size= 4 by 2,vertical sep=1.5cm },height=\barchartheight,width=4.5cm,cycle list name=barplot cycle list,
        area legend, 
        ybar=0pt]
        \nextgroupplot[legend style={at={(0.38,-1)}, anchor=south}, ylabel={Execution\\ Time in s}, xtick=\empty, xticklabels=\empty,xmin=-3, xmax=3, ymin=0, ymax=100, bar width=10pt, cycle list name=barplot cycle list,   area legend,
        nodes near coords={\pgfkeys{/pgf/fpu}\pgfmathparse{pow(e,\pgfplotspointmeta)}\pgfmathprintnumber[precision=2,zerofill,fixed]{\pgfmathresult}}, 
        log ticks with fixed point,
        ymode=log,
        log origin=infty,  
        every node near coord/.append style={rotate=90, anchor=west, /pgf/number format/precision=2, xshift=0.00cm},  
        ]
        \addplot+[
        error bars/.cd,
        y dir=both,
        y explicit
        ]
        table[
        x=id,
        y=Mean,
        y error=Stdev,
        col sep=semicolon
        ]
        {cpu_processor_cpu_optimized.projection-query-2-high-selectivity.report.csv};	 \label{plots:cpu_optimized}
        \addplot+[
        nodes near coords={{\pgfkeys{/pgf/fpu}\pgfmathparse{pow(e,\pgfplotspointmeta)}\pgfmathprintnumber[precision=2,zerofill,fixed]{\pgfmathresult}}},
        error bars/.cd,
        y dir=both,
        y explicit
        ]
        table[
        x=id,
        y=Mean,
        y error=Stdev,
        col sep=semicolon
        ]
        {cpu_processor_igpu_optimized.projection-query-2-high-selectivity.report.csv}; \label{plots:igpu_optimized}	
        \addplot+[
        nodes near coords={{\pgfkeys{/pgf/fpu}\pgfmathparse{pow(e,\pgfplotspointmeta)}\pgfmathprintnumber[precision=2,zerofill,fixed]{\pgfmathresult}}},
        error bars/.cd,
        y dir=both,
        y explicit
        ]
        table[
        x=id,
        y=Mean,
        y error=Stdev,
        col sep=semicolon
        ]
        {cpu_processor_dgpu_optimized.projection-query-2-high-selectivity.report.csv}; \label{plots:dgpu_optimized}
        \addplot+[
        nodes near coords={{\pgfkeys{/pgf/fpu}\pgfmathparse{pow(e,\pgfplotspointmeta)}\pgfmathprintnumber[precision=2,zerofill,fixed]{\pgfmathresult}}},
        error bars/.cd,
        y dir=both,
        y explicit
        ]
        table[
        x=id,
        y=Mean,
        y error=Stdev,
        col sep=semicolon
        ]
        {cpu_processor_phi_optimized.projection-query-2-high-selectivity.report.csv}; \label{plots:phi_optimized}
        \addplot+[
        error bars/.cd,
        y dir=both,
        y explicit
        ]
        coordinates
        {
          (0,0)
        };
        \addplot+[
        error bars/.cd,
        y dir=both,
        y explicit
        ]
        table[
        x=id,
        y=Mean,
        y error=Stdev,
        col sep=semicolon
        ]
        {cpu_processor_feature_wise.projection-query-2-high-selectivity.report.csv}; \label{plots:feature_wise_optimized}
        \coordinate (top) at (rel axis cs:0,1);
        \nextgroupplot[xtick=\empty, xticklabels=\empty,xmin=-3, xmax=3, ymin=0, ymax=50, bar width=10pt, cycle list name=barplot cycle list,   area legend,
        nodes near coords={\pgfkeys{/pgf/fpu}\pgfmathparse{pow(e,\pgfplotspointmeta)}\pgfmathprintnumber[precision=2,zerofill,fixed]{\pgfmathresult}}, 
        log ticks with fixed point,
        ymode=log,
        log origin=infty,                           
        every node near coord/.append style={rotate=90, anchor=west, /pgf/number format/precision=2, xshift=0.00cm},  
        ]
        \addplot+[
        nodes near coords={\rotatebox{-90}{\pgfkeys{/pgf/fpu}\pgfmathparse{pow(e,\pgfplotspointmeta)}\pgfmathprintnumber[precision=1,zerofill,fixed]{\pgfmathresult}}},
        error bars/.cd,
        y dir=both,
        y explicit
        ]
        table[
        x=id,
        y=Mean,
        y error=Stdev,
        col sep=semicolon
        ]
        {igpu_processor_cpu_optimized.projection-query-2-high-selectivity.report.csv};
        \addplot+[
        error bars/.cd,
        y dir=both,
        y explicit
        ]
        table[
        x=id,
        y=Mean,
        y error=Stdev,
        col sep=semicolon
        ]
        {igpu_processor_igpu_optimized.projection-query-2-high-selectivity.report.csv};
        \addplot+[
        error bars/.cd,
        y dir=both,
        y explicit
        ]
        table[
        x=id,
        y=Mean,
        y error=Stdev,
        col sep=semicolon
        ]
        {igpu_processor_dgpu_optimized.projection-query-2-high-selectivity.report.csv};  
        \addplot+[
        error bars/.cd,
        y dir=both,
        y explicit
        ]
        table[
        x=id,
        y=Mean,
        y error=Stdev,
        col sep=semicolon
        ]
        {igpu_processor_phi_optimized.projection-query-2-high-selectivity.report.csv};
        \addplot+[
        error bars/.cd,
        y dir=both,
        y explicit
        ]
        coordinates
        {
          (0,0)
        };
        \addplot+[
        error bars/.cd,
        y dir=both,
        y explicit
        ]
        table[
        x=id,
        y=Mean,
        y error=Stdev,
        col sep=semicolon
        ]
        {igpu_processor_feature_wise.projection-query-2-high-selectivity.report.csv};
        \nextgroupplot[xtick=\empty, xticklabels=\empty,xmin=-3, xmax=3, ymin=0, ymax=100, bar width=10pt, cycle list name=barplot cycle list,   area legend,
        nodes near coords={\pgfkeys{/pgf/fpu}\pgfmathparse{pow(e,\pgfplotspointmeta)}\pgfmathprintnumber[precision=2,zerofill,fixed]{\pgfmathresult}}, 
        log ticks with fixed point,
        ymode=log,
        log origin=infty,  
        every node near coord/.append style={rotate=90, anchor=west, /pgf/number format/precision=2, xshift=0.00cm},  
        ]
        \addplot+[
        nodes near coords={\rotatebox{-90}{\pgfkeys{/pgf/fpu}\pgfmathparse{pow(e,\pgfplotspointmeta)}\pgfmathprintnumber[precision=1,zerofill,fixed]{\pgfmathresult}}},
        error bars/.cd,
        y dir=both,
        y explicit
        ]
        table[
        x=id,
        y=Mean,
        y error=Stdev,
        col sep=semicolon
        ]
        {dgpu_processor_cpu_optimized.projection-query-2-high-selectivity.report.csv};
        \addplot+[
        error bars/.cd,
        y dir=both,
        y explicit
        ]
        table[
        x=id,
        y=Mean,
        y error=Stdev,
        col sep=semicolon
        ]
        {dgpu_processor_igpu_optimized.projection-query-2-high-selectivity.report.csv};
        \addplot+[
        error bars/.cd,
        y dir=both,
        y explicit
        ]
        table[
        x=id,
        y=Mean,
        y error=Stdev,
        col sep=semicolon
        ]
        {dgpu_processor_dgpu_optimized.projection-query-2-high-selectivity.report.csv};  
        \addplot+[
        error bars/.cd,
        y dir=both,
        y explicit
        ]
        table[
        x=id,
        y=Mean,
        y error=Stdev,
        col sep=semicolon
        ]
        {dgpu_processor_phi_optimized.projection-query-2-high-selectivity.report.csv};
        \addplot+[
        error bars/.cd,
        y dir=both,
        y explicit
        ]
        coordinates
        {
          (0,0)
        };
        \addplot+[
        error bars/.cd,
        y dir=both,
        y explicit
        ]
        table[
        x=id,
        y=Mean,
        y error=Stdev,
        col sep=semicolon
        ]
        {dgpu_processor_feature_wise.projection-query-2-high-selectivity.report.csv};
        \nextgroupplot[xtick=\empty, xticklabels=\empty, xmin=-3, xmax=3, ymin=0, ymax=10, bar width=10pt, cycle list name=barplot cycle list,   area legend,
        nodes near coords={\pgfkeys{/pgf/fpu}\pgfmathparse{pow(e,\pgfplotspointmeta)}\pgfmathprintnumber[precision=2,zerofill,fixed]{\pgfmathresult}}, 
        log ticks with fixed point,
        ymode=log,
        log origin=infty,                           
        every node near coord/.append style={rotate=90, anchor=west, /pgf/number format/precision=2, xshift=0.00cm},  
        ]
        \addplot+[
        nodes near coords={{\pgfkeys{/pgf/fpu}\pgfmathparse{pow(e,\pgfplotspointmeta)}\pgfmathprintnumber[precision=2,zerofill,fixed]{\pgfmathresult}}},
        error bars/.cd,
        y dir=both,
        y explicit
        ]
        table[
        x=id,
        y=Mean,
        y error=Stdev,
        col sep=semicolon
        ]
        {phi_processor_cpu_optimized.projection-query-2-high-selectivity.report.csv};
        \addplot+[
        nodes near coords={{\pgfkeys{/pgf/fpu}\pgfmathparse{pow(e,\pgfplotspointmeta)}\pgfmathprintnumber[precision=2,zerofill,fixed]{\pgfmathresult}}},
        error bars/.cd,
        y dir=both,
        y explicit
        ]
        table[
        x=id,
        y=Mean,
        y error=Stdev,
        col sep=semicolon
        ]
        {phi_processor_igpu_optimized.projection-query-2-high-selectivity.report.csv};
        \addplot+[
        error bars/.cd,
        y dir=both,
        y explicit
        ]
        table[
        x=id,
        y=Mean,
        y error=Stdev,
        col sep=semicolon
        ]
        {phi_processor_dgpu_optimized.projection-query-2-high-selectivity.report.csv};  
        \addplot+[
        error bars/.cd,
        y dir=both,
        y explicit
        ]
        table[
        x=id,
        y=Mean,
        y error=Stdev,
        col sep=semicolon
        ]
        {phi_processor_phi_optimized.projection-query-2-high-selectivity.report.csv}; 
        \addplot+[
        error bars/.cd,
        y dir=both,
        y explicit
        ]
        coordinates
        {
          (0,0)
        };
        \addplot+[
        nodes near coords={{\pgfkeys{/pgf/fpu}\pgfmathparse{pow(e,\pgfplotspointmeta)}\pgfmathprintnumber[precision=2,zerofill,fixed]{\pgfmathresult}}},
        error bars/.cd,
        y dir=both,
        y explicit
        ]
        table[
        x=id,
        y=Mean,
        y error=Stdev,
        col sep=semicolon
        ]
        {phi_processor_feature_wise.projection-query-2-high-selectivity.report.csv};
        \coordinate (bot) at (rel axis cs:1,0);
      \end{groupplot}
      \node[below = 0.1cm of my plots c1r1.south] {Executed on CPU};
      \node[below = 0.1cm of my plots c2r1.south] {Executed on iGPU};
      \node[below = 0.1cm of my plots c3r1.south] {Executed on dGPU};
      \node[below = 0.1cm of my plots c4r1.south] {Executed on MIC};               
      \path (top|-current bounding box.south)--
      coordinate(legendpos)
      (bot|-current bounding box.south);
      \matrix[
      matrix of nodes,
      anchor=south,
      draw,
      inner sep=0.2em,
      draw
      ]at([yshift=-5ex]legendpos)
      {
        \ref{plots:cpu_optimized}  & CPU Optimized  &[5pt]
        \ref{plots:igpu_optimized} & iGPU Optimized &[5pt]
        \ref{plots:dgpu_optimized} & dGPU Optimized &[5pt]
        \ref{plots:phi_optimized}  & MIC Optimized  &[5pt]
        \ref{plots:feature_wise_optimized} & Learned\\};
    \end{tikzpicture}

%% file: grouped_aggregation_query_1.tex
    \begin{tikzpicture}[every text node part/.style={align=center}, font=\barchartfontsize,        
      bar width=10pt,
      ]
      \begin{groupplot}[group style={group name=my plots,group size= 4 by 2,vertical sep=1.5cm },height=\barchartheight,width=4.5cm,cycle list name=barplot cycle list,
        area legend, 
        ybar=0pt]
        \nextgroupplot[legend style={at={(0.38,-1)}, anchor=south}, ylabel={Execution\\ Time in s}, xtick=\empty, xticklabels=\empty,xmin=-3, xmax=3, ymin=0, ymax=110, bar width=10pt, cycle list name=barplot cycle list,   area legend,
        nodes near coords={\pgfkeys{/pgf/fpu}\pgfmathparse{pow(e,\pgfplotspointmeta)}\pgfmathprintnumber[precision=2,zerofill,fixed]{\pgfmathresult}}, 
        log ticks with fixed point,
        ymode=log,
        log origin=infty,  
        every node near coord/.append style={rotate=90, anchor=west, /pgf/number format/precision=2, xshift=0.00cm},  
        ]
        \addplot+[
        error bars/.cd,
        y dir=both,
        y explicit
        ]
        table[
        x=id,
        y=Mean,
        y error=Stdev,
        col sep=semicolon
        ]
        {cpu_processor_cpu_optimized.grouped_aggregation_query_1.report.csv};	 \label{plots:cpu_optimized}
        \addplot+[
        nodes near coords={{\pgfkeys{/pgf/fpu}\pgfmathparse{pow(e,\pgfplotspointmeta)}\pgfmathprintnumber[precision=2,zerofill,fixed]{\pgfmathresult}}},
        error bars/.cd,
        y dir=both,
        y explicit
        ]
        table[
        x=id,
        y=Mean,
        y error=Stdev,
        col sep=semicolon
        ]
        {cpu_processor_igpu_optimized.grouped_aggregation_query_1.report.csv}; \label{plots:igpu_optimized}	
        \addplot+[
        nodes near coords={{\pgfkeys{/pgf/fpu}\pgfmathparse{pow(e,\pgfplotspointmeta)}\pgfmathprintnumber[precision=2,zerofill,fixed]{\pgfmathresult}}},
        error bars/.cd,
        y dir=both,
        y explicit
        ]
        table[
        x=id,
        y=Mean,
        y error=Stdev,
        col sep=semicolon
        ]
        {cpu_processor_dgpu_optimized.grouped_aggregation_query_1.report.csv}; \label{plots:dgpu_optimized}
        \addplot+[
        nodes near coords={{\pgfkeys{/pgf/fpu}\pgfmathparse{pow(e,\pgfplotspointmeta)}\pgfmathprintnumber[precision=2,zerofill,fixed]{\pgfmathresult}}},
        error bars/.cd,
        y dir=both,
        y explicit
        ]
        table[
        x=id,
        y=Mean,
        y error=Stdev,
        col sep=semicolon
        ]
        {cpu_processor_phi_optimized.grouped_aggregation_query_1.report.csv}; \label{plots:phi_optimized}
        \addplot+[
        error bars/.cd,
        y dir=both,
        y explicit
        ]
        coordinates
        {
          (0,0)
        };
        \addplot+[
        error bars/.cd,
        y dir=both,
        y explicit
        ]
        table[
        x=id,
        y=Mean,
        y error=Stdev,
        col sep=semicolon
        ]
        {cpu_processor_feature_wise.grouped_aggregation_query_1.report.csv}; \label{plots:feature_wise_optimized}
        \coordinate (top) at (rel axis cs:0,1);
        \nextgroupplot[xtick=\empty, xticklabels=\empty, xmin=-3, xmax=3, ymin=0, ymax=400, bar width=10pt, cycle list name=barplot cycle list, area legend,
        nodes near coords={\pgfkeys{/pgf/fpu}\pgfmathparse{pow(e,\pgfplotspointmeta)}\pgfmathprintnumber[precision=2,zerofill,fixed]{\pgfmathresult}}, 
        log ticks with fixed point,
        ymode=log,
        log origin=infty, 
        every node near coord/.append style={rotate=90, anchor=west, /pgf/number format/precision=2, xshift=0.00cm},  
        ]
        \addplot+[
        nodes near coords={\rotatebox{-90}{\pgfkeys{/pgf/fpu}\pgfmathparse{pow(e,\pgfplotspointmeta)}\pgfmathprintnumber[precision=1,zerofill,fixed]{\pgfmathresult}}},
        error bars/.cd,
        y dir=both,
        y explicit,
        ]
        table[
        x=id,
        y=Mean,
        y error=Stdev,
        col sep=semicolon
        ]
        {igpu_processor_cpu_optimized.grouped_aggregation_query_1.report.csv};
        \addplot+[
        error bars/.cd,
        y dir=both,
        y explicit
        ]
        table[
        x=id,
        y=Mean,
        y error=Stdev,
        col sep=semicolon
        ]
        {igpu_processor_igpu_optimized.grouped_aggregation_query_1.report.csv};
        \addplot+[
        error bars/.cd,
        y dir=both,
        y explicit
        ]
        table[
        x=id,
        y=Mean,
        y error=Stdev,
        col sep=semicolon
        ]
        {igpu_processor_dgpu_optimized.grouped_aggregation_query_1.report.csv};  
        \addplot+[
        error bars/.cd,
        y dir=both,
        y explicit
        ]
        table[
        x=id,
        y=Mean,
        y error=Stdev,
        col sep=semicolon
        ]
        {igpu_processor_phi_optimized.grouped_aggregation_query_1.report.csv};     
        \addplot+[
        error bars/.cd,
        y dir=both,
        y explicit
        ]
        coordinates
        {
          (0,0)
        };
        \addplot+[
        error bars/.cd,
        y dir=both,
        y explicit
        ]
        table[
        x=id,
        y=Mean,
        y error=Stdev,
        col sep=semicolon
        ]
        {igpu_processor_feature_wise.grouped_aggregation_query_1.report.csv};
        \nextgroupplot[xtick=\empty, xticklabels=\empty,xmin=-3, xmax=3, ymin=0, ymax=400, bar width=10pt, cycle list name=barplot cycle list,   area legend,
        nodes near coords={\pgfkeys{/pgf/fpu}\pgfmathparse{pow(e,\pgfplotspointmeta)}\pgfmathprintnumber[precision=2,zerofill,fixed]{\pgfmathresult}}, 
        log ticks with fixed point,
        ymode=log,
        log origin=infty,                           
        every node near coord/.append style={rotate=90, anchor=west, /pgf/number format/precision=2, xshift=0.00cm},  
        ]
        \addplot+[
        nodes near coords={\rotatebox{-90}{\pgfkeys{/pgf/fpu}\pgfmathparse{pow(e,\pgfplotspointmeta)}\pgfmathprintnumber[precision=1,zerofill,fixed]{\pgfmathresult}}},
        error bars/.cd,
        y dir=both,
        y explicit
        ]
        table[
        x=id,
        y=Mean,
        y error=Stdev,
        col sep=semicolon
        ]
        {dgpu_processor_cpu_optimized.grouped_aggregation_query_1.report.csv};
        \addplot+[
        error bars/.cd,
        y dir=both,
        y explicit
        ]
        table[
        x=id,
        y=Mean,
        y error=Stdev,
        col sep=semicolon
        ]
        {dgpu_processor_igpu_optimized.grouped_aggregation_query_1.report.csv};
        \addplot+[
        error bars/.cd,
        y dir=both,
        y explicit
        ]
        table[
        x=id,
        y=Mean,
        y error=Stdev,
        col sep=semicolon
        ]
        {dgpu_processor_dgpu_optimized.grouped_aggregation_query_1.report.csv};  
        \addplot+[
        error bars/.cd,
        y dir=both,
        y explicit
        ]
        table[
        x=id,
        y=Mean,
        y error=Stdev,
        col sep=semicolon
        ]
        {dgpu_processor_phi_optimized.grouped_aggregation_query_1.report.csv};    
        \addplot+[
        error bars/.cd,
        y dir=both,
        y explicit
        ]
        coordinates
        {
          (0,0)
        };
        \addplot+[
        error bars/.cd,
        y dir=both,
        y explicit
        ]
        table[
        x=id,
        y=Mean,
        y error=Stdev,
        col sep=semicolon
        ]
        {dgpu_processor_feature_wise.grouped_aggregation_query_1.report.csv};
        \nextgroupplot[xtick=\empty, xticklabels=\empty, xmin=-3, xmax=3, ymin=0, ymax=100, bar width=10pt, cycle list name=barplot cycle list,   area legend,
        nodes near coords={\pgfkeys{/pgf/fpu}\pgfmathparse{pow(e,\pgfplotspointmeta)}\pgfmathprintnumber[precision=2,zerofill,fixed]{\pgfmathresult}}, 
        log ticks with fixed point,
        ymode=log,
        log origin=infty, 
        every node near coord/.append style={rotate=90, anchor=west, /pgf/number format/precision=2, xshift=0.00cm},  
        ]
        \addplot+[
        nodes near coords={{\pgfkeys{/pgf/fpu}\pgfmathparse{pow(e,\pgfplotspointmeta)}\pgfmathprintnumber[precision=1,zerofill,fixed]{\pgfmathresult}}},
        error bars/.cd,
        y dir=both,
        y explicit,
        ]
        table[
        x=id,
        y=Mean,
        y error=Stdev,
        col sep=semicolon
        ]
        {phi_processor_cpu_optimized.grouped_aggregation_query_1.report.csv};
        \addplot+[
        nodes near coords={{\pgfkeys{/pgf/fpu}\pgfmathparse{pow(e,\pgfplotspointmeta)}\pgfmathprintnumber[precision=2,zerofill,fixed]{\pgfmathresult}}},
        error bars/.cd,
        y dir=both,
        y explicit
        ]
        table[
        x=id,
        y=Mean,
        y error=Stdev,
        col sep=semicolon
        ]
        {phi_processor_igpu_optimized.grouped_aggregation_query_1.report.csv};
        \addplot+[
        nodes near coords={{\pgfkeys{/pgf/fpu}\pgfmathparse{pow(e,\pgfplotspointmeta)}\pgfmathprintnumber[precision=2,zerofill,fixed]{\pgfmathresult}}},
        error bars/.cd,
        y dir=both,
        y explicit
        ]
        table[
        x=id,
        y=Mean,
        y error=Stdev,
        col sep=semicolon
        ]
        {phi_processor_dgpu_optimized.grouped_aggregation_query_1.report.csv};  
        \addplot+[
        error bars/.cd,
        y dir=both,
        y explicit
        ]
        table[
        x=id,
        y=Mean,
        y error=Stdev,
        col sep=semicolon
        ]
        {phi_processor_phi_optimized.grouped_aggregation_query_1.report.csv}; 
        \addplot+[
        error bars/.cd,
        y dir=both,
        y explicit
        ]
        coordinates
        {
          (0,0)
        };
        \addplot+[
        error bars/.cd,
        y dir=both,
        y explicit
        ]
        table[
        x=id,
        y=Mean,
        y error=Stdev,
        col sep=semicolon
        ]
        {phi_processor_feature_wise.grouped_aggregation_query_1.report.csv};
        \coordinate (bot) at (rel axis cs:1,0);
      \end{groupplot}
      \node[below = 0.1cm of my plots c1r1.south] {Executed on CPU};
      \node[below = 0.1cm of my plots c2r1.south] {Executed on iGPU};
      \node[below = 0.1cm of my plots c3r1.south] {Executed on dGPU};
      \node[below = 0.1cm of my plots c4r1.south] {Executed on MIC};               
      \path (top|-current bounding box.south)--
      coordinate(legendpos)
      (bot|-current bounding box.south);
      \matrix[
      matrix of nodes,
      anchor=south,
      draw,
      inner sep=0.2em,
      draw
      ]at([yshift=-5ex]legendpos)
      {
        \ref{plots:cpu_optimized}  & CPU Optimized  &[5pt]
        \ref{plots:igpu_optimized} & iGPU Optimized &[5pt]
        \ref{plots:dgpu_optimized} & dGPU Optimized &[5pt]
        \ref{plots:phi_optimized}  & MIC Optimized &[5pt]
        \ref{plots:feature_wise_optimized} & Learned\\};
    \end{tikzpicture}

%% file: grouped_aggregation_query_2.tex
    \begin{tikzpicture}[every text node part/.style={align=center}, font=\barchartfontsize,        
      bar width=10pt,
      ]
      \begin{groupplot}[group style={group name=my plots,group size= 4 by 2,vertical sep=1.5cm },height=\barchartheight,width=4.5cm,cycle list name=barplot cycle list,
        area legend, 
        ybar=0pt]
        \nextgroupplot[legend style={at={(0.38,-1)}, anchor=south}, ylabel={Execution\\ Time in s}, xtick=\empty, xticklabels=\empty,xmin=-3, xmax=3, ymin=0, ymax=100, bar width=10pt, cycle list name=barplot cycle list, area legend,
        nodes near coords={\pgfkeys{/pgf/fpu}\pgfmathparse{pow(e,\pgfplotspointmeta)}\pgfmathprintnumber[precision=2,zerofill,fixed]{\pgfmathresult}}, 
        log ticks with fixed point,
        ymode=log,
        log origin=infty, 
        every node near coord/.append style={rotate=90, anchor=west, /pgf/number format/precision=2, xshift=0.00cm},  
        ]
        \addplot+[
        error bars/.cd,
        y dir=both,
        y explicit
        ]
        table[
        x=id,
        y=Mean,
        y error=Stdev,
        col sep=semicolon
        ]
        {cpu_processor_cpu_optimized.grouped_aggregation_query_2_many_groups.report.csv};	 \label{plots:cpu_optimized}
        \addplot+[
        error bars/.cd,
        y dir=both,
        y explicit
        ]
        table[
        x=id,
        y=Mean,
        y error=Stdev,
        col sep=semicolon
        ]
        {cpu_processor_igpu_optimized.grouped_aggregation_query_2_many_groups.report.csv}; \label{plots:igpu_optimized}	
        \addplot+[
        error bars/.cd,
        y dir=both,
        y explicit
        ]
        table[
        x=id,
        y=Mean,
        y error=Stdev,
        col sep=semicolon
        ]
        {cpu_processor_dgpu_optimized.grouped_aggregation_query_2_many_groups.report.csv}; \label{plots:dgpu_optimized}
        \addplot+[
        error bars/.cd,
        y dir=both,
        y explicit
        ]
        table[
        x=id,
        y=Mean,
        y error=Stdev,
        col sep=semicolon
        ]
        {cpu_processor_phi_optimized.grouped_aggregation_query_2_many_groups.report.csv}; \label{plots:phi_optimized}
        \addplot+[
        error bars/.cd,
        y dir=both,
        y explicit
        ]
        coordinates
        {
          (0,0)
        };
        \addplot+[
        error bars/.cd,
        y dir=both,
        y explicit
        ]
        table[
        x=id,
        y=Mean,
        y error=Stdev,
        col sep=semicolon
        ]
        {cpu_processor_feature_wise.grouped_aggregation_query_2_many_groups.report.csv}; \label{plots:feature_wise_optimized}
        \coordinate (top) at (rel axis cs:0,1);
        \nextgroupplot[xtick=\empty, xticklabels=\empty,xmin=-3, xmax=3, ymin=0.1, ymax=100, bar width=10pt, cycle list name=barplot cycle list,   area legend,
        nodes near coords={\pgfkeys{/pgf/fpu}\pgfmathparse{pow(e,\pgfplotspointmeta)}\pgfmathprintnumber[precision=2,zerofill,fixed]{\pgfmathresult}}, 
        log ticks with fixed point,
        ymode=log,
        log origin=infty, 
        every node near coord/.append style={rotate=90, anchor=west, /pgf/number format/precision=2, xshift=0.00cm},  
        ]
        \addplot+[
        nodes near coords={\rotatebox{-90}{\pgfkeys{/pgf/fpu}\pgfmathparse{pow(e,\pgfplotspointmeta)}\pgfmathprintnumber[precision=1,zerofill,fixed]{\pgfmathresult}}},
        error bars/.cd,
        y dir=both,
        y explicit
        ]
        table[
        x=id,
        y=Mean,
        y error=Stdev,
        col sep=semicolon
        ]
        {igpu_processor_cpu_optimized.grouped_aggregation_query_2_many_groups.report.csv};
        \addplot+[
        error bars/.cd,
        y dir=both,
        y explicit
        ]
        table[
        x=id,
        y=Mean,
        y error=Stdev,
        col sep=semicolon
        ]
        {igpu_processor_igpu_optimized.grouped_aggregation_query_2_many_groups.report.csv};
        \addplot+[
        error bars/.cd,
        y dir=both,
        y explicit
        ]
        table[
        x=id,
        y=Mean,
        y error=Stdev,
        col sep=semicolon
        ]
        {igpu_processor_dgpu_optimized.grouped_aggregation_query_2_many_groups.report.csv};  
        \addplot+[
        error bars/.cd,
        y dir=both,
        y explicit
        ]
        table[
        x=id,
        y=Mean,
        y error=Stdev,
        col sep=semicolon
        ]
        {igpu_processor_phi_optimized.grouped_aggregation_query_2_many_groups.report.csv};
        \addplot+[
        error bars/.cd,
        y dir=both,
        y explicit
        ]
        coordinates
        {
          (0,0)
        };
        \addplot+[
        error bars/.cd,
        y dir=both,
        y explicit
        ]
        table[
        x=id,
        y=Mean,
        y error=Stdev,
        col sep=semicolon
        ]
        {igpu_processor_feature_wise.grouped_aggregation_query_2_many_groups.report.csv};
        \nextgroupplot[xtick=\empty, xticklabels=\empty,xmin=-3, xmax=3, ymin=0, ymax=100, bar width=10pt, cycle list name=barplot cycle list,   area legend,
        nodes near coords={\pgfkeys{/pgf/fpu}\pgfmathparse{pow(e,\pgfplotspointmeta)}\pgfmathprintnumber[precision=2,zerofill,fixed]{\pgfmathresult}}, 
        log ticks with fixed point,
        ymode=log,
        log origin=infty,                           
        every node near coord/.append style={rotate=90, anchor=west, /pgf/number format/precision=2, xshift=0.00cm},  
        ]
        \addplot+[
        nodes near coords={\rotatebox{-90}{\pgfkeys{/pgf/fpu}\pgfmathparse{pow(e,\pgfplotspointmeta)}\pgfmathprintnumber[precision=1,zerofill,fixed]{\pgfmathresult}}},
        error bars/.cd,
        y dir=both,
        y explicit
        ]
        table[
        x=id,
        y=Mean,
        y error=Stdev,
        col sep=semicolon
        ]
        {dgpu_processor_cpu_optimized.grouped_aggregation_query_2_many_groups.report.csv};
        \addplot+[
        error bars/.cd,
        y dir=both,
        y explicit
        ]
        table[
        x=id,
        y=Mean,
        y error=Stdev,
        col sep=semicolon
        ]
        {dgpu_processor_igpu_optimized.grouped_aggregation_query_2_many_groups.report.csv};
        \addplot+[
        error bars/.cd,
        y dir=both,
        y explicit
        ]
        table[
        x=id,
        y=Mean,
        y error=Stdev,
        col sep=semicolon
        ]
        {dgpu_processor_dgpu_optimized.grouped_aggregation_query_2_many_groups.report.csv};  
        \addplot+[
        error bars/.cd,
        y dir=both,
        y explicit
        ]
        table[
        x=id,
        y=Mean,
        y error=Stdev,
        col sep=semicolon
        ]
        {dgpu_processor_phi_optimized.grouped_aggregation_query_2_many_groups.report.csv};    
        \addplot+[
        error bars/.cd,
        y dir=both,
        y explicit
        ]
        coordinates
        {
          (0,0)
        };
        \addplot+[
        error bars/.cd,
        y dir=both,
        y explicit
        ]
        table[
        x=id,
        y=Mean,
        y error=Stdev,
        col sep=semicolon
        ]
        {dgpu_processor_feature_wise.grouped_aggregation_query_2_many_groups.report.csv};
        \nextgroupplot[xtick=\empty, xticklabels=\empty, xmin=-3, xmax=3, ymin=0, ymax=100, bar width=10pt, cycle list name=barplot cycle list,   area legend,
        nodes near coords={\pgfkeys{/pgf/fpu}\pgfmathparse{pow(e,\pgfplotspointmeta)}\pgfmathprintnumber[precision=2,zerofill,fixed]{\pgfmathresult}}, 
        log ticks with fixed point,
        ymode=log,
        log origin=infty,                           
        every node near coord/.append style={rotate=90, anchor=west, /pgf/number format/precision=2, xshift=0.00cm},  
        ]
        \addplot+[
        nodes near coords={{\pgfkeys{/pgf/fpu}\pgfmathparse{pow(e,\pgfplotspointmeta)}\pgfmathprintnumber[precision=1,zerofill,fixed]{\pgfmathresult}}},
        error bars/.cd,
        y dir=both,
        y explicit
        ]
        table[
        x=id,
        y=Mean,
        y error=Stdev,
        col sep=semicolon
        ]
        {phi_processor_cpu_optimized.grouped_aggregation_query_2_many_groups.report.csv};
        \addplot+[
        error bars/.cd,
        y dir=both,
        y explicit
        ]
        table[
        x=id,
        y=Mean,
        y error=Stdev,
        col sep=semicolon
        ]
        {phi_processor_igpu_optimized.grouped_aggregation_query_2_many_groups.report.csv};
        \addplot+[
        error bars/.cd,
        y dir=both,
        y explicit
        ]
        table[
        x=id,
        y=Mean,
        y error=Stdev,
        col sep=semicolon
        ]
        {phi_processor_dgpu_optimized.grouped_aggregation_query_2_many_groups.report.csv};  
        \addplot+[
        error bars/.cd,
        y dir=both,
        y explicit
        ]
        table[
        x=id,
        y=Mean,
        y error=Stdev,
        col sep=semicolon
        ]
        {phi_processor_phi_optimized.grouped_aggregation_query_2_many_groups.report.csv}; 
        \addplot+[
        error bars/.cd,
        y dir=both,
        y explicit
        ]
        coordinates
        {
          (0,0)
        };
        \addplot+[
        error bars/.cd,
        y dir=both,
        y explicit
        ]
        table[
        x=id,
        y=Mean,
        y error=Stdev,
        col sep=semicolon
        ]
        {phi_processor_feature_wise.grouped_aggregation_query_2_many_groups.report.csv};
        \coordinate (bot) at (rel axis cs:1,0);
      \end{groupplot}
      \node[below = 0.1cm of my plots c1r1.south] {Executed on CPU};
      \node[below = 0.1cm of my plots c2r1.south] {Executed on iGPU};
      \node[below = 0.1cm of my plots c3r1.south] {Executed on dGPU};
      \node[below = 0.1cm of my plots c4r1.south] {Executed on MIC};               
      \path (top|-current bounding box.south)--
      coordinate(legendpos)
      (bot|-current bounding box.south);
      \matrix[
      matrix of nodes,
      anchor=south,
      draw,
      inner sep=0.2em,
      draw
      ]at([yshift=-5ex]legendpos)
      {
        \ref{plots:cpu_optimized}  & CPU Optimized  &[5pt]
        \ref{plots:igpu_optimized} & iGPU Optimized &[5pt]
        \ref{plots:dgpu_optimized} & dGPU Optimized &[5pt]
        \ref{plots:phi_optimized}  & MIC Optimized &[5pt]
        \ref{plots:feature_wise_optimized} & Learned\\};
    \end{tikzpicture}

%% file: tpch_query1.tex
    \begin{tikzpicture}[every text node part/.style={align=center}, font=\barchartfontsize,        
      bar width=10pt,
      ]
      \begin{groupplot}[group style={group name=my plots,group size= 4 by 2,vertical sep=1.5cm },height=\barchartheight,width=4.5cm,cycle list name=barplot cycle list,
        area legend, 
        ybar=0pt]
        \nextgroupplot[legend style={at={(0.38,-1)}, anchor=south}, ylabel={Execution\\ Time in s}, xtick=\empty, xticklabels=\empty,xmin=-3, xmax=3, ymin=0, ymax=400, bar width=10pt, cycle list name=barplot cycle list,   area legend,
        nodes near coords={\pgfkeys{/pgf/fpu}\pgfmathparse{pow(e,\pgfplotspointmeta)}\pgfmathprintnumber[precision=2,zerofill,fixed]{\pgfmathresult}}, 
        log ticks with fixed point,
        ymode=log,
        log origin=infty,  
        every node near coord/.append style={rotate=90, anchor=west, /pgf/number format/precision=2, xshift=0.00cm},  
        ]
        \addplot+[
        error bars/.cd,
        y dir=both,
        y explicit
        ]
        table[
        x=id,
        y=Mean,
        y error=Stdev,
        col sep=semicolon
        ]
        {cpu_processor_cpu_optimized.tpch_query1.report.csv};	 \label{plots:cpu_optimized}
        \addplot+[
        nodes near coords={{\pgfkeys{/pgf/fpu}\pgfmathparse{pow(e,\pgfplotspointmeta)}\pgfmathprintnumber[precision=2,zerofill,fixed]{\pgfmathresult}}},
        error bars/.cd,
        y dir=both,
        y explicit
        ]
        table[
        x=id,
        y=Mean,
        y error=Stdev,
        col sep=semicolon
        ]
        {cpu_processor_igpu_optimized.tpch_query1.report.csv}; \label{plots:igpu_optimized}	
        \addplot+[
        nodes near coords={{\pgfkeys{/pgf/fpu}\pgfmathparse{pow(e,\pgfplotspointmeta)}\pgfmathprintnumber[precision=2,zerofill,fixed]{\pgfmathresult}}},
        error bars/.cd,
        y dir=both,
        y explicit
        ]
        table[
        x=id,
        y=Mean,
        y error=Stdev,
        col sep=semicolon
        ]
        {cpu_processor_dgpu_optimized.tpch_query1.report.csv}; \label{plots:dgpu_optimized}
        \addplot+[
        nodes near coords={{\pgfkeys{/pgf/fpu}\pgfmathparse{pow(e,\pgfplotspointmeta)}\pgfmathprintnumber[precision=2,zerofill,fixed]{\pgfmathresult}}},
        error bars/.cd,
        y dir=both,
        y explicit
        ]
        table[
        x=id,
        y=Mean,
        y error=Stdev,
        col sep=semicolon
        ]
        {cpu_processor_phi_optimized.tpch_query1.report.csv}; \label{plots:phi_optimized}
        \addplot+[
        error bars/.cd,
        y dir=both,
        y explicit
        ]
        coordinates
        {
          (0,0)
        };
        \addplot+[
        error bars/.cd,
        y dir=both,
        y explicit
        ]
        table[
        x=id,
        y=Mean,
        y error=Stdev,
        col sep=semicolon
        ]
        {cpu_processor_feature_wise.tpch_query1.report.csv}; \label{plots:feature_wise_optimized}
        \coordinate (top) at (rel axis cs:0,1);
        \nextgroupplot[xtick=\empty, xticklabels=\empty, xmin=-3, xmax=3, ymin=0, ymax=900, bar width=10pt, cycle list name=barplot cycle list, area legend,
        nodes near coords={\pgfkeys{/pgf/fpu}\pgfmathparse{pow(e,\pgfplotspointmeta)}\pgfmathprintnumber[precision=2,zerofill,fixed]{\pgfmathresult}}, 
        log ticks with fixed point,
        ymode=log,
        log origin=infty, 
        every node near coord/.append style={rotate=90, anchor=west, /pgf/number format/precision=2, xshift=0.00cm},  
        ]
        \addplot+[
        nodes near coords={\rotatebox{-90}{\pgfkeys{/pgf/fpu}\pgfmathparse{pow(e,\pgfplotspointmeta)}\pgfmathprintnumber[precision=1,zerofill,fixed]{\pgfmathresult}}},
        error bars/.cd,
        y dir=both,
        y explicit,
        ]
        table[
        x=id,
        y=Mean,
        y error=Stdev,
        col sep=semicolon
        ]
        {igpu_processor_cpu_optimized.tpch_query1.report.csv};
        \addplot+[
        error bars/.cd,
        y dir=both,
        y explicit
        ]
        table[
        x=id,
        y=Mean,
        y error=Stdev,
        col sep=semicolon
        ]
        {igpu_processor_igpu_optimized.tpch_query1.report.csv};
        \addplot+[
        error bars/.cd,
        y dir=both,
        y explicit
        ]
        table[
        x=id,
        y=Mean,
        y error=Stdev,
        col sep=semicolon
        ]
        {igpu_processor_dgpu_optimized.tpch_query1.report.csv};  
        \addplot+[
        error bars/.cd,
        y dir=both,
        y explicit
        ]
        table[
        x=id,
        y=Mean,
        y error=Stdev,
        col sep=semicolon
        ]
        {igpu_processor_phi_optimized.tpch_query1.report.csv};     
        \addplot+[
        error bars/.cd,
        y dir=both,
        y explicit
        ]
        coordinates
        {
          (0,0)
        };
        \addplot+[
        error bars/.cd,
        y dir=both,
        y explicit
        ]
        table[
        x=id,
        y=Mean,
        y error=Stdev,
        col sep=semicolon
        ]
        {igpu_processor_feature_wise.tpch_query1.report.csv};
        \nextgroupplot[xtick=\empty, xticklabels=\empty,xmin=-3, xmax=3, ymin=0, ymax=800, bar width=10pt, cycle list name=barplot cycle list,   area legend,
        nodes near coords={\pgfkeys{/pgf/fpu}\pgfmathparse{pow(e,\pgfplotspointmeta)}\pgfmathprintnumber[precision=2,zerofill,fixed]{\pgfmathresult}}, 
        log ticks with fixed point,
        ymode=log,
        log origin=infty,                           
        every node near coord/.append style={rotate=90, anchor=west, /pgf/number format/precision=2, xshift=0.00cm},  
        ]
        \addplot+[
        nodes near coords={\rotatebox{-90}{\pgfkeys{/pgf/fpu}\pgfmathparse{pow(e,\pgfplotspointmeta)}\pgfmathprintnumber[precision=1,zerofill,fixed]{\pgfmathresult}}},
        error bars/.cd,
        y dir=both,
        y explicit
        ]
        table[
        x=id,
        y=Mean,
        y error=Stdev,
        col sep=semicolon
        ]
        {dgpu_processor_cpu_optimized.tpch_query1.report.csv};
        \addplot+[
        error bars/.cd,
        y dir=both,
        y explicit
        ]
        table[
        x=id,
        y=Mean,
        y error=Stdev,
        col sep=semicolon
        ]
        {dgpu_processor_igpu_optimized.tpch_query1.report.csv};
        \addplot+[
        error bars/.cd,
        y dir=both,
        y explicit
        ]
        table[
        x=id,
        y=Mean,
        y error=Stdev,
        col sep=semicolon
        ]
        {dgpu_processor_dgpu_optimized.tpch_query1.report.csv};  
        \addplot+[
        error bars/.cd,
        y dir=both,
        y explicit
        ]
        table[
        x=id,
        y=Mean,
        y error=Stdev,
        col sep=semicolon
        ]
        {dgpu_processor_phi_optimized.tpch_query1.report.csv};    
        \addplot+[
        error bars/.cd,
        y dir=both,
        y explicit
        ]
        coordinates
        {
          (0,0)
        };
        \addplot+[
        error bars/.cd,
        y dir=both,
        y explicit
        ]
        table[
        x=id,
        y=Mean,
        y error=Stdev,
        col sep=semicolon
        ]
        {dgpu_processor_feature_wise.tpch_query1.report.csv};
        \nextgroupplot[xtick=\empty, xticklabels=\empty, xmin=-3, xmax=3, ymin=0, ymax=700, bar width=10pt, cycle list name=barplot cycle list,   area legend,
        nodes near coords={\pgfkeys{/pgf/fpu}\pgfmathparse{pow(e,\pgfplotspointmeta)}\pgfmathprintnumber[precision=2,zerofill,fixed]{\pgfmathresult}}, 
        log ticks with fixed point,
        ymode=log,
        log origin=infty, 
        every node near coord/.append style={rotate=90, anchor=west, /pgf/number format/precision=2, xshift=0.00cm},  
        ]
        \addplot+[
        nodes near coords={{\pgfkeys{/pgf/fpu}\pgfmathparse{pow(e,\pgfplotspointmeta)}\pgfmathprintnumber[precision=1,zerofill,fixed]{\pgfmathresult}}},
        error bars/.cd,
        y dir=both,
        y explicit,
        ]
        table[
        x=id,
        y=Mean,
        y error=Stdev,
        col sep=semicolon
        ]
        {phi_processor_cpu_optimized.tpch_query1.report.csv};
        \addplot+[
        nodes near coords={{\pgfkeys{/pgf/fpu}\pgfmathparse{pow(e,\pgfplotspointmeta)}\pgfmathprintnumber[precision=2,zerofill,fixed]{\pgfmathresult}}},
        error bars/.cd,
        y dir=both,
        y explicit
        ]
        table[
        x=id,
        y=Mean,
        y error=Stdev,
        col sep=semicolon
        ]
        {phi_processor_igpu_optimized.tpch_query1.report.csv};
        \addplot+[
        nodes near coords={{\pgfkeys{/pgf/fpu}\pgfmathparse{pow(e,\pgfplotspointmeta)}\pgfmathprintnumber[precision=2,zerofill,fixed]{\pgfmathresult}}},
        error bars/.cd,
        y dir=both,
        y explicit
        ]
        table[
        x=id,
        y=Mean,
        y error=Stdev,
        col sep=semicolon
        ]
        {phi_processor_dgpu_optimized.tpch_query1.report.csv};  
        \addplot+[
        error bars/.cd,
        y dir=both,
        y explicit
        ]
        table[
        x=id,
        y=Mean,
        y error=Stdev,
        col sep=semicolon
        ]
        {phi_processor_phi_optimized.tpch_query1.report.csv}; 
        \addplot+[
        error bars/.cd,
        y dir=both,
        y explicit
        ]
        coordinates
        {
          (0,0)
        };
        \addplot+[
        error bars/.cd,
        y dir=both,
        y explicit
        ]
        table[
        x=id,
        y=Mean,
        y error=Stdev,
        col sep=semicolon
        ]
        {phi_processor_feature_wise.tpch_query1.report.csv};
        \coordinate (bot) at (rel axis cs:1,0);
      \end{groupplot}
      \node[below = 0.1cm of my plots c1r1.south] {Executed on CPU};
      \node[below = 0.1cm of my plots c2r1.south] {Executed on iGPU};
      \node[below = 0.1cm of my plots c3r1.south] {Executed on dGPU};
      \node[below = 0.1cm of my plots c4r1.south] {Executed on MIC};               
      \path (top|-current bounding box.south)--
      coordinate(legendpos)
      (bot|-current bounding box.south);
      \matrix[
      matrix of nodes,
      anchor=south,
      draw,
      inner sep=0.2em,
      draw
      ]at([yshift=-5ex]legendpos)
      {
        \ref{plots:cpu_optimized}  & CPU Optimized  &[5pt]
        \ref{plots:igpu_optimized} & iGPU Optimized &[5pt]
        \ref{plots:dgpu_optimized} & dGPU Optimized &[5pt]
        \ref{plots:phi_optimized}  & MIC Optimized &[5pt]
        \ref{plots:feature_wise_optimized} & Learned\\};
    \end{tikzpicture}

%% file: query_ssb41.tex
    \begin{tikzpicture}[every text node part/.style={align=center}, font=\barchartfontsize,        
      bar width=10pt,
      ]
      \begin{groupplot}[group style={group name=my plots,group size= 4 by 2,vertical sep=1.5cm },height=\barchartheight,width=4.5cm,cycle list name=barplot cycle list,
        area legend, 
        ybar=0pt]
        \nextgroupplot[legend style={at={(0.38,-1)}, anchor=south}, ylabel={Execution\\ Time in s}, xtick=\empty, xticklabels=\empty,xmin=-3, xmax=3, ymin=0, ymax=110, bar width=10pt, cycle list name=barplot cycle list,   area legend,
        nodes near coords={\pgfkeys{/pgf/fpu}\pgfmathparse{pow(e,\pgfplotspointmeta)}\pgfmathprintnumber[precision=2,zerofill,fixed]{\pgfmathresult}}, 
        log ticks with fixed point,
        ymode=log,
        log origin=infty,  
        every node near coord/.append style={rotate=90, anchor=west, /pgf/number format/precision=2, xshift=0.00cm},  
        ]
        \addplot+[
        error bars/.cd,
        y dir=both,
        y explicit
        ]
        table[
        x=id,
        y=Mean,
        y error=Stdev,
        col sep=semicolon
        ]
        {cpu_processor_cpu_optimized.ssb41.report.csv};	 \label{plots:cpu_optimized}
        \addplot+[
        nodes near coords={{\pgfkeys{/pgf/fpu}\pgfmathparse{pow(e,\pgfplotspointmeta)}\pgfmathprintnumber[precision=2,zerofill,fixed]{\pgfmathresult}}},
        error bars/.cd,
        y dir=both,
        y explicit
        ]
        table[
        x=id,
        y=Mean,
        y error=Stdev,
        col sep=semicolon
        ]
        {cpu_processor_igpu_optimized.ssb41.report.csv}; \label{plots:igpu_optimized}	
        \addplot+[
        nodes near coords={{\pgfkeys{/pgf/fpu}\pgfmathparse{pow(e,\pgfplotspointmeta)}\pgfmathprintnumber[precision=2,zerofill,fixed]{\pgfmathresult}}},
        error bars/.cd,
        y dir=both,
        y explicit
        ]
        table[
        x=id,
        y=Mean,
        y error=Stdev,
        col sep=semicolon
        ]
        {cpu_processor_dgpu_optimized.ssb41.report.csv}; \label{plots:dgpu_optimized}
        \addplot+[
        nodes near coords={{\pgfkeys{/pgf/fpu}\pgfmathparse{pow(e,\pgfplotspointmeta)}\pgfmathprintnumber[precision=2,zerofill,fixed]{\pgfmathresult}}},
        error bars/.cd,
        y dir=both,
        y explicit
        ]
        table[
        x=id,
        y=Mean,
        y error=Stdev,
        col sep=semicolon
        ]
        {cpu_processor_phi_optimized.ssb41.report.csv}; \label{plots:phi_optimized}
        \addplot+[
        error bars/.cd,
        y dir=both,
        y explicit
        ]
        coordinates
        {
          (0,0)
        };
        \addplot+[
        error bars/.cd,
        y dir=both,
        y explicit
        ]
        table[
        x=id,
        y=Mean,
        y error=Stdev,
        col sep=semicolon
        ]
        {cpu_processor_feature_wise.ssb41.report.csv}; \label{plots:feature_wise_optimized}
        \coordinate (top) at (rel axis cs:0,1);
        \nextgroupplot[xtick=\empty, xticklabels=\empty, xmin=-3, xmax=3, ymin=0, ymax=400, bar width=10pt, cycle list name=barplot cycle list, area legend,
        nodes near coords={\pgfkeys{/pgf/fpu}\pgfmathparse{pow(e,\pgfplotspointmeta)}\pgfmathprintnumber[precision=2,zerofill,fixed]{\pgfmathresult}}, 
        log ticks with fixed point,
        ymode=log,
        log origin=infty, 
        every node near coord/.append style={rotate=90, anchor=west, /pgf/number format/precision=2, xshift=0.00cm},  
        ]
        \addplot+[
        nodes near coords={\rotatebox{-90}{\pgfkeys{/pgf/fpu}\pgfmathparse{pow(e,\pgfplotspointmeta)}\pgfmathprintnumber[precision=1,zerofill,fixed]{\pgfmathresult}}},
        error bars/.cd,
        y dir=both,
        y explicit,
        ]
        table[
        x=id,
        y=Mean,
        y error=Stdev,
        col sep=semicolon
        ]
        {igpu_processor_cpu_optimized.ssb41.report.csv};
        \addplot+[
        error bars/.cd,
        y dir=both,
        y explicit
        ]
        table[
        x=id,
        y=Mean,
        y error=Stdev,
        col sep=semicolon
        ]
        {igpu_processor_igpu_optimized.ssb41.report.csv};
        \addplot+[
        error bars/.cd,
        y dir=both,
        y explicit
        ]
        table[
        x=id,
        y=Mean,
        y error=Stdev,
        col sep=semicolon
        ]
        {igpu_processor_dgpu_optimized.ssb41.report.csv};  
        \addplot+[
        error bars/.cd,
        y dir=both,
        y explicit
        ]
        table[
        x=id,
        y=Mean,
        y error=Stdev,
        col sep=semicolon
        ]
        {igpu_processor_phi_optimized.ssb41.report.csv};     
        \addplot+[
        error bars/.cd,
        y dir=both,
        y explicit
        ]
        coordinates
        {
          (0,0)
        };
        \addplot+[
        error bars/.cd,
        y dir=both,
        y explicit
        ]
        table[
        x=id,
        y=Mean,
        y error=Stdev,
        col sep=semicolon
        ]
        {igpu_processor_feature_wise.ssb41.report.csv};
        \nextgroupplot[xtick=\empty, xticklabels=\empty,xmin=-3, xmax=3, ymin=0, ymax=600, bar width=10pt, cycle list name=barplot cycle list,   area legend,
        nodes near coords={\pgfkeys{/pgf/fpu}\pgfmathparse{pow(e,\pgfplotspointmeta)}\pgfmathprintnumber[precision=2,zerofill,fixed]{\pgfmathresult}}, 
        log ticks with fixed point,
        ymode=log,
        log origin=infty,                           
        every node near coord/.append style={rotate=90, anchor=west, /pgf/number format/precision=2, xshift=0.00cm},  
        ]
        \addplot+[
        nodes near coords={\rotatebox{-90}{\pgfkeys{/pgf/fpu}\pgfmathparse{pow(e,\pgfplotspointmeta)}\pgfmathprintnumber[precision=1,zerofill,fixed]{\pgfmathresult}}},
        error bars/.cd,
        y dir=both,
        y explicit
        ]
        table[
        x=id,
        y=Mean,
        y error=Stdev,
        col sep=semicolon
        ]
        {dgpu_processor_cpu_optimized.ssb41.report.csv};
        \addplot+[
        error bars/.cd,
        y dir=both,
        y explicit
        ]
        table[
        x=id,
        y=Mean,
        y error=Stdev,
        col sep=semicolon
        ]
        {dgpu_processor_igpu_optimized.ssb41.report.csv};
        \addplot+[
        error bars/.cd,
        y dir=both,
        y explicit
        ]
        table[
        x=id,
        y=Mean,
        y error=Stdev,
        col sep=semicolon
        ]
        {dgpu_processor_dgpu_optimized.ssb41.report.csv};  
        \addplot+[
        error bars/.cd,
        y dir=both,
        y explicit
        ]
        table[
        x=id,
        y=Mean,
        y error=Stdev,
        col sep=semicolon
        ]
        {dgpu_processor_phi_optimized.ssb41.report.csv};    
        \addplot+[
        error bars/.cd,
        y dir=both,
        y explicit
        ]
        coordinates
        {
          (0,0)
        };
        \addplot+[
        error bars/.cd,
        y dir=both,
        y explicit
        ]
        table[
        x=id,
        y=Mean,
        y error=Stdev,
        col sep=semicolon
        ]
        {dgpu_processor_feature_wise.ssb41.report.csv};
        \nextgroupplot[xtick=\empty, xticklabels=\empty, xmin=-3, xmax=3, ymin=0, ymax=100, bar width=10pt, cycle list name=barplot cycle list,   area legend,
        nodes near coords={\pgfkeys{/pgf/fpu}\pgfmathparse{pow(e,\pgfplotspointmeta)}\pgfmathprintnumber[precision=2,zerofill,fixed]{\pgfmathresult}}, 
        log ticks with fixed point,
        ymode=log,
        log origin=infty, 
        every node near coord/.append style={rotate=90, anchor=west, /pgf/number format/precision=2, xshift=0.00cm},  
        ]
        \addplot+[
        nodes near coords={{\pgfkeys{/pgf/fpu}\pgfmathparse{pow(e,\pgfplotspointmeta)}\pgfmathprintnumber[precision=1,zerofill,fixed]{\pgfmathresult}}},
        error bars/.cd,
        y dir=both,
        y explicit,
        ]
        table[
        x=id,
        y=Mean,
        y error=Stdev,
        col sep=semicolon
        ]
        {phi_processor_cpu_optimized.ssb41.report.csv};
        \addplot+[
        nodes near coords={{\pgfkeys{/pgf/fpu}\pgfmathparse{pow(e,\pgfplotspointmeta)}\pgfmathprintnumber[precision=2,zerofill,fixed]{\pgfmathresult}}},
        error bars/.cd,
        y dir=both,
        y explicit
        ]
        table[
        x=id,
        y=Mean,
        y error=Stdev,
        col sep=semicolon
        ]
        {phi_processor_igpu_optimized.ssb41.report.csv};
        \addplot+[
        nodes near coords={{\pgfkeys{/pgf/fpu}\pgfmathparse{pow(e,\pgfplotspointmeta)}\pgfmathprintnumber[precision=2,zerofill,fixed]{\pgfmathresult}}},
        error bars/.cd,
        y dir=both,
        y explicit
        ]
        table[
        x=id,
        y=Mean,
        y error=Stdev,
        col sep=semicolon
        ]
        {phi_processor_dgpu_optimized.ssb41.report.csv};  
        \addplot+[
        error bars/.cd,
        y dir=both,
        y explicit
        ]
        table[
        x=id,
        y=Mean,
        y error=Stdev,
        col sep=semicolon
        ]
        {phi_processor_phi_optimized.ssb41.report.csv}; 
        \addplot+[
        error bars/.cd,
        y dir=both,
        y explicit
        ]
        coordinates
        {
          (0,0)
        };
        \addplot+[
        error bars/.cd,
        y dir=both,
        y explicit
        ]
        table[
        x=id,
        y=Mean,
        y error=Stdev,
        col sep=semicolon
        ]
        {phi_processor_feature_wise.ssb41.report.csv};
        \coordinate (bot) at (rel axis cs:1,0);
      \end{groupplot}
      \node[below = 0.1cm of my plots c1r1.south] {Executed on CPU};
      \node[below = 0.1cm of my plots c2r1.south] {Executed on iGPU};
      \node[below = 0.1cm of my plots c3r1.south] {Executed on dGPU};
      \node[below = 0.1cm of my plots c4r1.south] {Executed on MIC};               
      \path (top|-current bounding box.south)--
      coordinate(legendpos)
      (bot|-current bounding box.south);
      \matrix[
      matrix of nodes,
      anchor=south,
      draw,
      inner sep=0.2em,
      draw
      ]at([yshift=-5ex]legendpos)
      {
        \ref{plots:cpu_optimized}  & CPU Optimized  &[5pt]
        \ref{plots:igpu_optimized} & iGPU Optimized &[5pt]
        \ref{plots:dgpu_optimized} & dGPU Optimized &[5pt]
        \ref{plots:phi_optimized}  & MIC Optimized &[5pt]
        \ref{plots:feature_wise_optimized} & Learned\\};
    \end{tikzpicture}

%% file: table_ssbm_sf10_measurements_generic.tex
\setlength{\tabcolsep}{1pt}
\begin{table*}
\begin{center}
{
\begin{tabular}{ | c | c || >{\columncolor{myblue2}}  c  c  c  >{\columncolor{mygreen}} c | c   >{\columncolor{myblue2}}c  c >{\columncolor{mygreen}}c | c  c  >{\columncolor{myblue2}}c >{\columncolor{mygreen}}c |  }
\hline
& Hyper &  \multicolumn{4}{c|}{Executed on CPU} & \multicolumn{4}{c|}{Executed on dGPU} & \multicolumn{4}{c|}{Executed on MIC}\\
& (CPU) & cpu-o&dgpu-o&mic-o&per-q&cpu-o&dgpu-o&mic-o&per-q&cpu-o&dgpu-o&mic-o&per-q\\
\hline
\hline 
Q1.1 & 0.149 & 0.186 & 0.441 & 0.342 & 0.189 & 0.067 & 0.015 & 0.015 & 0.015 & 0.055 & 0.062 & 0.057 & 0.057\\
Q1.2 & 0.099 & 0.113 & 0.271 & 0.272 & 0.114 & 0.052 & 0.047 & 0.047 & 0.013 & 0.046 & 0.061 & 0.06 & 0.064\\
Q1.3 & 0.092 & 0.111 & 0.25 & 0.248 & 0.109 & 0.052 & 0.022 & 0.023 & 0.013 & 0.047 & 0.056 & 0.049 & 0.051\\
\hline 
Q3.2 & 0.2 & 0.21 & 2.258 & 0.885 & 0.206 & 56.697 & 0.191 & 1.724 & 0.138 & 5.021 & 0.247 & 0.221 & 0.155\\
Q3.3 & 0.146 & 0.115 & 1.467 & 0.61 & 0.114 & 53.543 & 0.073 & 0.472 & 0.052 & 3.781 & 0.14 & 0.14 & 0.114\\
Q3.4 & 0.146 & 0.111 & 1.468 & 0.615 & 0.114 & 53.646 & 0.073 & 0.471 & 0.053 & 3.795 & 0.132 & 0.128 & 0.109\\
\hline 
Q4.1 & 0.654 & 0.567 & 2.186 & 1.559 & 0.567 & 77.701 & 0.743 & 5.188 & 0.25 & 11.146 & 0.423 & 0.397 & 0.417\\
Q4.2 & 0.588 & 0.444 & 1.758 & 1.272 & 0.45 & 78.042 & 0.523 & 1.704 & 0.111 & 8.552 & 0.341 & 0.322 & 0.351\\
Q4.3 & 0.316 & 0.195 & 2.421 & 1.073 & 0.212 & 58.718 & 0.764 & 4.816 & 0.286 & 4.709 & 0.435 & 0.415 & 0.343\\
\hline 
\hline 
Q5 & 0.857 & 0.934 & 5.105 & 4.095 & 1.033 & 73.572 & 7.091 & 10.605 & 0.261 & 10.874 & 0.905 & 0.907 & 0.838\\
Q6 & 0.147 & 0.185 & 0.257 & 0.258 & 0.195 & 0.063 & 0.009 & 0.009 & 0.011 & 0.033 & 0.037 & 0.036 & 0.036\\
\hline 
\end{tabular}
}
\end{center}
	\caption{{Execution times in seconds of variants optimized for CPU (cpu-o), dGPU (dgpu-o), and MIC (mic-o) for selected queries of the star schema and TPC-H benchmark (Scale Factor 10), executed on a CPU, a dedicated GPU, and a MIC processor. 
}}
        \label{table:selected_ssb_queries}
\end{table*}